\documentclass[12pt,dviwindows]{article}

\usepackage{color,epsf,graphicx}
\usepackage{amsmath,amssymb}
\usepackage{psfrag}
\usepackage[colorlinks=true,citecolor=blue,linkcolor=red]{hyperref}

\newcounter{comment}

{\refstepcounter{comment}%
\begin{quote}
\ttfamily\small$\blacksquare$ \textbf{\underline{Comment} $\sharp$\thecomment:}}%
{\end{quote}}

{
\begin{quote}
\ttfamily\small$\blacktriangleright$ \textbf{\underline{Reply} $\sharp$\thecomment:}}%
{\end{quote}}

\newcommand{\req}[1]{(\ref{#1})}

\newcommand{\insertfig}[2]{\mbox{\epsfxsize=#1cm \epsfbox{#2.eps}}}

\newcommand{\GeV}{{\rm GeV}}
\font\cmss=cmss12 
\def\1{\hbox{{1}\kern-.25em\hbox{l}}}
\def\bfZ{\relax{\hbox{\cmss Z\kern-.4em Z}}}

\begin{document}

\begin{titlepage}

\centerline{\large \bf Towards a fitting procedure for deeply virtual Compton scattering
} \vspace{2mm}
\centerline{\large \bf  at next-to-leading order and beyond }

\vspace{7mm}

\centerline{\bf   K.~Kumeri{\v c}ki$^{a,b}$,
D.~M\"uller$^{b,c,d}$, and K.~Passek-Kumeri{\v c}ki$^{b,e}$}

\vspace{7mm} \centerline{\it $^a$Department of Physics, Faculty of
Science, University of Zagreb} \centerline{\it P.O.B. 331,
HR-10002 Zagreb, Croatia}

\vspace{4mm} \centerline{\it $^b$Institut f\"ur Theoretische
Physik, Universit\"at Regensburg} \centerline{\it D-93040
Regensburg, Germany}

\vspace{4mm} \centerline{\it $^c$Department of Physics and
Astronomy, Arizona State University} \centerline{\it Tempe, AZ
85287-1504, USA}

\vspace{4mm} \centerline{\it $^d$Institut f\"ur Theoretische Physik
II, Ruhr-Universit\"at Bochum} \centerline{\it D-44780 Bochum,
Germany}

\vspace{5mm} \centerline{\it $^e$Theoretical Physics Division,
Rudjer Bo{\v s}kovi{\'c} Institute} \centerline{\it P.O.Box 180,
HR-10002 Zagreb, Croatia}

\vspace{5mm}

\centerline{\bf Abstract}

\vspace{0.5cm}

\noindent
Combining dispersion and operator product expansion
techniques, we derive the conformal partial wave decomposition of
the virtual Compton scattering amplitude in terms of complex
conformal spin to twist-two accuracy.   The perturbation theory
predictions for the deeply virtual Compton scattering (DVCS)
amplitude are presented in next-to-leading order for both
conformal  and modified minimal subtraction scheme. Within a
conformal subtraction scheme, where we exploit predictive power of
conformal symmetry, the radiative corrections are presented up to
next-to-next-to-leading order accuracy. Here, because of the trace
anomaly, the mixing of conformal moments of generalized parton
distributions (GPD) at the three-loop level remains unknown.
Within a new proposed parameterization for GPDs, we then study the
convergence of perturbation theory and demonstrate that our
formalism is suitable for a fitting procedure of DVCS observables.

\vspace*{12mm}


\vspace{3mm}


\end{titlepage}

\tableofcontents

\newpage

\section{Introduction}

The partonic content of the nucleon, other hadrons, and nuclei
has been studied within hard inclusive processes over almost four decades.
Since partons are confined by the strong force, their response to,
e.g., an electromagnetic probe cannot be directly accessed in
experiments; the long-range interaction in their environment
remains essential. Fortunately, based on factorization theorems
\cite{Muel89}, the short- and the long-distance interaction can
often be separated. The former depends on the specific process and
can be systematically computed using perturbation theory, while
the latter is encoded in process-independent nonperturbative
quantities, e.g., parton densities. These densities have a
semiclassical interpretation within the parton picture, which
is, however, not independent of the conventions used in the
evaluation of short-distance physics. Together with an
increasing amount of experimental data, the perturbative
factorization approach leads to a deeper and more precise insight
into the hadronic world, which is essential even for the search of
new physics.

Most importantly, the factorization approach is
a tool that relates observables measured in various inclusive
processes and thus has a predictive power.
The improvement of quantitative predictions
requires, besides precise experimental measurements, also a refinement
of the theoretical approach. This includes both the perturbative
evaluation of radiative corrections to the short-distance physics
at higher orders and an understanding of the so-called power
suppressed corrections, which alter the factorization theorems. In
particular, the effort in the perturbative sector, which reached
the three-loop level, led to a quantitative understanding of the
inclusive QCD physics at the level of a few percent. In practice,
parton densities have for a long time been extracted via a theoretically
motivated functional ansatz, see, e.g., Ref.\ \cite{BroBurSch94},
depending on a number of parameters, that is globally fitted to experimental data
\cite{MarStiTho06,Ale05,Pumetal02,GluReyVog98,LeaSidSta05,BluBot02,GluReyStrVog00,GehSti95}.
In such fits, based on traditional $\chi^2$ minimization, the
error estimation using the hypothesis of linear error propagation
has become standard in the last few years
\cite{MarStiTho06,Ale05,Pumetal02,BluBot02}. To overcome drawbacks
of this traditional method, alternative frameworks of statistical inference
\cite{GieKel98} and neural network parameterization \cite{ForGarLatPic02,Roj06} have been
proposed.

In contrast to inclusive processes, the understanding
and the theoretical description of exclusive ones remain
poor. In general, there is a lack of experimental data
and so the underlying theoretical framework, e.g.,
the factorization of hadronic amplitudes in terms of
distribution amplitudes
\cite{BroLep80,BroLep81,EfrRad80,DunMue80,CheZhi84},
cannot be quantitatively tested. Moreover, the theoretical
framework at the next-to-leading order (NLO) is developed
for only a few processes
and the applicability of the
factorization approach at accessible scales is
controversially discussed in the literature \cite{IsgLle84,IsgLle89,Rad84}.
The old problem of whether all valence partons take part in the
short-distance process or only the active one
(Feynman mechanism) remains open. A way out of this
theoretical dilemma might be provided by the use of
light-cone sum rules \cite{BalBraKol89,BraFil90}.

Some years ago a new nonperturbative generalized distribution
amplitude\footnote{This notion, proposed in Ref.\
\cite{MueRobGeyDitHor94}, stood for expectation values of
light-ray operators sandwiched between vacuum and final hadronic
or between initial and final hadronic states. Today, it denotes the
distribution corresponding to the former expectation value,
while the distribution corresponding to the latter one
is called generalized parton distribution.}
was proposed as a means to access the partonic
content of hadrons \cite{MueRobGeyDitHor94,GeyDitHorMueRob88}. In
particular, it was theoretically studied in connection with deeply
virtual Compton scattering (DVCS), where the partonic content of a
hadron is probed by two photons. To leading order, the underlying
picture portrays a parton which is probed by a virtual photon, travels
near the light cone, emits a real photon, and remains a constituent
of the probed hadron. The partonic probability amplitude for such
a process is given by the generalized parton distribution (GPD)
\cite{Ji96,Ji96a,Rad96}. It has soon been realized that
diffractive vector meson electroproduction
\cite{BroFraGunMueStr94}, measured in the collider experiments H1
and ZEUS at DESY (see, e.g., Refs.\
\cite{Breetal98,Adletal99,Breetal99}),  can be described in terms of
GPDs \cite{Rad96a,ColFraStr96}, too.
The usefulness of the GPDs
has also been widely realized in connection with the spin problem,
since they encode the angular
momentum carried by the individual parton species, as explicated by the Ji's
sum rule \cite{Ji96}. During the last
decade they have become an attractive object for theoretical and
experimental investigations. Today, they are considered
as a new concept that provides a link between different fields:
exclusive and inclusive processes, perturbative and
nonperturbative physics (e.g., lattice simulations
\cite{Hagetal03,Hagetal04,Hagetal04a,Gocetal03,Gocetal05,Edwetal06} and model building
\cite{JiMelSon97,DieFelJakKro00,BroDieHwa00,ScoVen02,BofPasTra02,ChoJiKis01,TibMil02,NogTheVen02,AhmHonLiuTan06}).
For comprehensive reviews, see Refs.\ \cite{Die03a,BelRad05}.

In contrast to parton densities, which depend on the longitudinal
momentum fraction of the probed parton and on the resolution
scale, GPDs encode also transversal degrees of freedom. This
allows a three-dimensional probabilistic interpretation of the
parton distribution either in the infinite momentum
\cite{RalPir01,Bur00,Bur02,Die02,BelMue02}  or in the rest frame
\cite{BelJiFen03}. Indeed, to some extent this information can
already be extracted from present experiments. The main
theoretical complication arises from the fact that GPDs depend on
two longitudinal momentum fractions: $x$ and skewness $\eta$,
which are related to the $s$- and $t$-channel exchanges,
respectively. The former one  is either integrated out during the
convolution (in the real part of the scattering amplitude) or
identified with the latter one (in the imaginary part of the
scattering amplitude). Even if we precisely knew the modulus
and the phase of hard exclusive leptoproduction scattering
amplitudes%
\footnote{We remark that the scan of the GPD shape in a certain region
of the momentum fraction plane is possible only in the so-called
double DVCS process \cite{GuiVan02,BelMue02a,BelMue03} by
variation of the incoming photon virtuality and the lepton pair
mass.}, a complete reconstruction of the GPD would not be uniquely
possible, except with ideal data and employing evolution. We note
that it was pointed out that already the Fourier transform of
the DVCS amplitude with respect to the skewness parameter provides
an image of the target with respect to the longitudinal degrees of
freedom \cite{BroChaHarMukVar06,BroChaHarMukVar06a}. The problem
of deconvolution can be overcome to some extent either by a
realistic model, which we at present do not have, or by a
hypothesis about the form of skewness dependence. In the case of
those GPDs that in the forward kinematics reduce to parton
densities, such a hypothesis  can be tested by statistical
analysis of inclusive and exclusive data, e.g., by means of the
$\chi^2$ criteria. Certainly, if the skewness problem could be
solved, the extrapolation to the $\eta\to 0$ case would be simple and
would be a step towards the experimental access of both Ji's spin
sum rule and a three-dimensional picture of the proton.

Among processes which enable us to access GPDs at present
experiments, the DVCS is considered the theoretically cleanest
one. Indeed, the first experimental DVCS data on the beam spin
asymmetry in fixed target experiments \cite{Airetal01,Steetal01},
or the cross section, measured by the H1 and ZEUS collaborations
\cite{Adletal01,Chekanov:2003ya,Aktas:2005ty}, could be
successfully understood even in terms of oversimplified GPD
ans\"atze \cite{BelMueKir01}. In contrast, the normalization of
the cross sections of vector-meson electroproduction, predicted to LO
in the collinear factorization approach, in general overshoots the
H1 and ZEUS data. This process is widely believed to be
affected by power-suppressed contributions. They are mainly
related to the transversal size of the meson and appear so
separately as factorized contributions, which might be modelled
by themselves \cite{FraKoeStr95,GolKro05}. On the other hand, perturbative
higher-order corrections for this process are large and reduce
the size of the predicted cross sections \cite{IvaSzyKra04}.
However, to the best of our knowledge, a global NLO analysis of all
available experimental data has not been achieved so far. In our
opinion, this is the only possible way of confronting the collinear
approach with experimental findings and before this is done, a
judgement about this approach can hardly be drawn.

With increasing amount and precision of experimental DVCS data
\cite{Airetal06,Zhe06,Cheetal06,Cametal06}, there arises a need
for a better theoretical understanding of this process. Certainly,
for the phenomenology of GPDs  it is essential to include and estimate
perturbative and power-suppressed contributions as well
as to introduce a flexible parameterization of them. Besides a
qualitative or a semianalytic understanding of observables in
dependence on GPDs, see here, for instance, Refs.
\cite{BelMueKir01,KirMue03}, also fast and stable numerical
routines are needed for a fitting procedure.

So far the perturbative contributions to the DVCS
have been worked out to NLO accuracy
\cite{BelMue97a,ManPilSteVanWei97,JiOsb97,JiOsb98,BelMue00},
including the evolution \cite{BelMue98c,BelFreMue99,BelFreMue00}.
There exist numerical routines in the momentum fraction space
\cite{FreMcD01b,Vin06,Sch06}. In this space the GPD ansatz is given in
terms of a spectral function \cite{MueRobGeyDitHor94,Rad97}. Here
one has to model the functional dependencies on two momentum
fraction variables and the momentum transfer squared.
Alternatively, one can work with the conformal moments of GPDs
\cite{BelGeyMueSch97}, which diagonalize the LO evolution
equation. This offers the possibility of a flexible GPD
parameterization, which covers the complete set of degrees of
freedom and makes direct contact with lattice results. The
problem here is rather to find an appropriate and physically
motivated truncation in the parameter space. A partial wave
expansion with respect to the angular momentum, the so-called
dual GPD parameterization, has been proposed \cite{PolShu02},
where the partial wave amplitudes are those of the mesonic
exchanges in the $t$-channel with given angular momentum. This
also allows one to use the angular momentum as an expansion
parameter \cite{GuzTec06} and Regge phenomenology as a
guideline towards realistic GPD ans\"atze.

Unfortunately, if one works with discrete conformal moments, the
GPDs are expanded in terms of (mathematical) distributions, which
live only in the so-called central region ($-\eta\le x\le \eta$)
of the whole support ($-1\le x\le 1$). This is quite analogous to
the expansion of parton densities with respect to the Dirac function
and its derivatives, which live at the point $x=0$. A variety of
approaches have been proposed to resum this formal series and thus
to restore the correct support of GPDs: smearing the expansion
\cite{BelGeyMueSch97,PolShu02}, an integral transformation
\cite{Shu99,Nor00}, Fourier transformation in the light-cone position space
\cite{BalBra89,KivMan99b,ManKirSch05,KirManSch05a}, and a
Sommerfeld--Watson transformation \cite{MueSch05}. Note that all
these approaches should be mathematically equivalent and, in
particular, those of Refs.\ \cite{ManKirSch05} and \cite{MueSch05}
lead to essentially the same representation.  So far only the
smearing method has been extended to NLO accuracy in the
$\overline{\rm MS}$ scheme; however, speed and stability
remain restricted \cite{BelMueNieSch98,BelMueNieSch98a}.

We believe that the GPD formalism that is based on the Sommerfeld-Watson
transformation and Mellin--Barnes integral representation \cite{MueSch05} is suitable
to satisfy the requirements which we spelled out. Moreover, we can
employ the power of conformal symmetry to investigate the
convergence of the perturbative series up to NNLO
\cite{Mue05a,KumMuePasSch06}.
Therefore, the fivefold goal of this article is
\begin{itemize}
\item
to present a detailed derivation,
based on analyticity and short-distance operator product expansion,
of the Mellin--Barnes integral representation for the twist-two Compton
form factors (CFFs),
\item
to present the radiative corrections up to NNLO
obtained using the predictive power of conformal symmetry,
as well as to present the radiative corrections in the standard
$\overline{\rm MS}$ scheme up to NLO accuracy, including the evolution,
\item to propose an intuitive ansatz for conformal GPD moments,
\item to investigate the convergency properties of the perturbative
      expansion,
\item to demonstrate the usefulness of the GPD fitting procedure
based on this formalism, as well as to discuss the
requirements for a GPD ansatz.
\end{itemize}
The outline is as follows. In Sect.\ \ref{Sec-GenFor} we
introduce the Compton scattering tensor and its decomposition in
leading twist-two Compton form factors (CFFs). Employing their
analytic properties, we derive  dispersion relations for CFFs. In
Sect.\ \ref{Sec-OPE} these dispersion relations together with the
short-distance operator product expansion are used to represent
the CFFs as Mellin--Barnes integrals. To diagonalize the
evolution operator, we then introduce the conformal partial wave
expansion of CFFs. In Sect.\ \ref{Sec-PreConSym} we employ
conformal symmetry in the  perturbative QCD sector to find the
Wilson coefficients for CFFs, expanded up to NNLO in the
coupling. We also employ the conformal partial wave expansion for
the representation of the NLO corrections in the $\overline{\rm
MS}$ scheme. An intuitive ansatz for the conformal GPD moments is introduced in Sect.\
\ref{Sec-GPDpar}. It is based on the internal duality of GPDs and consists of an SO(3)
partial wave expansion in the $t$-channel.
In Sect.\ \ref{Sec-PerCorCrosSec} we discuss the
convergency of the perturbative series. We then demonstrate in
Sect.\ \ref{Sec-FitPro} on hand of the DVCS cross section,
measured at high energies by the H1 and ZEUS collaborations, that
our formalism is suitable for a fitting procedure that is used to
extract GPD information from experimental data. Finally, we
summarize and conclude. Three appendices contain details on our
conventions and the evaluation of conformal moments.

\section{Compton scattering tensor and Compton form factors}
\label{Sec-GenFor}

\subsection{General formalism}

We are interested in the perturbative description of hard photon
leptoproduction off a hadronic target, e.g., a proton.
Besides the Bethe-Heitler bremsstrahlungs process, parameterized
by electromagnetic form factors, the DVCS process contributes
\cite{MueRobGeyDitHor94, Ji96, Rad96}. The amplitude of the latter
is expressed by the Compton tensor. This tensor is defined  in
terms of the time-ordered product of two electromagnetic currents,
sandwiched between the initial and final hadronic state,
\begin{eqnarray}
\label{Def-ComScaTen} T_{\mu\nu} (q, P, \Delta) = \frac{i}{e^2}
\int\! d^4x\, e^{i x\cdot q} \langle P_2, S_2 | T j_\mu (x/2)
j_\nu (-x/2) | P_1, S_1 \rangle,
\end{eqnarray}
where $q = (q_1 + q_2)/2$ ($\mu$  and $q_2$ refer to the outgoing
real photon), while $P = P_1 + P_2$ and $\Delta=P_2-P_1$. The
incoming photon has a large virtuality $q_1^2=-{\cal Q}^2$ and we
require that, in the limit $-q^2 = Q^2\to \infty$, the scaling
variables
\begin{eqnarray}
\xi = \frac{Q^2}{P\cdot q}\,,\qquad \eta = -\frac{\Delta\cdot
q}{P\cdot q}\,,
\end{eqnarray}
and the momentum transfer squared $\Delta^2$  are fixed. The
dominant contributions arise then from the light-cone
singularities of the time-ordered product. This kinematics is a
generalization of the Bjorken limit, well-known from deep
inelastic scattering (DIS). In particular, if the final photon is
on-shell, i.e., for the process we are interested in, the skewness
parameter $\eta$ and the Bjorken-like scaling parameter $\xi$ are
equal to twist-two accuracy, i.e., $\eta = \xi + {\cal O}(1/{\cal
Q}^2) $. Note that in the following we assume $\Delta\cdot P =
P_2^2-P_1^2=0$.

In the generalized Bjorken limit, we can employ the
OPE to evaluate the Compton scattering tensor
(\ref{Def-ComScaTen}). Its dominant part is given by matrix
elements of leading twist-two operators. The use of the
light-cone expansion, a resummed version of the short-distance
operator product expansion, allows a straightforward evaluation
\cite{MueRobGeyDitHor94}. On the other hand, one might employ the
short-distance expansion, which yields a Taylor expansion of the
Compton scattering tensor with respect to the variable $1/\xi$.
Since this expansion converges only in the unphysical region,
i.e., $\xi >1$, we need in addition a dispersion relation that
connects the Mellin moments in the physical region with the
short-distance expansion. This technique well-known from deep inelastic
scattering has been adopted for nonforward kinematics, e.g., in
Ref.\ \cite{Che97}. Below we use it within  a special
short-distance OPE, namely, the conformally covariant one.

Let us introduce a parameterization of the Compton tensor in terms
of the so-called Compton form factors (CFFs), which has been employed
for the evaluation of the differential cross section
\cite{BelMueNieSch00,BelMueKir01}. To leading twist-two accuracy
it reads
\begin{eqnarray}
\label{decom-T} T_{\mu\nu} (q,P,\Delta) &=& - \widetilde{g}^{\rm
T}_{\mu\nu} \frac{q_\sigma V^\sigma}{P\cdot q} - i
\widetilde{\epsilon}_{\mu \nu q P} \frac{q_\sigma
A^\sigma}{(P\cdot q)^2} + \cdots ,
\end{eqnarray}
where $\widetilde{g}^{\rm T}_{\mu\nu}$ and
$\widetilde{\epsilon}_{\mu \nu q P} \equiv
\widetilde{\epsilon}_{\mu \nu \alpha \beta} q^\alpha P^\beta$  are
the transversal part of the metric%
\footnote{The transversal part of the metric tensor reads ${g}^{\rm
T}_{\mu\nu}={g}_{\mu\nu}-n_\mu n'_{\nu} -n_\nu n'_{\mu}$, where
$n_\mu=(1,0,0,\pm1)/\sqrt{2}$ and $n'_\mu=(1,0,0,\mp1)/\sqrt{2}$ and at twist-2 we
have $n_\mu n'_{\nu}=q_\mu P_\nu/(q\cdot P)$.} and the Levi-Civita
tensor, respectively, which are contracted $\widetilde{X}_{\mu\nu}
\equiv {\cal P}_{\mu\rho}\, X^{\rho\sigma}\, {\cal P}_{\sigma\nu}$
with projectors
\begin{eqnarray}
{\cal P}^{\alpha \beta} = g^{\alpha\beta} -\frac{q_1^\alpha
q^\beta_2}{q_1\cdot q_2}
\end{eqnarray}
to ensure current conservation
\cite{BelMueNieSch00}. The ellipsis indicate
terms that are finally power suppressed%
\footnote{Here we have neglected a further twist-two Compton form
factor that can occur for virtual photons that are longitudinal polarized.
Obviously, it does not contribute to the DVCS amplitude, where the
final photon state is of course transversally polarized.} in the
DVCS amplitude, or are determined by the gluon transversity GPD.
The latter is a twist-two contribution that enters at NLO,
evaluated in \cite{HooJi98,BelMue00,Die01}, and is tied to a
specific azimuthal angular dependence in the cross section
\cite{DieGouPirRal97,BelMueKir01}. Hence,  using an appropriate
definition of observables, it can be separated from the other
twist-two (and twist-three) contributions. For the time being we
consider its perturbative corrections beyond NLO more as an
academic issue and will not evaluate it here.

In the parity even sector the vector
\begin{eqnarray}
\label{dec-FF-V} V^{\sigma} = \overline{U} (P_2, S_2) \left( {\cal
H} \gamma^\sigma + {\cal E} \frac{i\sigma^{\sigma\rho}
\Delta_\rho}{2M} \right) U (P_1, S_1) + \cdots\, ,
\end{eqnarray}
is decomposed into the target helicity  conserving
CFF ${\cal H}$ and the helicity flip one ${\cal E}$.
Analogously, the axial-vector
\begin{eqnarray}
\label{dec-FF-A} A^{\sigma} = \overline{U} (P_2, S_2) \left(
\widetilde{\cal H} \gamma^\sigma\gamma_5 + \widetilde{\cal E}
\frac{\Delta^\sigma \gamma_5}{2M} \right) U (P_1, S_1) + \cdots ,
\end{eqnarray}
is parameterized in terms of $\widetilde{\cal
H}$ and $\widetilde{\cal E}$, where again higher-twist
contributions are neglected. The normalization of the spinors is
$\overline{U} (P, S) \gamma^\sigma U (P, S) = 2 P^\sigma$.

It is convenient to introduce the nomenclature
\begin{eqnarray}
\label{Def-CFFs}
 {\cal F} = \{{\cal F}^{\rm V},{\cal F}^{\rm
A}\}\,,\qquad {\cal F}^{\rm V} = \{{\cal H},{\cal E}\}\,,\quad
{\cal F}^{\rm A} = \{\widetilde{\cal H},\widetilde{\cal E}\}\,,
\end{eqnarray}
with the following choice of arguments: ${\cal F}(\nu, \vartheta,
\Delta^2, Q^2)$. The variable
\begin{eqnarray} \nu = \frac{P\cdot q}{\sqrt{P^2}}  = \frac{1}{\xi}
\frac{Q^2}{\sqrt{P^2}} =
\frac{W^2-u}{2\sqrt{P^2}}\quad\mbox{with}\quad W^2= (P_1+q_1)^2\,,\;
u= (P_1-q_2)^2
\end{eqnarray}
is in the Breit frame simply given by  $q_0$, the conjugate
variable of the time $x_0$. It is an appropriate generalization of
the photon energy in the rest frame for DIS. Instead of the
skewness variable $\eta$, depending on $\nu$, we will employ the
virtual photon mass asymmetry
\begin{eqnarray}
\label{Def-vartheta}
\vartheta= \frac{q_1^2-q_2^2}{q_1^2+q_2^2} = \frac{\eta}{\xi} +
{\cal O}(\Delta^2/Q^2)\,.
\end{eqnarray}
Obviously, $\vartheta=0$ in the forward case and $\vartheta=1$ for
DVCS.

\subsection{Analyticity and dispersion relations}

\begin{figure}[t]
\begin{center}
\mbox{
\begin{picture}(450,150)(0,0)
\put(-15,20){\insertfig{7.4}{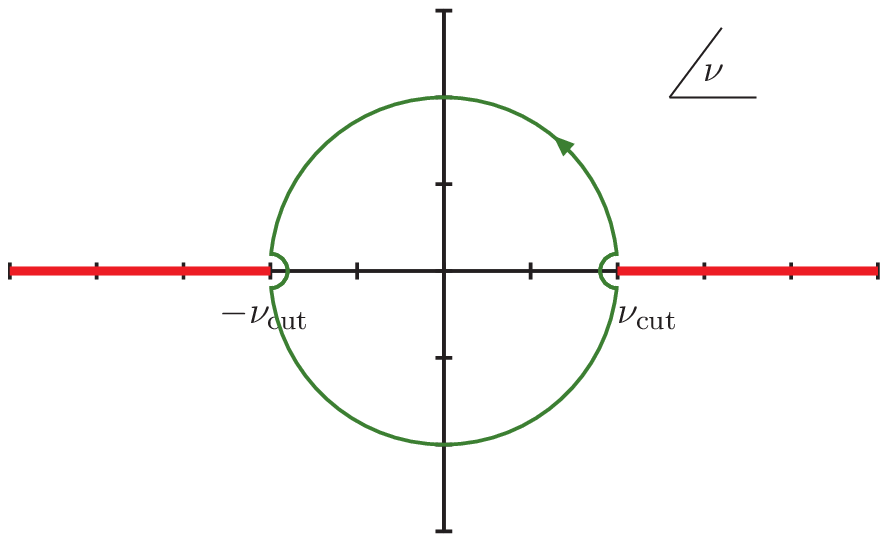}}
\put(220,0){\insertfig{8.7}{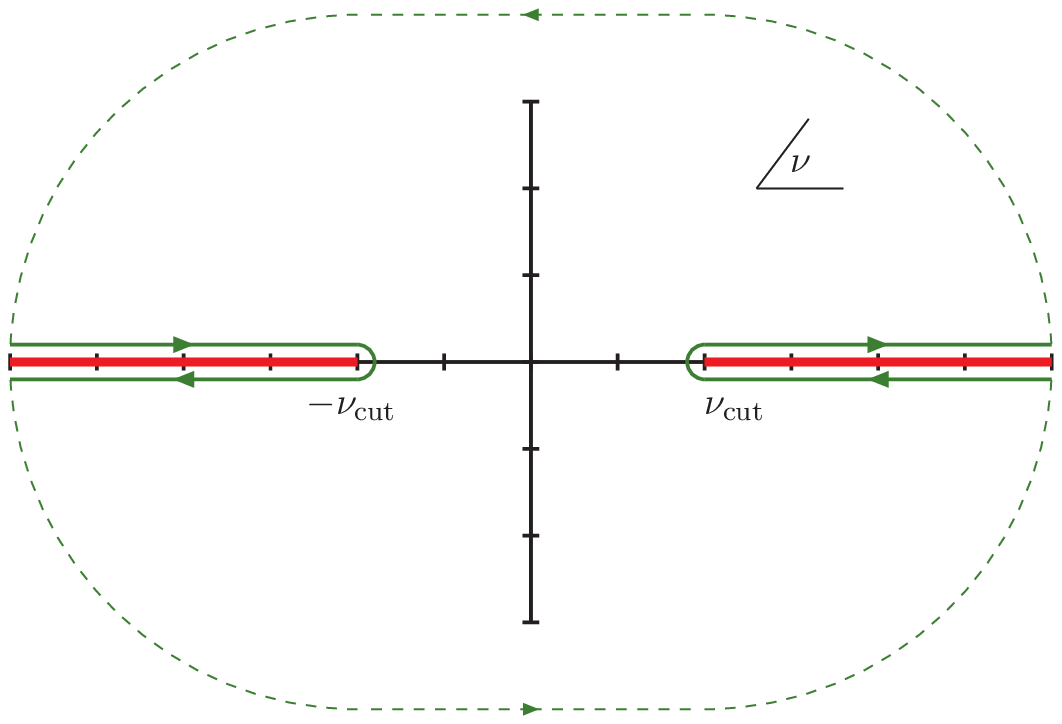}}
\end{picture}
}
\end{center}
\caption{ \label{Fig-Cou} The integration contour used in  Eq.\
(\ref{Def-DisRel-0}) and its deformation, employed to derive the
dispersion relation (\ref{Def-DisRel}), are sketched in the left
and right panel, respectively. Contribution of dashed segments vanishes.}
\end{figure}

Based on the analyticity of the Compton tensor
(\ref{Def-ComScaTen}), now we derive a dispersion relation for the
CFFs, introduced in Eqs.\ (\ref{dec-FF-V}) and (\ref{dec-FF-A}).
{F}rom the definition of the Compton tensor (\ref{Def-ComScaTen}),
we read off for complex valued $q$ the following properties, see,
for instance, Refs.\ \cite{BjoDreBoo,RomBoo},
\begin{eqnarray}
\label{T-SchRefPri} \left[T_{\mu\nu} (q, P,
\Delta;S_1,S_2)\right]^\ast &\!\!\!=\!\!\!& T_{\nu\mu} (q^\ast, P,
-\Delta;S_2,S_1)\,,
\\
\label{T-BosSym} T_{\mu\nu} (q, P, \Delta;S_1,S_2) &\!\!\!=\!\!\!&
T_{\nu\mu} (-q, P, \Delta;S_1,S_2)\,.
\end{eqnarray}
The former of these equations can be viewed as a generalization of
the Schwarz reflection principle,
whereas the latter equation stems from crossing
symmetry under the Bose exchange $(q_1,q_2,\mu,\nu)
\leftrightarrow (-q_2,-q_1,\nu,\mu)$. Furthermore, in the spacelike
region, i.e., $0 < Q^2$, and for $0\le  -\Delta^2 \le
-\Delta^2_{\rm Max}$, where the upper limit is constrained by
kinematics, the Compton tensor is holomorphic in $\nu$
except for the branch cuts along the real axis.
Its absorptive part
\begin{eqnarray}
\label{Def-AbsT} T_{\mu\nu}(q, P, \Delta;S_1,S_2)-
\left[T_{\nu\mu}(q, P, -\Delta;S_2,S_1)\right]^\ast \equiv 4\pi i
W_{\mu\nu}(q, P, \Delta;S_1,S_2)\,,
\end{eqnarray}
is given by the commutator of the electromagnetic currents
\begin{eqnarray}
\label{Def-TenW} W_{\mu\nu}= \frac{1}{4\pi e^2}\int d^4x e^{i
x\cdot q} \langle P_2, S_2 | \left[j_\mu (x/2), j_\nu (-x/2)
\right]| P_1, S_1 \rangle\,.
\end{eqnarray}
Finally, using the generalized Schwarz reflection principle
(\ref{T-SchRefPri}), we have for the discontinuity, now expressed
in terms of the more appropriate variable $\nu$,
\begin{eqnarray}
\label{Def-DisT} {\rm Disc}\, T_{\mu\nu} \equiv T_{\mu\nu}(\nu+i
\epsilon,\cdots) - T_{\mu\nu}(\nu-i \epsilon,\cdots) = 4\pi i
W_{\mu\nu}\,.
\end{eqnarray}
The spectral representation of the hadronic tensor $ W_{\mu\nu}$
and baryon number conservation tell us that this discontinuity
does not vanish for $|\nu| \geq \nu_{\rm cut}$ with $\nu_{\rm
cut}=(4 Q^2 + \Delta^2)/4\sqrt{P^2}$, see Fig.\ \ref{Fig-Cou}.

Obviously, these analytic properties hold for the CFFs, too. The
crossing relation (\ref{T-BosSym}) implies the symmetry relations
\begin{eqnarray}
\label{Sym-Rel} {\cal F}^{\rm V}(\nu, \cdots) = {\cal F}^{\rm
V}(-\nu,\cdots)\,,\qquad {\cal F}^{\rm A}(\nu,\cdots) =  -{\cal
F}^{\rm A}(-\nu,\cdots)\,.
\end{eqnarray}
To fix the form of the dispersion relation, we should estimate the
high-energy behavior of the CFFs from the common arguments of
Regge phenomenology. The leading meson trajectories give rise to a
$\nu^{\alpha_{M}(\Delta^2)}$ behavior with $\alpha_{M}(\Delta^2)
\le \alpha_{M}(0) \simeq 1/2$ for $\Delta^2 < 0$.
Thus, for the meson-exchange induced part it is
appropriate to use Cauchy's theorem
with one subtraction, i.e., cf. Fig.\ \ref{Fig-Cou}(a),
\begin{eqnarray}
\label{Def-DisRel-0} {\cal F}(\nu,\vartheta,\Delta^2,Q^2) =
\frac{1}{2 \pi i} \oint^{\nu_{\rm cut} -0}_{-\nu_{\rm cut} +0}
\frac{d\nu^\prime}{\nu^\prime} \frac{\nu}{\nu^\prime-\nu} {\cal
F}(\nu^\prime,\vartheta,\Delta^2,Q^2) + {\cal
C}(\vartheta,\Delta^2,Q^2)\,.
\end{eqnarray}
In the axial-vector case, the subtraction constant ${\cal
C}(\cdots)\equiv {\cal F}(\nu=0,\cdots)$ is zero for symmetry
reasons, see Eq.\ (\ref{Sym-Rel}).  In the flavor singlet sector
of parity even CFFs ${\cal F}^V$ the pomeron exchange appears. It
induces a stronger power-like growth with an exponent
$\alpha_{\mathbb{P}}(\Delta^2)$, where
$\alpha_{\mathbb{P}}(\Delta^2) \le \alpha_{\mathbb{P}}(0) \simeq
1$. For this contribution a second subtraction should in principle
be introduced in Eq.\ (\ref{Def-DisRel-0}), giving rise to a
$\nu\, {\cal C}^\prime $ term. However, since a parity even
CFFs is a even function in $\nu$, this constant ${\cal C}^\prime $
is zero.
Hence, the analytic properties allow us to derive a singly
subtracted dispersion relation. Inflating the integration path to
infinity, pictured in Fig.\ \ref{Fig-Cou}(b), we are left with a
path that encircles the cuts on the real axis. Picking up the
discontinuity and employing now the symmetry (\ref{Sym-Rel}), we
find for all CFFs and $|\xi|>1$ the dispersion relation%
\footnote{Although the function ${\cal F}$ is now expressed in
terms of $\xi$ rather than $\nu$, for simplicity we  do not
introduce a new notation.}
\begin{eqnarray}
\label{Def-DisRel} {\cal F}(\xi,\vartheta,\Delta^2,Q^2) =
\frac{1}{\pi}\int_0^{1}\! d\xi^\prime
\left(\frac{1}{\xi-\xi^\prime} \mp \frac{1}{\xi+\xi^\prime}
\right) \Im{\rm m}{\cal F}(\xi^\prime-i 0,\vartheta,\Delta^2,Q^2)
+{\cal C}(\vartheta,\Delta^2,Q^2) \,.
\end{eqnarray}
Here, the upper and lower sign belong to the vector (V) and
axial-vector (A) case, respectively, where in the latter case the
subtraction constant $\cal C$ vanishes.  We have also used the
fact that the imaginary part $\Im{\rm m}{\cal
F}(\xi^\prime,\cdots)$ has the support $|\xi^\prime| \le 1/(1+
\Delta^2/4 Q^2)\lesssim 1$.

The dispersion relation  (\ref{Def-DisRel}) will be used to
relate the short-distance OPE, given as a Taylor series with respect
to the variable $\omega= 1/\xi < 1$, to the imaginary part of the
CFFs in the physical region. If we choose to write the Taylor
expansion of the CFFs with respect to $\omega$ as
\begin{eqnarray}
\label{eq:Taylor-in-omega} {{\cal F}}(\xi,\vartheta,\Delta^2,Q^2)
=
\sum_{j=0}^\infty \left[1\mp (-1)^j\right] \omega^{j+1}\; {{\cal
F}_j}(\vartheta,\Delta^2,Q^2) + {{\cal
C}}(\vartheta,\Delta^2,Q^2)\, ,
\end{eqnarray}
taking into account the symmetry relation (\ref{Sym-Rel}), we
arrive at
\begin{eqnarray}
\label{Def-ShoDisMom} {\cal F}_j(\cdots) \equiv \frac{1}{2(j+1)!}
\frac{d^{j+1}}{d\omega^{j+1}} {\cal
F}(1/\omega,\cdots)\Big|_{\omega=0} = \frac{1\mp
(-1)^j}{2\pi}\int_{0}^{1}\!d\xi\; \xi^j\; \Im{\rm m}{\cal F}(\xi-i
0,\cdots)\,.
\end{eqnarray}
This relation is valid for $ 0 \le j$, where the subtraction term
does not contribute. The latter  enters for $j=-1$, for which the
first term in Eq.\ (\ref{Def-DisRel}) drops out, and so we have
the trivial identity
\begin{eqnarray}
\label{Def-ShoDisMomInv}
{\cal C}(\vartheta,\Delta^2,Q^2) \equiv {\cal F}^{\rm
V}(1/\omega,\vartheta,\Delta^2,Q^2)\Big|_{\omega=0} \,.
\end{eqnarray}

\section{Operator product expansion to twist-two accuracy}
\label{Sec-OPE}

Using the conformal OPE, presented in Sect.\
\ref{SubSec-OPE-GenFor}, and the dispersion relation, we derive in
Sect.\ \ref{SubSec-OPE-DisRel} a Mellin--Barnes representation for
CFFs. Then we give a detailed discussion of the subtraction
constant, which appears in a certain set of CFFs. On hand of a
simple toy GPD model we  investigate some mathematical aspects of
the Mellin--Barnes representation. In Sect.\
\ref{SubSec-OPE-ConMom} we  present  the CFFs as conformal partial
wave expansion in terms of conformal GPD moments, depending on the
complex conformal spin. We also provide the rotation of common
Mellin GPD moments to the conformal ones.

\subsection{General form of the operator product expansion at twist-two level}
\label{SubSec-OPE-GenFor}

The perturbative QCD predictions for the CFFs (\ref{Def-CFFs})
might be derived by means of the OPE for the time-ordered product
of two electromagnetic currents, which is then plugged into the
Compton tensor (\ref{Def-ComScaTen}). The predictive power of
conformal symmetry allows for an economical evaluation and is the
key for the perturbative QCD predictions of the DVCS amplitude
beyond NLO, which will be presented in Sect.\ \ref{Sec-PreConSym}.
Thus, we choose an operator basis in which this symmetry can be
manifestly implemented. The evaluation of the CFFs in this
so-called conformal OPE is straightforward and can be found in
Refs.\
\cite{FerGriGat71,FerGriGat72,FerGriGat72a,FerGriPar73,Mue97a,BelMue97a,BraKorMue03}.
In this section we mainly present our conventions and give a short
insight into several aspects of the conformal OPE.

The time-ordered product of two electromagnetic currents is
expanded in the basis of so-called conformal operators. The
twist-two  operators that are built of quark fields read at tree
level
\begin{eqnarray}
\label{Def-ConOpe-Q} \left\{ {^{a}\! {\cal O}}^{\rm V} \atop
{^{a}\! {\cal O}}^{\rm A} \right\}_j\! =
\frac{\Gamma(3/2)\Gamma(1+j)}{2^{j} \Gamma(3/2+j)}\, (i
\partial_+)^j\; \bar{\psi} \lambda^a \left\{ { \gamma_+   \atop
\gamma_+ \gamma_5 }\right\} \, C_j^{3/2}\!\left(\!
\frac{\stackrel{\leftrightarrow}{D}_{+}}{\partial_+}\!
\right)\psi\,,
\end{eqnarray}
where, e.g., for three active quarks the nonsinglet and singlet
flavor matrices are
\begin{eqnarray} \lambda^{\rm NS}= \left(
\begin{array}{ccc}
2 &0 &0\\
0&-1&0\\
0&0&-1
\end{array}\right)\quad\mbox{and}\quad
\lambda^{\rm \Sigma}= \left(
\begin{array}{ccc}
1 &0 &0\\
0&1&0\\
0&0&1
\end{array}\right).
\end{eqnarray}
The gluon operators are expressed by the field strength tensor
$G^a_{\mu\nu} =
\partial_\mu B^a_\nu - \partial_\nu B^a_\mu + g
f^{abc}B^b_\mu B^c_\nu$
\begin{eqnarray}
\label{Def-ConOpe-G}
 \left\{{ {^{\rm G}\! {\cal O}}^{\rm V} \atop  {^{\rm G}\!
{\cal O}}^{\rm A} }\right\}_j\! =
\frac{\Gamma(5/2)\Gamma(j)}{2^{j-2} \Gamma(3/2+j)}\,  (i
\partial_+)^{j-1}\; G^{\phantom{+}\mu}_{+}
\left\{ { g_{\mu\nu}   \atop i\epsilon_{\mu\nu -+} }\right\} \,
C_{j-1}^{5/2}\!\left(\!
\frac{\stackrel{\leftrightarrow}{D}_{+}}{\partial_+}\!
\right)G^\nu_{\phantom{\nu} +}\,.
\end{eqnarray}
The covariant derivatives either in the fundamental or adjoint
representation contracted with the light-like vector $n$ are
denoted as $\stackrel{\leftrightarrow}{D}_+ =
\stackrel{\rightarrow}{D}_+ -\stackrel{\leftarrow}{D}_+$ and
$\partial_+ = \stackrel{\rightarrow}{\partial}_+ +
\stackrel{\leftarrow}{\partial}_+$ for the total derivative.
$C_j^\nu$ is the Gegenbauer polynomial of order $j$ with index
$\nu$. The ``$-$'' component is obtained by contraction with the
dual light-like vector $n'$, i.e., $n\cdot n'=1$. The
operators \req{Def-ConOpe-Q} and \req{Def-ConOpe-G} are the ground
states of conformal multiplets (towers), which are labelled by the
conformal spin $j+2$. Their Lorentz spin is $J=j+1$ and higher states are
obtained by multiplying the ground state with total derivatives $i
\partial_+ $. Hence, their spin $J$ is increased by the number of
total derivatives.

The advantage of such a basis is that in the  hypothetical
conformal limit conformal symmetry guarantees  that the
renormalization procedure can be performed in such a way that only
operators with the same conformal spin will mix. Beyond LO, this statement is
not true in an arbitrary scheme, and in particular not in the
$\overline{\rm MS}$ one. Suppose we rely on a scheme in which
the conformal spin is a good quantum number. Then the only mixing
problem occurs in the flavor singlet sector, where quark and
gluon operators with the same conformal spin will mix \cite{Mue97a}. The
$2\times 2$ anomalous dimension matrix can be simply diagonalized
by the transformation
\begin{eqnarray}
\label{Def-Bas-Tra}
\left({ {^+\! {\cal O}^{\rm I}} \atop  {^-\! {\cal O}^{\rm I}} } \right)_j =
\mbox{\boldmath $U$}_j^{\rm I}(\alpha_s(\mu)) \left({ {^{\rm \Sigma}\!
{\cal O}^{\rm I}} \atop  {^{\rm G}\! {\cal O}^{\rm I}} } \right)_j\,,
\end{eqnarray}
where the $2\times 2$ matrices $\mbox{\boldmath $U$}_j^{\rm I}$ are
expressed in terms of the anomalous dimensions. Here $\{+,-\}$
labels the eigenfunctions of the renormalization group equation in
this sector and ${\rm I}\in\{V,A\}$.
Consequently, the renormalization group equation for
all operators in question is then simply
\begin{eqnarray}
\label{Def-RGE} \mu\frac{d}{d\mu} {^a\! {\cal O}^{\rm I}}_{j}(\mu) =-
{^a\!\gamma_j^{\rm I}}(\alpha_s(\mu))\;
{^a\! {\cal O}^{\rm I}}_{j}(\mu)
\end{eqnarray}
for $a\in\{{\rm NS},+,-\}$. Moreover, if conformal symmetry is
implemented in such a manifest way, it can be employed to predict
the Wilson coefficients of the OPE
\cite{FerGriGat71,FerGriGat72,FerGriGat72a,FerGriPar73,Mue97a,BelMue97a,BraKorMue03}.

With respect to the mixing properties of operators, it is
convenient to perform also a decomposition in flavor nonsinglet
(NS) and singlet (S) CFFs and express the latter  by the two
eigenmodes $a\in\{+,-\}$ of the evolution operator:
\begin{eqnarray}
 {\cal F} = Q_{\rm NS}^2\,
{^{\rm NS}\!{\cal F}} +  Q_{\rm S}^2\, {^{\rm S}\! {\cal F}} \,,
\qquad  {^{\rm S}\!{\cal F}} ={^{\Sigma}\!{\cal F}} + {^{\rm
G}\!{\cal F}}
 ={^{+}\!{\cal F}} + {^{-}\!{\cal F}}
\, .
\end{eqnarray}
The fractional charge factors are for three [four] active light
quarks%
\footnote{See Appendix \ref{App-NorWilCoe} for more details.
Basically, the squared fractional charges $Q_{\rm NS}^2$ and
$Q_{\rm \Sigma}^2$ are obtained from the decomposition $Q_{\rm
NS}^2\,  {^{\rm NS}\!{\cal F}} + Q_{\rm \Sigma}^2\, {^{\rm \Sigma
}\! {\cal F}}= Q_{\rm u}^2\,  {^{\rm u}\!{\cal F}} + Q_{\rm d}^2\,
{^{\rm d}\! {\cal F}} + Q_{\rm s}^2\, {^{\rm s}\! {\cal F}}
\left[+ Q_{\rm c}^2\, {^{\rm c}\! {\cal F}} \right] $.}
\begin{eqnarray}
\label{Def-SqaChaFac} {Q}^{2}_{\rm NS}   =
\frac{1}{9}\left[\frac{1}{6}\right]\,, \qquad {Q}^{2}_{\rm
S}={Q}^{2}_{\rm \Sigma}={Q}^{2}_{G} ={Q}_\pm^{2} = \frac{2}{9}
\left[\frac{5}{18}\right]\,.
\end{eqnarray}
The general form of the short-distance OPE is governed by Lorentz
invariance and is given as sum over the irreducible
representations of local operators. In our conformal operator basis
this expansion leads  to the following leading twist approximation
of the CFFs
\begin{eqnarray}
\label{Def-ConParDecInt} {\cal F} = \sum_{a={\rm NS},\pm} Q^2_a \;
{^a\! {\cal F}}\,,\quad {^a\! {\cal F}}\simeq\sum_{j=0}^\infty [1\mp
(-1)^j]\; {\xi}^{-j-1}\; {^a\! C_j^{\rm I}}(\vartheta,
Q^2/\mu^2,\alpha_s(\mu))\, {^a\! F_j} (\eta, \Delta^2,\mu^2)\,.
\end{eqnarray}
Here, they are factorized in the perturbatively calculable Wilson
coefficients ${^a\! C_j^{\rm I}}(\vartheta, Q^2/\mu^2,\alpha_s(\mu))$ and
nonperturbative reduced matrix elements ${^a\! F_j}
(\eta, \Delta^2,\mu^2)$, defined below in  Eqs.\
(\ref{Def-ConMomVec}) and (\ref{Def-ConMomAxiVec}). These matrix
elements carry conformal spin $j+2$ and are nothing but the
conformal moments of GPDs (see Appendix
\ref{App-ConfMom}). In Eq.\ (\ref{Def-ConParDecInt}) the sum
runs over the conformal spin and for I=V (A) only  its odd (even)
values contribute
(in the following we will lighten our
notation by suppressing the superscript ${\rm I}\in\left\{{\rm
V,A}\right\}$).

The $\vartheta=\eta/\xi$ dependence of the Wilson coefficients
${^a\! C_j}$ encodes information on how operators with the same
spin, but a different number of total derivatives, or different
conformal spin projection, are
arranged%
\footnote{This can be easily seen from the Taylor expansion of the
Wilson coefficients in powers of $\eta/\xi$. Taking the spin $J$
as summation label, the entire expansion is given in terms of
inverse powers $\xi^{-J}$. For each given value $J$ there appear
$J=\{1,2,3\cdots \}$  matrix elements $\eta^n
F_{J-n-1}(\eta,\Delta^2,\mu^2)$ with $n=\{0,1,\cdots, J-1\}$.  }.
This arrangement is indeed governed in the hypothetical conformal
limit by conformal symmetry. For instance, to LO approximation
this symmetry predicts the form of the Wilson coefficients, that
are expressed by hypergeometric functions, see e.g., Ref.\
\cite{Mue97a}, as
\begin{eqnarray}
\label{Def-ResCOPEMomSpa} {^a C}^{\rm tree}_j(\vartheta) = {^a\!
c_j^{\rm tree}}\; {_2F_1}\left({(1+j)/2, (2+ j )/2 \atop (5+2
j)/2}\Big|\vartheta^2 \right)\,.
\end{eqnarray}
The normalization ${^a\! c_j^{\rm tree}}$ is not fixed. However,
crucial for our following results is the fact that it can
be borrowed from deeply inelastic scattering. Namely,
setting $\vartheta=0$, we realize that
${^a\! c_j^{\rm tree}}$ coincides with the forward Wilson
coefficients
\begin{eqnarray}
{^a\! c_j^{\rm tree}} \equiv {^a C}^{\rm
tree}_j(\vartheta=0)=1\quad\mbox{for}\quad a=u,d,s,(c)\,.
\end{eqnarray}
Such correspondence is also valid in all orders of perturbation
theory, see next section and Appendix~\ref{App-NorWilCoe}.

The conformal moments $F_j=
\{H_j,E_j,\widetilde{H}_j,\widetilde{E}_j\}$ arise from the form
factor decomposition of the matrix elements of conformal operators
(\ref{Def-ConOpe-Q}) and (\ref{Def-ConOpe-G}). The reduced matrix
elements
\begin{eqnarray}
\label{Def-ConMomVec} \frac{1}{ P_+^{j
}}\langle P_2, S_2 \big|
{^a\!{\cal O}^{\rm V}_j} \big|P_1, S_1 \rangle =
\overline{U} (P_2, S_2) \!\left(\!
{^a\!H_j}(\eta,\Delta^2,\mu^2) \gamma_+ +
 {^a\!E_j}(\eta,\Delta^2,\mu^2) \frac{i\sigma_{+\nu}
 \Delta^\nu}{2M}\!
\right)\! U (P_1, S_1)  \,, \\
\label{Def-ConMomAxiVec} \frac{1}{ P_+^{j
}} \langle P_2, S_2
\big| {^a\! {\cal O}^{\rm A}_j} \big|P_1, S_1 \rangle =
\overline{U} (P_2, S_2)\! \left( {^a\!{\widetilde
H}_j}(\eta,\Delta^2,\mu^2) \gamma_+ \gamma_5 + {^a\!{\widetilde
E}_j}(\eta,\Delta^2,\mu^2) \frac{\Delta_+ \gamma_5}{2M}\!
\right)\! U (P_1, S_1) \,,
\end{eqnarray}
of these operators  are defined in such a way that in the forward
limit we encounter the standard Mellin moments
of parton densities
\begin{eqnarray}
{^a\!q_j}(\mu^2) = \lim_{\Delta\to 0}
{^a\!H_j}(\eta,\Delta^2,\mu^2)\,\quad\mbox{and}\quad \Delta
{^a\!q_j}(\mu^2) = \lim_{\Delta\to 0} {^a\!{\widetilde
H}_j}(\eta,\Delta^2,\mu^2)\,.
\end{eqnarray}
Note that the matrix elements of conformal operators are tensors
of rank $j+1$, contracted with the light-like vector $n$. Hence
the reduced ones (\ref{Def-ConMomVec}) and
(\ref{Def-ConMomAxiVec})  are polynomials in $\eta$ of order
$j+1$.
In particular, $H_j$ and $E_j$ are of order $j+1$, whereas
$\widetilde H_j$, $\widetilde E_j$ and $H_j+E_j$ are of order $j$,
see, e.g., Ref.\ \cite{Ji98}. Moreover, time reversal invariance
and hermiticity ensure that all these polynomials are even under
reflection $\eta \to -\eta$. This then implies that for odd $j$
conformal moments, such as those appearing in the OPE of ${\cal
F}^{\rm V}$ form factors, the combination $H_j+E_j$ is actually of
order $j-1$.

\subsection{Employing the dispersion relation}
\label{SubSec-OPE-DisRel}

In this section, we combine the dispersion and the OPE
techniques to restore both the imaginary and real part of the CFFs
in the physical region. We also give an extended discussion of the
subtraction term.

\subsubsection{Mellin--Barnes representation for CFFs}
\label{SubSubSec--MelBarRepGPD}

We express the Taylor expansion of the CFFs with respect to
$\omega=1/\xi$ in the convenient form analogous to that introduced
in Eq. \req{eq:Taylor-in-omega}
\begin{eqnarray}
\label{eq:Taylor-in-omega2} {^a\!{\cal
F}}(\xi,\vartheta,\Delta^2,Q^2) = \sum_{j=0}^\infty \left[1\mp
(-1)^j\right] \omega^{j+1}\; {^a\!{\cal
F}_j}(\vartheta,\Delta^2,Q^2) + {^a\!{\cal
C}}(\vartheta,\Delta^2,Q^2)\,.
\end{eqnarray}
Note that the CFFs  and so  the Taylor coefficients are physical
quantities that are independent of the
renormalization/factorization scheme
 that is used for their evaluation.

First, we evaluate the Taylor coefficients ${^a\!{\cal F}_j}$, as
defined in Eq.\ (\ref{Def-ShoDisMom}), from the conformal OPE (COPE) result
(\ref{Def-ConParDecInt}) taking $\eta/\xi=\eta \omega=\vartheta$
as an independent variable:
\begin{eqnarray}
\label{Def-DerF} {^a\! {\cal F}_j} \simeq \sum_{{n=0\atop {\rm
even}}}^\infty {^a\! C_{j+n}}(\vartheta,
Q^2/\mu^2,\alpha_s(\mu))\; \vartheta^n \;
{^a\!F_{j+n}^{(n)}}(\Delta^2,\mu^2)\,,\quad F_{j}^{(l)} =
\frac{1}{l!} \frac{d^l}{d\eta^l}
F_{j}(\eta,\Delta^2,\mu^2)\Big|_{\eta=0}\,,
\end{eqnarray}
where $j$ is (even) odd in the (axial-)vector case. Here we used
the fact that all considered $F_{j}$ are even polynomials
in $\eta$ of the order $j$ or $j+1$ and so $ F_{j}^{(l)} = 0$ for
$l> j$ or  $l> j+1$. Consequently, the sum on the r.h.s.\ runs
only over even $n$.

Next, we use the analytic properties of the CFFs which ensure that
the Taylor coefficients (\ref{Def-DerF}) are given by the
Mellin moments of the imaginary part. Corresponding to Eq.\
(\ref{Def-ShoDisMom}), they then
 read for general kinematics
 \begin{eqnarray}
\label{Def-MelMomImaParF} \int_{0}^\infty\!d\xi\, \xi^{j}\,\Im{\rm
m}{^a\!{\cal F}}(\xi,\vartheta,\Delta^2,Q^2)= \pi\; {^a\!{\cal
F}_j}(\vartheta,\Delta^2,Q^2) \,.
\end{eqnarray}
For technical reasons, we extend here and in the following the
integration region to infinity using
$\Im{\rm m}\,{^a\!{\cal F}}(\xi,\cdots) =0 $  for $\xi> 1$.

Now, we can combine Eqs.\ \req{Def-DerF} and \req{Def-MelMomImaParF}.
For the DVCS kinematics
($\vartheta=1$)  we have with the notation ${\cal
F}(\xi,\Delta^2,{\cal Q}^2)\equiv {\cal
F}(\xi,\vartheta=1,\Delta^2,Q^2)$
\begin{eqnarray}
\label{Def-MelMomImaParFDVCS} \int_{0}^\infty\!d\xi\,
\xi^{j}\,\Im{\rm m}{^a\!{\cal F}}(\xi,\Delta^2,{\cal Q}^2)\simeq \pi
\sum_{{n=0\atop {\rm even}}}^\infty  {^a\! C_{j+n}}( {\cal
Q}^2/\mu^2,\alpha_s(\mu))\, {^a\!F_{j+n}^{(n)}}
(\Delta^2,\mu^2)\,,
\end{eqnarray}
where we denote the DVCS Wilson coefficients as $C_{j}({\cal
Q}^2/\mu^2,\alpha_s(\mu)) \equiv
C_{j}(\vartheta=1,Q^2/\mu^2,\alpha_s(\mu))$ and make use of the
photon virtuality ${\cal Q}^2\simeq  2 Q^2$. It is worth
mentioning that the Mellin moments (\ref{Def-MelMomImaParFDVCS})
are directly measurable in single spin asymmetries. They contain
the sum of all reduced matrix elements built by operators with
conformal spin equal or larger than $j+2$ and so its spin is
larger than $j$.

\begin{figure}[t]
\begin{center}
\mbox{
\begin{picture}(600,150)(0,0)
\put(20,-7){\insertfig{5.5}{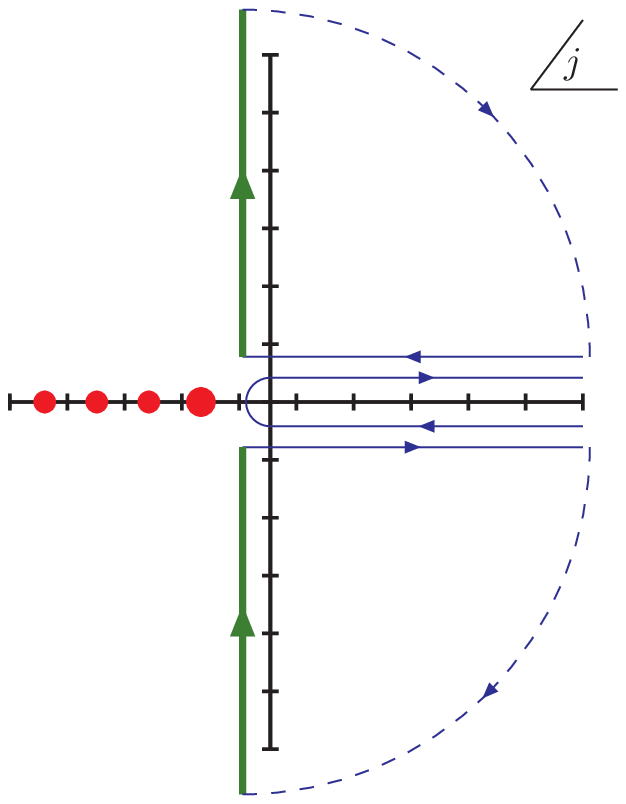}}
\put(110,-15){$(a)$}
\put(265,5){\insertfig{5.6}{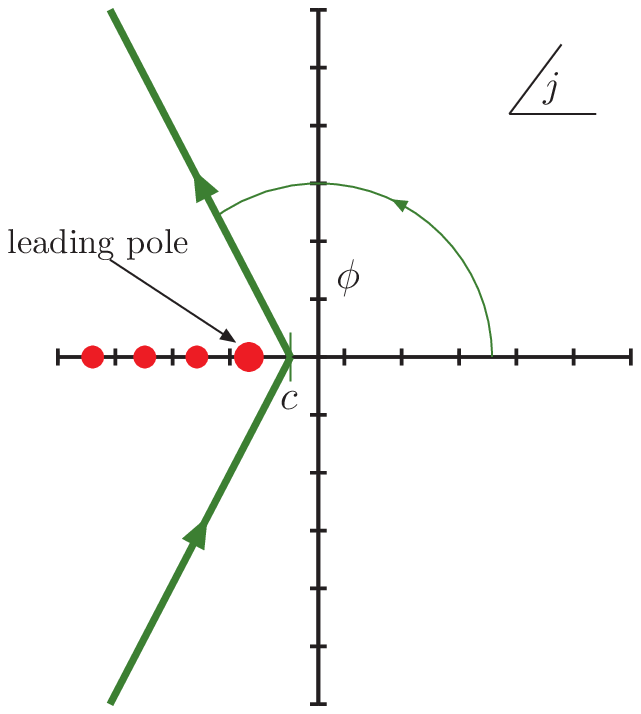}}
\put(335,-15){$(b)$}
\end{picture}
}
\end{center}
\caption{\label{FigCou2} $(a)$ The integration contour in Eq.\
(\ref{Res-ImCFFSer}) parallel to the imaginary axis  in the complex
$j$-plane. Adding semicircles results in an integration path which
encloses  the real axis. Contribution of dashed segments vanishes.
$(b)$ The integration contour that will be used below
in the numerical evaluation of CFFs.}
\end{figure}

In the next step  we  invert the Eq.  (\ref{Def-MelMomImaParFDVCS})
following the standard mathematical procedure for inverse Mellin transform.
To do so we analytically continue the l.h.s.\ with respect
to the conformal spin, which is trivially
done by considering $j$ as complex valued.
The Mellin moments are holomorphic functions in $j$, as long as the
$\xi$-integral containing the weight $\xi^j (\ln\xi)^N $ for
$N=0,1,2,...$ exists, and the condition $\Re{\rm e}\, j  \ge c$,
with an appropriate constant $c$, is satisfied.
The choice of $c$ should exclude the singularities in the
complex $j$-plane arising from the endpoint at $\xi=0$.
Notice also that in the limit $\Re{\rm e}\, j\to \infty$  the
Mellin moments of a regular spectral function vanish, except for
a elastic pole contribution, i.e., a  $\delta(\xi-1)$ term. The
inverse Mellin transform reads
\begin{eqnarray}
\label{Res-ImCFFSer} \Im{\rm m}{^a {\cal F}}(\xi, \vartheta,
\Delta^2,Q^2)= \frac{1}{2i}\int_{c-i\infty}^{c+i\infty}\! dj\,
\xi^{-j-1}\;
 {^a\!{\cal F}_j}(\vartheta,\Delta^2,Q^2)\,.
\end{eqnarray}
Here all singularities of the integrand in the complex $j$-plane
lie to the left of integration contour, which is parallel to the
imaginary axis, see Fig.\ \ref{FigCou2} (a). The rightmost lying
singularity might be related to the leading Regge trajectory,
and so we have for $c\equiv \Re{\rm e}\, j$ the condition%
\footnote{The leading Regge trajectory corresponds to
$\xi^{-\alpha(\Delta^2)}$ behavior of ${^a {\cal F}}$,
i.e., $\xi^{j-\alpha(\Delta^2)}$ behavior of the integrand
in \req{Def-MelMomImaParFDVCS}, and the pole in $j=\alpha(\Delta^2)-1$.
Note that in perturbation theory the Wilson coefficients and
anomalous dimensions possess additional poles at negative integer $j$.
Consequently, if ${^{a}\!{\cal F}_j}$  is perturbatively expanded
we have the condition
${\rm max}(\alpha(\Delta^2)-1,-1) < c$.  Moreover,
the flavor singlet anomalous dimension matrix in the parity even sector has
poles at $j=0$. These poles could lie right to the pomeron pole
$j=\alpha_{\mathbb{P}}(\Delta^2)-1$, which appear in the
nonperturbative ansatz of GPD moments. Hence, we have in this sector the
condition  ${\rm max}(\alpha_{\mathbb{P}}(\Delta^2)-1,0) < c$.
\label{FooNot-Pol}}
$\alpha(\Delta^2)-1 < c$.

There is one comment in order. The analytic continuation of the
integrand in Eq.\ (\ref{Res-ImCFFSer}), known for non-negative
integer $j$, is not unique. According to Carlson's theorem
\cite{Car14,EdeOliPol66}, this ambiguity is resolved by specifying
the behavior at $j\to \infty$. Consequently, the analytic
continuation of the Mellin moments must be done in such a way
that they vanish (or at least are bounded) in the limit $\Re{\rm
e} \, j\to \infty$. Especially, one has to avoid an exponential
growth, e.g., by a phase factor $e^{i j \pi}$. Obviously,  then
for $\xi > 1$ the integrand in Eq.\ (\ref{Res-ImCFFSer}) does not
contribute on the arc with infinite radius that surrounds the
first and fourth quadrants. Completing the integration path to a
semicircle, Cauchy's theorem states that the imaginary part of the
CFFs vanishes for $\xi > 1$ as it should be.

Finally, to restore the real part of the CFFs from the Mellin moments, we plug the
imaginary part (\ref{Res-ImCFFSer}) into the dispersion relation
(\ref{Def-DisRel}), extending the integration interval to $0\le
\xi^\prime \le \infty$, and perform the $\xi^\prime$ integration
using the principal value prescription. The existence of
the integral is ensured by appropriate bounds for $\Re{\rm e}\, j$.
The lower one is ${\rm max}(\alpha(\Delta^2)-1,-1) < c$ or   ${\rm
max}(\alpha_{\mathbb{P}}(\Delta^2)-1,0) < c$, see footnote
\ref{FooNot-Pol}, and in addition for the upper one we have $ c < 1$ $(c <
0)$ in the (axial-)vector case. The integration yields
an additional weight factor in the Mellin--Barnes integral
(\ref{Res-ImCFFSer}) which is given by $\tan(\pi j/2)$ and
$-\cot(\pi j/2)$ for the vector and axial-vector case,
respectively. The whole amplitude for $\xi>0$ reads
\begin{eqnarray}
\label{Def-MelBarSer} {^a\!{\cal F}}(\xi,\vartheta,\Delta^2,Q^2)
&\!\!\! =\!\!\! &  \frac{1}{2 i}\int_{c-i \infty}^{c+ i \infty}\!
dj\; \xi^{-j-1} \left[i \pm \left\{{\tan\atop
\cot}\right\}\!\left(\!\frac{\pi j}{2}\right)\! \right]
{ ^a\!{\cal F}_j} (\vartheta,\Delta^2,Q^2) + { ^a {\cal
C}}(\vartheta,\Delta^2,Q^2)\,.
\end{eqnarray}
where the Mellin moments are defined by the series
(\ref{Def-DerF}). The analogous expression for $\xi<0$ follows
from the symmetry relations \req{Sym-Rel}. We stress again that
analyticity ties the real and imaginary part in a unique way. This is
obvious in the axial-vector case, where the subtraction constant
is zero for symmetry reasons. The vector case will be discussed in
the next section.

By comparing \req{Def-MelBarSer} with \req{eq:Taylor-in-omega2}
one sees that with the help of dispersion relation technique the
analytic continuation of the CFFs in the unphysical region, given
by the series \req{Def-MelBarSer}, yields the complex
Mellin--Barnes integral \req{eq:Taylor-in-omega2}. The
corresponding Mellin moments ${^a\! {\cal F}_j}$ are
perturbatively predicted by another series (\ref{Def-DerF}), which
depends on the GPD moments ${^a\! F}_{j+n}^{n}$. This
remaining summation we postpone for later.

\subsubsection{Discussing the subtraction constant}
\label{SubSubSec-subtraction-cte}

The evaluation of the subtraction constant $\cal C$ from the limit
$\omega\to 0$ deserves some additional comments. The definition
(\ref{Def-ShoDisMomInv}) of this constant suggests that it is
entirely determined by short-distance physics. The relevant local
operator has  dimension two (or spin zero) and simply does not
exist as a gauge invariant one%
\footnote{Such an operator might be interpreted as the expectation
value of a gauge invariant non-local operator that carries spin
zero. In the light cone gauge such an operator is for instance
given by the local two gluon operator $g^{\mu\nu}B^a_{\mu}(0)
B^a_{\nu}(0)$. However, in a gauge invariant scheme its Wilson
coefficient is zero.}. So on the first sight one would expect
that, as in the forward case, the subtraction constant is zero
in the (generalized) Bjorken limit.
Surprisingly,  a more careful look at the OPE result
(\ref{Def-ConParDecInt}) shows that all (gauge invariant local)
operators contribute, because the highest possible order in $\eta=
\vartheta/\omega$, e.g., $j+1$, of their conformal moments, cancels
the suppression factor $\omega^{j+1}$ in front of the Wilson
coefficients. Hence, the subtraction constant is in fact
determined by light-cone physics and resumming all terms
proportional to $F_{j}^{(j+1)}$ leads to:
\begin{eqnarray}
\label{Pre-SubConC}
{^a\!{\cal C}}(\vartheta,\Delta^2,Q^2) \simeq 2 \sum_{{n=2 \atop \rm
even}}^\infty {^a\! C_{n-1}}(\vartheta, Q^2/\mu^2,\alpha_s(\mu))
\vartheta^{n} \,{^a\!F_{n-1}^{(n)}}(\Delta^2,\mu^2)\,,\quad
{^a\!{\cal C}}(\vartheta=0,\Delta^2,Q^2) \simeq 0\,.
\end{eqnarray}
Since $F_{j}^{(j+1)}$   appears only for $j+1$ even, this constant
vanishes for symmetry reasons in the axial-vector case --- as it
musts --- and is absent in the combination ${\cal H}+{\cal E}$.
Moreover, it is predicted to be zero if the virtualities of the
photons are equal, i.e., for $\vartheta=0$. The expression
(\ref{Pre-SubConC}) for the subtraction constant has been also
derived in the momentum fraction representation at tree level
\cite{Ter05}, where it was expressed by a so-called $D$-term
\cite{PolWei99}. This term  entirely  lives in the central GPD
region and collects the highest possible order terms in $\eta$.
Hence, its  conformal moments provide us $F_{j}^{(j+1)}$, e.g., for
$\eta >0$,
\begin{eqnarray}
\int_{-\eta}^{\eta}\! dx\,
\eta^j C_j^{3/2}\left(\frac{x}{\eta}\right) D\left(\frac{x}{\eta}\right) =
F_{j}^{(j+1)} \eta^{j+1}\,.
\end{eqnarray}

In the following we discuss the subtraction constant for the CFF ${\cal E}$,
rather than ${\cal H}$. The reason
for doing so is that $E_j^{(j+1)}$ originally arises from a
nonperturbatively induced helicity-flip, as we will
see in Sect.\ 5.
We note that such nonperturbative effects
are absent in linear combination $H_j + E_j$.

A subtraction constant is usually viewed as additional
information that cannot be obtained from the imaginary part.
However, if $E_{j}^{(j+1)}$ is the analytic continuation of
$E_{j}^{(n)}$, we should indeed be able to express the subtraction
constant by the imaginary part of the CFF ${\cal E}$.

Let us demonstrate such a possible cross talk between subtraction
constant and imaginary part in the most obvious manner. To do so
we first make the mathematical assumption that ${^a\!{\cal E}}(\xi,
\vartheta,\ldots)$ vanishes for $\xi\to 0$. In this case the
subtraction in the dispersion relation (\ref{Def-DisRel}) has been
overdone. Taking the $\xi\to 0$ limit in the dispersion relation
indeed leads to the desired relation between subtraction
constant and imaginary part:
\begin{eqnarray}
\label{Cal-SubCon-2} { ^a {\cal C}}(\vartheta,\ldots) =
\frac{2}{\pi}\int_{0}^\infty\!d\xi\, \xi^{-1}\, \Im{\rm m}
{^a\!{\cal E}}(\xi,\vartheta,\ldots)\,.
\end{eqnarray}
Since the subtraction constant in the equal photon kinematics is
vanishing, we derive the sum rule
\begin{eqnarray}
\label{Cal-SubCon-4} { ^a {\cal C}}(\vartheta=0,\ldots) =
\frac{2}{\pi}\int_{0}^\infty\!d\xi\, \xi^{-1}\, \Im{\rm m} {
^a\!{\cal E}}(\xi,\vartheta=0,\ldots) =0\,.
\end{eqnarray}
We remark, however, that this  equality is {\em not} of physical
interest, because Regge phenomenology suggests that the small $\xi$
behavior of ${^a\!{\cal E}}$ arises from the pomeron pole and so
it will approximately grow with $1/\xi$ in the high-energy
asymptotics. Hence, our mathematical
assumption was purely academic and the sum rule
(\ref{Cal-SubCon-4}) does not apply.

Let us now consider a realistic behavior of the CFF $\cal E$.
Comparing the OPE prediction (\ref{Pre-SubConC}) for the
subtraction constant with the Mellin moments (\ref{Def-DerF}),
one realizes that for $j\to -1$ the only difference is
the first term in the series (\ref{Def-DerF}). Thus, removing this
term and taking the limit leads to a formal definition of the  subtraction constant
\begin{eqnarray}
\label{Cal-SubCon0} {^a\! {\cal C}}(\vartheta,\Delta^2,Q^2)  \simeq
2\lim_{j\to -1} \left[{^a\! {\cal E}_{j}}(\vartheta,\Delta^2,Q^2)
- {^a\! C_{j}}(\vartheta, Q^2/\mu^2,\alpha_s(\mu))\;
{^a\!E_{j}^{(0)}}(\Delta^2,\mu^2) \right]\,.
\end{eqnarray}
Here the separate terms on the r.h.s.\ can be singular and so the
analytic continuation plays the role of a regularization. We note
that, corresponding to  the high-energy ($\xi\to 0$) behavior,
there are also poles and perhaps branch cuts in the complex
$j$-plane, which have to be surrounded first, to arrive at
$\Re{\rm e} j=-1$. This is certainly possible if the
Mellin moments are meromorphic functions. The situation is more
tricky if ${^a\! {\cal E}_{j}}$ contains singularities at $j=-1$ and the
subtraction procedure is maybe not uniquely defined. In
particular, the appearance of a fixed singularity at $j=-1$, e.g.,
a $\delta_{j,-1}$ proportional contribution, might spoil the
relation between the subtraction constant and the imaginary part
of the corresponding Compton form factor.

Let us explore the consequences in the case that both scheme
dependent quantities ${^a\! C_{j=-1}}$ and ${^a\!E_{j=-1}^{(0)}}$
can be defined by analytic continuation. The GPD moment, appearing
on the r.h.s.\ of Eq.\ (\ref{Cal-SubCon0}), might be expressed by
the kinematical forward quantity
\begin{eqnarray}
{^a\!E_{j}^{(0)}}(\Delta^2,\mu^2)\simeq \frac{{^a\! {\cal
E}_{j}}(\vartheta=0,\Delta^2,Q^2)}{{{^a\!
C_{j}}(\vartheta=0,\Delta^2,Q^2/\mu^2,\alpha_s(\mu))}} \,,
\nonumber
\end{eqnarray}
see Eq.\ (\ref{Def-DerF}) with $\vartheta=0$. This provides an
important relation, which formally reads
\begin{eqnarray}
\label{Cal-SubCon} {^a\! {\cal C}}(\vartheta,\Delta^2,Q^2)  =
2\lim_{j\to -1}\left[ {^a\! {\cal E}_{j}}(\vartheta,\Delta^2,Q^2)-
\frac{{^a\!
C_{j}}(\vartheta,\Delta^2,Q^2/\mu^2,\alpha_s(\mu))}{{^a\!
C_{j}}(\vartheta=0,\Delta^2,Q^2/\mu^2,\alpha_s(\mu))} {^a\! {\cal
E}_{j}}(\vartheta=0,\Delta^2,Q^2) \right]
\,.
\end{eqnarray}
Remarkably, only the ratio of Wilson coefficients appearing among physical quantities in
Eq.\ (\ref{Cal-SubCon}) is potentially
plagued by ambiguities. Hence, we must  conclude that in the
$j \to -1$ limit this ratio is scale and
scheme independent and has the same value in any order of
perturbation theory. This value can thus be read off from the tree
level coefficients (\ref{Def-ResCOPEMomSpa}):
\begin{eqnarray}
\label{Cal-SubCon-RatWC} \lim_{j\to -1} \frac{{^a\!
C_{j}}(\vartheta,\Delta^2,Q^2/\mu^2,\alpha_s(\mu))}{{^a\!
C_{j}}(\vartheta=0,\Delta^2,Q^2/\mu^2,\alpha_s(\mu))}=1\,.
\end{eqnarray}
This formula has been checked to NLO accuracy. It turns out that
radiative $\alpha_s$ corrections blow up in the limit
$j\to -1$; however, the leading singularities in $j-1$ are
$\vartheta$ independent.

Relying on the validity of the prescription (\ref{Cal-SubCon}), we
can calculate the subtraction constant  from the knowledge of the
imaginary part. Taking into account Eq. (\ref{Cal-SubCon}), we
formally write the desired relation as
\begin{eqnarray}
\label{Cal-SubCon-1} { ^a {\cal C}}(\vartheta,\ldots) =
\frac{2}{\pi} \lim_{j\to -1} \left\{\int_0^\infty\! d\xi\; \xi^{j}
\,\Im{\rm m } \left[{ ^a\!{\cal E}}(\xi,\vartheta,\ldots) -
 { ^a\!{\cal E}}(\xi,\vartheta=0,\ldots) \right]\right\}_{\rm AC}\,,
\end{eqnarray}
where the integral is analytically regularized. That means that
one first has to evaluate the integral for ${\rm
max}(\alpha(\Delta^2)-1,-1)< \Re {\rm e} \, j$ and then perform
the analytic continuation to $j=-1$. Certainly, the prescription
(\ref{Cal-SubCon-1}) relies on analyticity. If this property is
absent, e.g., because of a fixed pole or singularity at $j=-1$
with a $\vartheta$-dependent residue, the subtraction constant ${
^a {\cal C}}$ is only partially  related to the imaginary part. We
remark that in such a case the $D$-term for GPDs \cite{PolWei99},
introduced as a separate contribution to cure the spectral representation
\cite{MueRobGeyDitHor94,Rad97}, seems to be justified on the first
view. Otherwise the prescription of Ref.\ \cite{BelMueKirSch00}
for the restoration of the highest possible order in $\eta$ is
more appropriate%
\footnote{The spectral representation is not uniquely defined
\cite{Ter01}. Indeed,  if one takes into account that
both $D$-term and spectral function are dependent -- they
arise from different projections of the same spectral function
that enters in the improved representation --
the choice of a `common' spectral representation
\cite{MueRobGeyDitHor94,Rad97} plus $D$-term is
equivalent to the improved ones \cite{BelMueKirSch00}.}.

Let us remind that the existence of  a fixed pole at angular
momentum $J=0$, i.e., $j=-1$ was already argued from Regge
phenomenology inspired arguments at the end of the sixties and it
was proposed to access it by the uses of finite energy sum rules
\cite{CreDrePas69}. The analyses of experimental measurements
\cite{DamGil69,DomFerSua70}, although not fully conclusive,
indicate a fixed pole at $J=0$. Interestingly, its residue might
be related to the Thomson limit of the Compton amplitude
\cite{Sch73}. On the theoretical side studies in the partonic
\cite{BroCloGun71,BroCloGun73} or light-cone \cite{Ban72,Fri73}
approach lead to a real term in the transversal part of the
(forward) Compton amplitude that  in the language of Reggeization
was interpreted as a $J=0$ fixed pole. Nevertheless,  via a
subtracted sum rule
\cite{CorCorNor70,CorCorNor71,BroCloGun71,BroCloGun73} it is
related to the imaginary part of the Compton amplitude. This would
imply that there is {\em no} separate $D$-term contribution in the
central GPD region.

\subsection{Representation of CFFs in terms of conformal moments}
\label{SubSec-OPE-ConMom}

In this section we derive a Mellin--Barnes representation
alternative to Eq.\ (\ref{Def-MelBarSer}), which expresses the
CFFs in terms of conformal GPD moments rather than the usual
Mellin moments. The advantage of such a representation is that the
CFFs are expanded in conformal partial waves, which diagonalize
the evolution operator to LO and facilitate direct application of
conformal symmetry beyond LO.   In Sect.\
\ref{SubSubSec-OPE-ToyExa} we consider the well understood
tree-level approximation to spell out the mathematical subtleties
that appear in the mapping of conformal moments of GPDs to the
Mellin $\xi$-moments (\ref{Def-DerF}) of CFFs. We propose then in
Sect.\ \ref{SubSubSec-SomWatTra} a Sommerfeld--Watson
transformation which allows us to derive the desired
Mellin--Barnes representation for CFFs, which has been already
derived previously by other methods \cite{MueSch05,Mue05a}.
Finally, we discuss the Mellin--Barnes representation of conformal
GPD moments themselves.

\subsubsection{Lessons from a toy example}
\label{SubSubSec-OPE-ToyExa}

Let us now have a closer look at the series representation (\ref{Def-MelMomImaParFDVCS})
of the Mellin moments $^{a}\!{\cal F}_j$, appearing in the Mellin--Barnes
representation (\ref{Def-MelBarSer}) for CFFs. To get some insight
into its mathematical properties we consider the OPE at tree
level. For simplicity, we will discard here the
$\Delta^2$-dependence, which appears only as a dummy variable, as well
as the flavor index. The Wilson coefficients of the COPE are
given in Eq.\ (\ref{Def-ResCOPEMomSpa}) and for DVCS kinematics,
i.e., $\vartheta=1$, we immediately have
\begin{eqnarray}
\label{Def-ResCOPEMomSpa-1} {C}^{\rm tree}_j(\vartheta=1) =
 {_2F_1}\left({(1+j)/2, (2+j )/2 \atop (5+2
j)/2}\Big|1 \right) = \frac{2^{j+1} \Gamma(5/2+j
)}{\Gamma(3/2)\Gamma(3+j)} \,.
\end{eqnarray}
If we form partonic matrix elements of the conformal operators,
e.g., for the CFF $\cal H$, the conformal moments $H_{j}^{\rm
tree}$  are simply given in terms of Gegenbauer polynomials
$\eta^j C_j^{3/2}(1/\eta)$. According to Eq.\ (\ref{Def-DerF}) the
corresponding  expression of $H_{j+n}^{(n){\rm tree}}$ reads
\begin{eqnarray}
\label{Def-h-TreLev} H^{\rm tree}_j(\eta) =
\frac{\Gamma(3/2)\Gamma(1+j)}{2^{j} \Gamma(3/2+j)}\, \eta^j\
C_j^{3/2}\!\left(\! \frac{1}{\eta}\right)\; \Rightarrow\;
 H_{j+n}^{(n){\rm tree}} =   \frac{(-1)^{n/2}\Gamma(3/2+j+n/2)
 \Gamma(1+j+n)}{2^{n}\Gamma(1+j)\Gamma(1+n/2)\Gamma(3/2+j+n)}\,.
\end{eqnarray}
Consequently, the Mellin moments (\ref{Def-DerF}) are for our toy
example given by the following series
\begin{eqnarray}
\label{Def-MelMomTree}
{\cal H}^{\rm tree}_j = \sum_{{n=0\atop {\rm even}}}^\infty
\frac{(-1)^{n/2} 2^{j+1} (3+2j+2n) \Gamma(3/2+j+n/2)}{
(1+j+n)(2+j+n) \Gamma(1/2)\Gamma(1+j)\Gamma(1+n/2)}.
\end{eqnarray}
As one realizes from the asymptotics of the $\Gamma$-functions,
for large $n$ with $n\gg j$ the summands behave as $\propto n^{j -
1/2}$ and for non-negative integer $j$ this series must be
resummed to arrive at a meaningful result, which in our case is
simply one. Hence, ${\cal H}^{\rm tree}_j=1$ and the inverse
Mellin transform yields a $\delta$-function. Note that the
exponentially growing factor $2^j$ in the Wilson coefficients
(\ref{Def-ResCOPEMomSpa-1}) is finally cancelled in the conformal
Mellin moments (\ref{Def-MelMomTree}) by an exponential suppression
factor in the conformal GPD moments (\ref{Def-h-TreLev}). This
cancellation ensures that the imaginary part of the CFF is
vanishing for $\eta=\xi>1$. We conclude that realistic conformal
GPD moments for $\eta \ge 1$ are exponentially suppressed by a
factor $2^{-j}$.

The real part of the CFF can be obtained from the dispersion relation
(\ref{Def-DisRel}), while the subtraction constant follows from
(\ref{Cal-SubCon})
\begin{eqnarray}
{\cal C}^{\rm tree} = \lim_{j\to-1}2\left[{\cal H}^{\rm tree}_j -
{C_j^{\rm tree}}(1) H_{j}^{(0){\rm tree}}  \right]=0\,.
\end{eqnarray}
This is in agreement with the fact that the highest order terms
possible in $\eta$ are absent for our toy conformal moments, i.e.,
$H_{n-1}^{(n){\rm tree}}=0$ for $n>0$, see Eq.\
(\ref{Def-h-TreLev}). Consequently,  the OPE prediction
(\ref{Pre-SubConC}) shows that the subtraction constant vanishes,
too. Thus, we removed the ambiguity caused by the
subtracted dispersion relation, and arrive at the unique result:
\begin{eqnarray}
\label{Res-CFF-Tre} \int_{0}^1\!d\xi\, \xi^{j}\,\Im{\rm m}{\cal
H}^{\rm tree}(\xi)= \pi\quad \Rightarrow\quad  \Im{\rm m}{\cal
H}^{\rm tree}(\xi) = \pi \delta(1-\xi)\,, \quad \Re{\rm e}{\cal
H}^{\rm tree}(\xi) = -\frac{2}{1-\xi^2} \,.
\end{eqnarray}
Of course, this example is trivial  and the findings coincide with
the well-known answer,
\begin{eqnarray}
\label{Def-ConFor-TreLev} {\cal H}^{\rm tree}(\xi,\vartheta=1) =
\int_{-1}^{1}\! dx\left( \frac{1}{\xi-x-i\epsilon}
-\frac{1}{\xi+x-i\epsilon}\right) H(x,\xi)\,,
\end{eqnarray}
arising from the evaluation of the hand-bag diagram  with a
partonic GPD $H(x,\eta) = \delta(1-x)$.


If evolution is included, an explicit resummation of the series
for the Mellin moments (\ref{Def-MelMomImaParFDVCS}), appearing
in the Mellin--Barnes integral (\ref{Def-MelBarSer}), cannot be
achieved, since each term is labelled by the conformal spin and
will evolve differently. There are following possibilities to
solve this problem:
\begin{itemize}
\item Numerical evaluation of the corresponding series.
\item Relying on an approximation of the series.
     \item Arrangement of the integrand in the Mellin--Barnes
     integral (\ref{Def-MelBarSer}) in such a way that the
     integration, instead of Mellin moment, runs over the complex conformal spin
\end{itemize}
The third possibility will be worked out in the next section. In
the following we shall study whether an approximation of the
series is appropriate. As we just saw, for complex $j$ with $\Re{\rm
e}\,j$ sufficiently small, the $\xi$ independent series
(\ref{Def-h-TreLev}) converges. If the first few terms are
numerically dominating, we might hope that these terms already
induce a good approximation of the CFFs. Certainly, if the $\eta$
dependence in the conformal moments is weak in the vicinity of
$\eta=0$ , i.e., if the $F_{j+n}^{(n)} (\Delta^2,\mu^2)$ are
numerically small for larger $n$, the series might be approximated
by a finite sum. This should induce a good approximation for the
CFFs for smaller values of $\xi$.

Let us shortly demonstrate how this ``expansion'' works at tree
level. We keep in the conformal moments (\ref{Def-h-TreLev}) only
the leading term in $\eta^2$ for $\eta\to 0$, i.e.,
$H_j^{(0)\rm{tree}} = 1$ and all terms with $n=2,4,\cdots $ are
neglected. Hence,  the Mellin moments (\ref{Def-MelMomTree}) read
in this approximation
\begin{eqnarray}
\label{Def-MelMomTreeApp} \int_{0}^\infty\!d\xi\, \xi^{j}\,\Im{\rm
m}{\cal H}^{\rm tree}(\xi)\simeq \pi \frac{2^{j+1}
\Gamma(5/2+j)}{\Gamma(3/2)\Gamma(3+j)}.
\end{eqnarray}
The inverse Mellin transform and the use of dispersion relation
leads then for $\xi\ge 0$ to
\begin{eqnarray}
\Im{\rm m}{\cal H}^{\rm tree}(\xi) \simeq \theta(2-\xi) \frac{
{\xi }^{\frac{3}{2}}}{{\sqrt{2 - \xi }}}, \qquad \Re{\rm e}{\cal
H}^{\rm tree}(\xi) \simeq  -2  + {\cal O}(\xi^{3/2})
\,.
\end{eqnarray}
The exponential factor $2^j$ in the approximated moments
(\ref{Def-MelMomTreeApp}) induces a wrong support. However, we
have still a useful approximation for small values of $\xi$, where
the deviation from the correct result starts at order $\xi^{3/2}$.
Naively, one might have expected that the accuracy is of order
$\xi^{2}$. We remark, however, that by taking into account the
next order in the expansion (\ref{Def-h-TreLev}), i.e., $n=2$ as
well, the $\xi^{3/2}$ proportional terms are annulled and the
result is then valid up to order ${\cal O}(\xi^{5/2})$. That the
expansion of the conformal moments with respect to $\eta$ induces
a systematic expansion of the CFFs in powers of $\xi$ is perhaps
not true in general.

\subsubsection{Complex conformal partial wave expansion}
\label{SubSubSec-SomWatTra}

We will now change the integration variable of the Mellin--Barnes integral
(\ref{Def-MelBarSer}) so that the
integration finally runs over the complex conformal spin and the
nonperturbative input is given in terms of conformal GPD
moments. As mentioned above, in such a basis the evolution
operator is diagonal with respect to this quantum number. If this
diagonality is not preserved in a given scheme, we can always
perform a scheme transformation so that the conformal GPD moments
evolve diagonally. Let us start with the Sommerfeld-Watson transformation
of the series for the Mellin moments ${\cal F}_j$ into a Mellin--Barnes
integral. To do this we rewrite the series
\begin{eqnarray}
{ ^a\!{\cal F}}_j(\vartheta=1,\Delta^2,Q^2)\equiv { ^a\!{\cal
F}}_j(\Delta^2,{\cal Q}^2) \simeq \sum_{{n=0 \atop {\rm even}}}^\infty
{ ^a C}_{j+n}({\cal Q}^2/\mu^2,\alpha_s(\mu))\;
 { ^a\!F}_{j+n}^{(n)} (\Delta^2,\mu^2)
\end{eqnarray}
as an contour integral that includes the real positive axis and
add two quarters of an infinite circle, see Fig.\ \ref{FigCou2}, and arrive so at the
Mellin--Barnes representation:
\begin{eqnarray}
{ ^a\!{\cal F}}_j(\Delta^2,{\cal Q}^2) \simeq
 \frac{i}{4} \int_{d-i\infty}^{d+i \infty}\!dn\;  \frac{1}{\sin(n \pi/2)}
{ ^a C}_{j+n}({\cal Q}^2/\mu^2,\alpha_s(\mu))\;
 { ^a\! \hat{F}}_{j+n}^{(n)}(\Delta^2,\mu^2)\,,
\end{eqnarray}
where  ${^a\!\hat{F}}_{j+n}^{(n)}$ is the analytic continuation of
$(-1)^{n/2}\,{ ^a\!F}_{j+n}^{(n)}$   and the integration contour
is chosen so that all singularities lie to the left of it. Note that for
$j=-1$ the expansion coefficient  ${ ^a\! \hat{F}}_{j}^{(j+1)}$ is
absent from this formula. Rather, it is contained in the subtraction
constant ${ ^a {\cal C}}$. Plugging this representation into Eq.\
(\ref{Def-MelBarSer}), we arrive after the shift of the integration
variable $j\to j-n$ at a Mellin--Barnes  representation for the
CFFs
\begin{eqnarray}
\label{Res-ImReCFF} { ^a\!{\cal F}}(\xi,\Delta^2,{\cal Q}^2)
\simeq \frac{1}{2i}\int_{c-i \infty}^{c+ i \infty}\!
dj\,\xi^{-j-1} \left[i +\left\{{\tan\atop
-\cot}\right\}\left(\frac{\pi j}{2}\right) \right] { ^a
C}_{j}({\cal Q}^2/\mu^2,\alpha_s(\mu)) { ^a\!F}_{j}
(\xi,\Delta^2,\mu^2)
 \,.
\end{eqnarray}
Here the conformal moments $F_{j}$ are  now written as a
Mellin--Barnes integral
\begin{eqnarray}
\label{Res-MelBarConMom}
 { ^a F}_{j} (\xi=\eta,\Delta^2,\mu^2) = \frac{i}{4 } \int_{d-i
\infty}^{d+i
 \infty}\! dn\;
 \frac{1 \pm  e^{i j \pi }}{1 \pm e^{i ( j - n)\pi }} (\eta-i \epsilon)^n
\frac{{^a \hat{F}}_{j}^{(n)}(\Delta^2,\mu^2)}{\sin(n \pi/2)}\,,
\end{eqnarray}
valid for $0\le \eta\le 1$. The (lower) upper sign corresponds to
the (axial-)vector case. We included in this definition a $-i
\epsilon$ prescription, which ensures that
\begin{eqnarray}
\frac{1 \pm  e^{i j \pi }}{1 \pm e^{i ( j - n)\pi }} (\eta-i
\epsilon)^n = \frac{1 \pm  e^{i j \pi }}{1 \pm e^{i ( j - n)\pi }}
e^{n (\ln\eta -i \epsilon) }
\end{eqnarray}
vanishes also for $n\to -i \infty$ in the complex $n$-plane
rather than generate an exponential $j$-dependent phase factor. We
remark that the exponential suppression in the integral
(\ref{Res-MelBarConMom}), arising from $1/\sin(n \pi/2)$ in the
limit $n\to \pm i \infty $, is in general needed to cancel an
exponential growing of ${^a \hat{F}}_{j}^{(n)}$.

The Mellin--Barnes representation (\ref{Res-ImReCFF}) for the
CFFs, which we will from now on use in all our numerical analysis,
has been already derived in two alternative ways in Refs.\
\cite{MueSch05,Mue05a}. As a gift of the present approach we found for
the conformal moments (\ref{Def-ConMomVec}) and
(\ref{Def-ConMomAxiVec}) the Mellin--Barnes integral
representation (\ref{Res-MelBarConMom}), which gives the analytic
continuation of the conformal moments. This issue has not been
completely resolved before.

We will discuss briefly the representation
(\ref{Res-MelBarConMom}) within our partonic toy GPD to shed light
on some tricky points. It can be shown that for the conformal
moments of GPD $H(x,\eta) = \delta(1-x)$ the integral
(\ref{Res-MelBarConMom}) exists for $0<\eta <1 $, where the
integrand can be read off from the second expression in Eq.\
(\ref{Def-h-TreLev}). After completing the integration path, it
reduces for complex valued $j$ to the function $H_{j}^{{\rm
tree}}$ in Eq.\ (\ref{Def-h-TreLev}) and an addenda that arises
from the additional phase factors in the definition
(\ref{Res-MelBarConMom}) of conformal moments. For $0\le \eta< 1$
we might express our improved findings in terms of hypergeometric
functions
\begin{eqnarray}
\label{Res-ConMom-tree} H^{\rm tree}_j(\eta) &\!\!\!=\!\!\!&
\frac{\Gamma(3/2)\Gamma(3 + j)}{2^{1+j}\Gamma(3/2 + j) } \eta^j\;
{_2\!F_1}\left( { -j, j+3\atop
2 }\Big|\frac{\eta-1}{2\eta}\right) \\
&&- \frac{4^{-j-1}\eta^{3+2j}}{i+\tan(\pi j/2)}
\frac{\Gamma(1+j)\Gamma(3 + j)}{\Gamma(3/2 + j)\Gamma(5/2 + j) }\;
{_2\!F_1}\left( {3/2+j/2, 2+j/2\atop 5/2+j }\Big|\eta^2-i
0\right)\,. \nonumber
\end{eqnarray}
Hence for positive odd integer  $j$ this formula reduces to the
Gegenbauer polynomials (times the conventional normalization
factor). The addenda is proportional to $1/(i+\tan(\pi j/2))$,
where this factor cancels the corresponding one in the conformal
partial wave expansion (\ref{Res-ImReCFF}). We can again close the
integration path by an infinite arc so that the first and forth
quadrant is encircled. It can be shown that the infinite arc does
not contribute. Since the integrand stemming from the addenda is a
holomorphic function in $j$, we conclude that the addenda finally
vanishes for $0\le\eta<1$. Consequently, we have found that the
appropriate analytic continuation of the conformal GPD moments is
given by the first term on the r.h.s.\ of Eq.\
(\ref{Res-ConMom-tree}), i.e., the analytic continuation of
Gegenbauer polynomials is given by the replacement
\begin{eqnarray}
\label{DefAnaConC}
\frac{\Gamma(3/2)\Gamma(1+j)}{2^{j} \Gamma(3/2+j)}\, \eta^j\
C_j^{3/2}\!\left(\! \frac{1}{\eta}\right)\;\; \Rightarrow\;\;
\frac{\Gamma(3/2)\Gamma(3 +j)}{2^{1+j}\Gamma(3/2 + j) } \eta^j\;
{_2\!F_1}\left( { -j,j+3\atop 2
}\Big|\frac{\eta-1}{2\eta}\right)\quad \mbox{for}\quad \eta\ge
0\,.
\end{eqnarray}
This result allows also for the extension of the $\eta$ region to
$\eta\ge 1$. Of course, symmetry under reflection allows for
negative values of $\eta$ by replacement $\eta\to -\eta$.

\subsubsection{Mellin--Barnes representation of conformal GPD moments}

First, notice that ${ ^a \hat{F}}_{j}^{(n)}$ from the
Mellin--Barnes representation (\ref{Res-MelBarConMom}) is not an
arbitrary function; rather it must be guaranteed that ${ ^a F}_j$
\begin{itemize}
\item[\phantom{ii}i)]  is holomorphic in the first and fourth
quadrant of the complex $j$-plane
\item[\phantom{i}ii)] reduces for
non-negative integers $j$ to a polynomial of order $j$ or $j+1$
\item[iii)] possesses for $j\to\infty$ and $\xi\to 1$ an exponential suppression
factor $2^{-j}$.
\end{itemize}
 The
requirement $i)$ is satisfied when  it holds for ${^a \!
\hat{F}}_{j}^{(n)}$.  To show that also the second requirement is
fulfilled, let us suppose that the analytic continuation of ${^ a
\! \hat{F}}_{j}^{(n)}$ is defined in such a way that it does not
grow exponentially for $n\to \infty$ with ${\rm arg }(n) \le
\pi/2$ and fixed $j$. Moreover, for simplicity let us assume that
${ ^a \! \hat{F}}_{j}^{(n)}$ is a meromorphic function with
respect to $n$. For $0<\eta<1$ we can then close the integration
contour in Eq.\ (\ref{Res-MelBarConMom}) so that the first and
forth quadrant are encircled. Cauchy's theorem states that ${^a \!
F}_{j}$ is given as a sum of the residues in these two quadrants
\begin{eqnarray}
\label{Res-MelBarConMom-Ser} {^a \! F}_{j}(\eta,\Delta^2,\mu^2) =
\frac{\pi}{2}  \sum_{\rm poles} {\rm Res} \left[\frac{1 \pm e^{i j
\pi }}{1 \pm e^{i ( j - n)\pi }} \frac{\eta^n}{\sin(n \pi/2)}
{^a\! \hat{F}}_{j}^{(n)}(\Delta^2,\mu^2)\right].
\end{eqnarray}
This result is in general a series, defining for fixed $\eta$ a
holomorphic function with respect to $j$. For non-negative integer
$j$, any contribution that does {\em not} arise from the poles at
$n=0,2,4,\cdots$ will drop out, due to the factor $ 1 \pm e^{i j
\pi }$. Consequently, in this case only the poles on the real
$n$-axis contribute and with the requirement ${^a \!
F}_{j}^{(n)}=0$ for $n\ge j+2$ we get the polynomial
\begin{eqnarray}
{^a\!F}_{j}(\eta,\Delta^2,\mu^2) = \sum_{n=0\atop {\rm
even}}^{j+p} \eta^n \; {^a\!
F}_{j}^{(n)}(\Delta^2,\mu^2)\quad\mbox{for}\quad
j=\{1,3,5,\cdots\}\;\mbox{or}\; j=\{0,2,4,\cdots\}\,,
\end{eqnarray}
where $p=\{0,1\}$. Note that for $p=1$ the highest possible term
of order $\eta^{j+1}$ is now included. Moreover, it is required
that for $\eta=\xi >1$ the  following expression, which appears as
integrand in the Mellin--Barnes integral (\ref{Res-ImReCFF}),
\begin{eqnarray}
\label{eq:condt}
\lim_{j\to\infty}
\frac{1}{\sqrt{j}}\left(\frac{\xi}{2}\right)^{-j-1} \left[i
+\left\{{\tan\atop -\cot}\right\}\left(\frac{\pi j}{2}\right)
\right] { ^a\! F}_{j}(\xi,\Delta^2,\mu^2)\; \ln^N j \,,
\end{eqnarray}
vanishes in the limit $j\to \infty$ for ${\rm arg}(j) \le \pi/2$.
Here we took into account that the Wilson coefficients in fixed order of
perturbation theory behave as
$\ln^N j/\sqrt{j}$ in the asymptotics we are considering. More precisely,
$1/\sqrt{j}$
arises from the factor $\Gamma(j+5/2)/\Gamma(3+j)$ in the normalization of
the Wilson coefficients,
see Eq.\  (\ref{Def-ResCOPEMomSpa-1}), and the logarithmical growing is
induced by radiative corrections.
The condition (\ref{eq:condt}) is necessary to ensure that the support of the
imaginary part of the CFFs is restricted to $|\xi|\le 1$. Namely,
for $\xi>1$ we can close the integration path to form a contour
that encircles the positive real axis. Obviously, the imaginary
part drops out and only the poles of the trigonometric functions
contribute, yielding the series given by the COPE
(\ref{Def-ConParDecInt}). For $\eta=\xi\to\infty$ we precisely
recover the subtraction constant ${^a {\cal C}}$, see Eq.\
(\ref{Pre-SubConC}), which shows that it is already contained in
the final representation (\ref{Res-ImReCFF}) of the CFFs in the
conformal moments (\ref{Res-MelBarConMom}).

Moreover, we can easily generalize the formula
(\ref{Res-MelBarConMom}) for the computation of the conformal
moments for a given GPD, which has definite symmetry with respect
to the momentum fraction $x$. In this case the analytic
continuation of the conformal moments are
\begin{eqnarray}
{^a\! F}_{j} (\eta,\Delta^2,\mu^2) &\!\!\!=\!\!\!&
 \frac{i \Gamma(1+j)}{4\Gamma(3/2+j)}\int_{d-i \infty}^{d+i
 \infty}\! dn\; \frac{1 \pm  e^{i j \pi }}{1 \pm e^{i ( j - n)\pi }}
 \frac{(\eta-i\epsilon)^n}{\sin(n \pi/2)} \frac{2^{-n}\Gamma(3/2+j-n/2)}{\Gamma(1+j-n)\Gamma(1+n/2)}
 \nonumber
 \\
 &&\times
 {\rm AC}\left[\int_{0}^1\!dx\;
 x^{j-n} \; {^a\! F}(x,\eta,\Delta^2,\mu^2) \right]\,.
\end{eqnarray}
Here the Mellin moments in the second line appear in the form
factor decomposition of the matrix elements of local operators
that are  built exclusively out of covariant derivatives. Precisely
these matrix elements are measured on the lattice
\cite{Hagetal03,Hagetal04,Hagetal04a,Gocetal03,Gocetal05,Edwetal06}. It is understood
that $\Re{\rm e }(j-n)$ is restricted to such values that the
integral exists. We note that the polynomiality of these moments
is ensured by the symmetry we assumed. So for instance for an
(anti-)symmetric GPD we have for even (odd) values of $n$
\begin{eqnarray}
\frac{1}{2}\int_{-1}^1\! dx\; x^n \; {^a\!
F}(x,\eta,\Delta^2,\mu^2) = \int_{0}^1\! dx\; x^n \; {^a\!
F}(x,\eta,\Delta^2,\mu^2) = \sum_{{i=0 \atop {\rm even}}}^{n+p}
{^a\! F}_{ni}(\Delta^2,\mu^2)\; \eta^i\,,
\end{eqnarray}
where $p$ takes the values $0$ and $1$ for even and odd $n$,
respectively. However,  irrespective of the symmetry $p=0$ holds
true for the combination $H+E$, $\widetilde H$, and $\widetilde
E$.

\section{DVCS amplitude up to next-to-next-to-leading order}
\label{Sec-PreConSym}

In Sect.\ \ref{SubSec-COPE-pre} we first discuss the
predictive power of conformal symmetry in a conformally invariant
theory for general kinematics. Here we rely on a special
subtraction scheme, the so-called conformal one, in which the
renormalized operators are covariant under collinear conformal
transformations. This allows us to predict the functional form of
the Wilson coefficients up to a normalization factor. This factor
is simply the Wilson coefficient that appears in DIS. To reduce the
massless QCD  to a conformally invariant theory, we intermediately
assume that the QCD has a non-trivial fixed-point so that the
trace anomaly vanishes. Then we move to the real world and
discuss the inclusion of the running coupling.
In Sect.\ \ref{SuSec-StrDVCS} we present the
results for the DVCS kinematics. The recent progress in the
evaluation of radiative corrections in DIS allows us to present
the CFFs to NNLO accuracy. However, the trace anomaly induces a
mixing of conformal GPD moments due to the evolution. Also note
that the forward anomalous dimensions of parity odd conformal GPD
moments are mostly unknown beyond NLO.
In Sect.\ \ref{SubSec-MS} we present the NLO
corrections to the twist-two CFFs in the $\overline{\rm MS}$
scheme and discuss the inclusion of mixing effects in the
Mellin--Barnes representation.

\subsection{Conformal OPE prediction and beyond}
\label{SubSec-COPE-pre}

In the hypothetical conformal limit, i.e., if the $\beta$ function
of QCD has a non-trivial fixed point  $\alpha_s^\ast \neq 0$, the
$\eta/\xi$ dependence of the Wilson coefficients is predicted by
conformal symmetry
\cite{FerGriGat71,FerGriGat72,FerGriGat72a,FerGriPar73}. For all the
cases we are considering, Wilson coefficients ${^a\! C}^{\rm I}$, with $a=\{{\rm
NS},+,-\}$ and ${\rm I}=\{{\rm V},{\rm A}\}$,
have the same general form. Thus, when there is no possibility of
confusion, we will simplify our notation by suppressing both of these
superscripts. The COPE prediction for general kinematics reads
\cite{Mue97a}
\begin{eqnarray}
\label{Def-ResCOPEMomSpa-2} C_j(\eta/\xi, Q^2/\mu^2,\alpha_s^\ast)
=  c_j(\alpha_s^\ast)\, {_2F_1}\!\!\left(\!\!{(2+2 j
+\gamma_j(\alpha_s^\ast))/4, (4+2 j + \gamma_j(\alpha_s^\ast))/4
\atop (5+2 j +
\gamma_j(\alpha_s^\ast))/2}\Big|\frac{\eta^2}{\xi^2}\! \right)
\left(\!\!\frac{\mu^2}{Q^2}\!\!\right)^{\gamma_j(\alpha_s^\ast)/2},
\end{eqnarray}
where the normalization $c_j$ remains so far unknown.  In the
kinematical forward limit ($\eta =0$) the Wilson coefficients
coincide with those known from DIS
\begin{eqnarray}
\label{Def-ResCOPEMomSpaFW} \lim_{\eta\to 0}
 C_j(\eta/\xi,
Q^2/\mu^2,\alpha_s^\ast) =  c_j(\alpha_s^\ast)
\left(\!\!\frac{\mu^2}{Q^2}\!\!\right)^{\gamma_j(\alpha_s^\ast)/2}\,,
\end{eqnarray}
see Appendix \ref{App-NorWilCoe}. Neither Wilson coefficients
(\ref{Def-ResCOPEMomSpa-2}) nor conformal GPD moments
(\ref{Def-ConMomVec}), (\ref{Def-ConMomAxiVec}) mix under
collinear conformal transformations. This implies that they
evolve autonomously with respect to a scale change,
\begin{eqnarray}
\label{Def-EvoEq-Ope}
 \mu \frac{d}{d\mu}  F_j(...,\mu^2) &\!\!\!=\!\!\!&
-\gamma_{j}(\alpha_s^\ast)\,  F_j \,
(...,\mu^2) \, ,
\\
\label{Def-EvoEq-WC} \mu \frac{d}{d\mu}
C_j(...,Q^2/\mu^2,\alpha_s^\ast) ]&\!\!\!=\!\!\!&
 C_j(...,Q^2/\mu^2,\alpha_s^\ast)\,
\gamma_{j}(\alpha_s^\ast)
\, .
\end{eqnarray}
Here the anomalous dimensions are the same as in DIS.  We also realize that
the conformal spin $j+2$ is a good quantum number.

Unfortunately, conformal symmetry is not manifest in a general
factorization scheme, and in particular not in the $\overline{\rm
MS}$ scheme. What is required is a special scheme, which ensures
that the {\em renormalized} conformal operators form an
irreducible representation of the collinear conformal group
SO(2,1) \cite{Mue97a}. The breaking of the covariant behavior with
respect to dilatation, e.g., shows up in the mixing of operators' expectation
values and of Wilson coefficients under renormalization:
\begin{eqnarray}
\label{Def-EvoEq-gen-Ope}
 \mu \frac{d}{d\mu}  F_j(...,\mu^2) &\!\!\!=\!\!\!&
-\sum_{k=0}^j\gamma_{jk}(\alpha_s(\mu))\, \eta^{j-k}\, F_k \,
(...,\mu^2) \, ,
\\
\label{Def-EvoEq-gen-WC} \mu \frac{d}{d\mu}
C_j(...,Q^2/\mu^2,\alpha_s(\mu)) &\!\!\!=\!\!\!&
\sum_{k=j}^\infty C_k(...,Q^2/\mu^2,\alpha_s(\mu))
\gamma_{kj}(\alpha_s(\mu))\left(\frac{\eta}{\xi}\right)^{k-j} \,
\, ,
\end{eqnarray}
respectively%
\footnote{Both these equations together ensure that the CFF
$\int\!\!\!\!\!\! {\scriptstyle \sum_j} \xi^{-j-1}
C_j(\cdots,Q^2/\mu^2,\alpha_s(\mu)) F_j(\cdots,\mu)$ is a
renormalization group invariant quantity.}. This mixing arises in
NLO and is induced by the breaking of special conformal symmetry
in LO. The mixing matrix, i.e., $\gamma_{jk}$ for $j> k$, contains
besides a $\beta$ proportional term also one that does not vanish
in the hypothetical conformal limit.
Knowing the form of the special conformal symmetry breaking
in the $\overline{\rm MS}$ scheme to
LO, it was shown that, after rotation, the diagrammatic evaluation
of NLO corrections to (flavor nonsinglet) anomalous dimensions
\cite{DitRad84,Sar84,MikRad85} and Wilson coefficients
\cite{ManPilSteVanWei97,JiOsb97,JiOsb98} in the $\overline{\rm
MS}$ scheme coincide, for $\beta=0$, with the conformal symmetry
predictions \cite{Mue94,Mue97a,BelMue97a}. A formal proof about
the restoration of conformal symmetry, valid to all orders of
perturbation theory, has been given in Ref.\ \cite{Mue97a}. Hence,
we have no doubt that conformal symmetry can be effectively used
for the evaluation of higher order perturbative corrections
\cite{MelMuePas02, Mue05a, KumMuePasSch06}. The interested reader
can find a comprehensive review of the uses of conformal symmetry
in QCD in Ref.\ \cite{BraKorMue03}.

Let us suppose that we employ a scheme in which conformal symmetry
is manifest in the hypothetical conformal limit. Such a scheme we
call a conformal subtraction (CS) scheme. We discuss now the
structure of the conformal OPE beyond this limit, i.e., the
implementation of the running coupling constant  in the
Wilson coefficients (\ref{Def-ResCOPEMomSpa-2}) and the evolution
equation. In the simplest case, i.e., forward kinematics, the
conformal representation is trivial, as there are only operators
without total derivatives left, so we can restore the
Wilson coefficients of the full perturbative theory
from the result (\ref{Def-ResCOPEMomSpaFW}) of the conformal limit.
Namely,
renormalization group invariance of the CFFs allows us to revive
the $\beta$-proportional term by expanding the exponential%
\footnote{We consider $c_j$ and $\gamma_j$ as $\beta$ independent
quantities and so their perturbative expansion in the fixed-point
and full theory looks the same. Certainly, both of them contain to
two-loop accuracy $\beta_0$ proportional terms and the anomalous
dimensions to three-loop accuracy might be also rewritten in terms
of the $\beta$ function expanded to two-loop accuracy.}
\begin{eqnarray}
c_j(\alpha_s^\ast)
\left(\frac{\mu^2}{Q^2}\right)^{\gamma_j(\alpha_s^\ast)/2}\qquad
\Rightarrow \qquad  c_j(\alpha_s(Q))\,
\exp\left\{\int^\mu_Q\frac{d\mu^\prime}{\mu^\prime}
\gamma_j(\alpha_s(\mu^\prime))\right\}
\end{eqnarray}
in terms of $\ln(Q^2/\mu^2)$. Of course, also the evolution of
Mellin moments of parton densities is governed by a diagonal
evolution equation, i.e., by Eq.\ (\ref{Def-EvoEq-gen-Ope}) with
$\eta=0$. Finally, we are using the normalization condition
\begin{eqnarray}
\label{NorCon-c}
c_j(\alpha_s) =c_j^{\overline{\rm MS}}(\alpha_s)
\end{eqnarray}
and thus we end up with the standard result for the
Wilson coefficients of DIS, evaluated in the $\overline{\rm MS}$
scheme.

For general kinematics the $\beta$ proportional term is not
automatically fixed, so different prescriptions might be used.
For a discussion of this issue see Ref.\ \cite{MelMuePas02}. We
will point out here two appealing possibilities.

Let us first assume that the conformal symmetry breaking can be
entirely incorporated in the running of the coupling so that the
evolution equation for conformal GPD moments is diagonal:
\begin{eqnarray}
\mu \frac{d}{d\mu}  F_j(\eta, \Delta^2,\mu^2) =
-\gamma_j(\alpha_s(\mu))\, F_j \, (\eta, \Delta^2,\mu^2)\,.
\end{eqnarray}
Then also the Wilson coefficients in the full theory are  simply
obtained by replacing the expression for the resummed evolution
logs. However, it is likely that the trace anomaly generates a non
vanishing $\beta$ proportional addenda, which will appear at
two-loop level. If it does, renormalization group invariance fixes
the scale dependence and so the Wilson coefficients in such a
scheme read for the full theory
\begin{eqnarray} \label{Def-ResCOPEMomSpaFull}
C^{\rm full}_j(\eta/\xi,Q^2/\mu^2,\alpha_s(\mu))
&\!\!\!=\!\!\!& \left[
C_j(\eta/\xi,1,\alpha_s(Q)) +\frac{\beta}{g}(\alpha_s(Q)) \Delta
C_j(\eta/\xi,1,\alpha_s(Q))\right]
\nonumber\\
&&\times\exp\left\{\int^\mu_Q\frac{d\mu^\prime}{\mu^\prime}
\gamma_j(\alpha_s(\mu^\prime))\right\}\,,
\end{eqnarray}
where $C_j(\eta/\xi,1,\alpha_s)$ is the COPE prediction
(\ref{Def-ResCOPEMomSpa-2}). In the forward limit we again require
that the Wilson coefficients coincide with the DIS ones,
evaluated in the $\overline{\rm MS}$. Thus, we have the condition
that the $\beta$ proportional addenda vanishes, i.e.,
\begin{eqnarray}
\lim_{\eta\to 0} \Delta  C_j(\eta/\xi,1,\alpha_s({\cal Q})) =0\,.
\end{eqnarray}

The disadvantage of such a conformal scheme is  that a consistent determination of
the $\beta$ proportional addenda $\Delta C_j$ requires a diagrammatical evaluation at
NNLO, e.g., in the $\overline{\rm MS}$ scheme. Both the knowledge of the
$\beta_0$-proportional terms in the two-loop corrections to the
Wilson coefficients and the off-diagonal three-loop corrections to the
anomalous dimensions are required. While the former are known in the
$\overline{\rm MS}$ scheme  \cite{BelSch98,MelNicPas02}, the calculation of the latter,
although reducible to a two-loop problem by means of conformal constraints,
still requires a great effort. Fortunately, to NNLO only the quark sector
suffers from this uncertainty. The best what we can do to NNLO
accuracy is to minimize the effects of this lack of knowledge.
To do so, we will perform a transformation into a new scheme,
called $\overline{\rm CS}$, where the normalization condition reads
\begin{eqnarray}
\label{Def-NorConDel}
 \Delta C_j(\eta/\xi,1,\alpha_s(Q)) =0\,,
\end{eqnarray}
while the mixing term is now shifted to the evolution equation. This
mixing term will be suppressed at the input scale by an
appropriate initial condition and we might expect that the
evolution effects are small, see discussion in Sect.\ \ref{SecCon2MSscheme} below.
Hence the Wilson coefficients read now
\begin{eqnarray}
\label{Def-ResCOPEMomSpaTru}
C_j(\eta/\xi,Q^2/\mu^2,\alpha_s(\mu))
&\!\!\!=\!\!\!& \sum_{k=j}^\infty  C_k(\eta/\xi,1,\alpha_s(Q))
\\
&&\times {\cal P}
\exp\left\{\int^\mu_Q\frac{d\mu^\prime}{\mu^\prime}
\left[\gamma_j(\alpha_s(\mu^\prime))
\delta_{kj}+\left(\frac{\eta}{\xi}\right)^{k-j} \frac{\beta}{g}
\Delta_{kj}(\alpha_s(\mu^\prime))\right]\right\}\,,
\nonumber
\end{eqnarray}
where $C_k(\eta/\xi,1,\alpha_s(Q))$ is given in Eq.\
(\ref{Def-ResCOPEMomSpa-2}), $\Delta_{kj}$ is determined by the
off-diagonal part of the anomalous dimension matrix, and ${\cal
P}$ denotes the path ordering operation%
\footnote{The path ordering operation defined by
\begin{displaymath}
{\cal P} \exp \left\{\int^\mu_Q\frac{d\mu^\prime}{\mu^\prime}
\gamma_{jk}(\alpha_s(\mu^\prime))\right\} = 1 +
\int^\mu_Q\frac{d\mu^\prime}{\mu^\prime}
\gamma_{jk}(\alpha_s(\mu^\prime)) +
\int^\mu_Q\frac{d\mu^\prime}{\mu^\prime} \int^{\mu
'}_Q\frac{d\mu^{\prime \prime}}{\mu^{\prime \prime}}
\sum_{m=k}^j\gamma_{jm}(\alpha_s(\mu^\prime)) \gamma_{mk}(\alpha_s(\mu^{\prime
\prime})) + \ldots
\end{displaymath}
enables the compact representation of the solution of the
non-diagonal renormalization group equations. Obviously, in
diagonal case, i.e. for $\gamma_{jk}=\delta_{jk} \gamma_j$, path
ordering converts the evolution operator in a simple exponential
function.}. The renormalization group equation for the conformal
moments reads
\begin{eqnarray}
\label{RGE-used} \mu \frac{d}{d\mu}  F_j(\cdots,\mu^2) =
-\gamma_j(\alpha_s(\mu))\,  F_j \, (\cdots,\mu^2) -
\frac{\beta(\alpha_s(\mu))}{g(\mu)} \sum_{k=0}^{j-2}
\eta^{j-k}\Delta_{jk} (\alpha_s(\mu))
F_k(\cdots,\mu^2)\,.
\end{eqnarray}
Since this mixing term appears for the first time at NNLO accuracy, and
 is suppressed at the input scale by the initial condition, we
might expect that its effects will be small.

Formulae (\ref{Def-ResCOPEMomSpaTru}) and (\ref{RGE-used}) are our
main result that can be used for the numerical investigation of
radiative corrections to DVCS up to NNLO.

\subsection{Perturbative expansion of Compton form factors}
\label{SuSec-StrDVCS}

For DVCS kinematics the general expression for the Wilson
coefficients (\ref{Def-ResCOPEMomSpaTru}) simplifies considerably.
Since to twist-two accuracy $\vartheta=\eta/\xi=1$ is valid, the
argument of hypergeometric functions is unity and they simplify to
products of $\Gamma$ functions.  Having also in mind that in DVCS
the photon virtuality ${\cal Q}^2 \simeq 2 Q^2$ is considered as
the relevant scale, the COPE prediction
(\ref{Def-ResCOPEMomSpa-2}) together with Eq.\
(\ref{Def-ResCOPEMomSpaTru}) leads to the following expression for
the DVCS Wilson coefficients
\begin{eqnarray}
\label{Def-WilCoeDVCScope}
C_{j}({\cal Q}^2/\mu^2,\alpha_s(\mu))) &\!\!\!\equiv\!\!\! &
C_{j}(\vartheta=1,{\cal Q}^2/(2\mu^2),\alpha_s(\mu))
\\
&\!\!\!=\!\!\! & c_{j}(\alpha_s({\cal Q}))
\frac{2^{j+1+\gamma_j(\alpha_s({\cal Q}))}
\Gamma\!\left(\frac{5}{2}+j+\frac{1}{2} \gamma_j(\alpha_s({\cal
Q}))\right)}{\Gamma\!\left(\frac{3}{2}\right)\Gamma\!\left(3+j+\frac{1}{2}
\gamma_j(\alpha_s({\cal Q}))\right)} \exp\left\{\int^\mu_{{\cal
Q}}\frac{d\mu^\prime}{\mu^\prime} \gamma_j(\alpha_s(\mu^\prime))
\right\}\,
\nonumber
\end{eqnarray}
at the considered NNLO accuracy. Note that the off-diagonal part
of the evolution operator does not appear in this approximation,
and that  we have slightly changed the normalization condition for
$\mu={\cal Q}$. In fact to arrive at the above result, Eq.\
(\ref{Def-ResCOPEMomSpaTru}) has to be multiplied by a factor
\begin{eqnarray}
\left(\sqrt{2}\right)^{\gamma_j(\alpha_s({\cal Q}))}
\exp\left\{-\int^{\cal Q}_{{\cal
Q}/\sqrt{2}}\frac{d\mu^\prime}{\mu^\prime}
{\gamma_j}(\alpha_s(\mu^\prime)) \right\} =
1+\frac{\alpha_s^2({\cal Q})}{(2\pi)^2} \frac{\beta_0}{8}
{\gamma_j^{{(0)}}} \ln^2(2) + {\cal O}(\alpha_s^3)\,,
\end{eqnarray}
and such terms can be compensated by a change of the normalization
condition (\ref{Def-NorConDel}). Here and in the following the
first expansion coefficient of $\beta(g)/g = (\alpha_s/4\pi)
\beta_0 + {\cal O}(\alpha_s^2)$ is $\beta_0 =
(2/3) n_f-11$, where $n_f$ is the number of active quarks.

We now use perturbation theory to determine $C_j({\cal
Q}^2/\mu^2,\alpha_s({\mu}))$ to NNLO accuracy. In contrast to
$F_j(\cdots,\mu^2)$, it is customary to express $C_j({\cal
Q}^2/\mu^2,\alpha_s(\mu))$ completely expanded in
$\alpha_s(\mu)$, i.e., without resummation of leading logs from
$\mu$ to $Q$. This actually corresponds to the result that one
would obtain from a diagrammatical calculation, without using the
renormalization group equation. Obviously, since the leading
logarithms in $F_j$ will be resumed and in $C_j$ not, one will end
up with a residual dependence on the factorization scale $\mu$. This
scale might also be used as the relevant scale in the running coupling
constant. However, for generality, another scale $\mu_r$ is
better suited for the expansion parameter $\alpha_s(\mu_r)$. The
truncation of perturbation theory then implies that our results
will depend also on this renormalization scale $\mu_r$.

The perturbative expansion of the DVCS Wilson coefficients
(\ref{Def-WilCoeDVCScope}) in terms of $\alpha_s(\mu_r)$ is
done in a straightforward manner by means of
\begin{eqnarray}
\alpha_s(\mu) = \alpha_s(\mu_r) + \beta_0
\frac{\alpha_s^2(\mu_r)}{4\pi} \ln\frac{\mu^2}{\mu_r^2}
\end{eqnarray}
and consequently we can expand the exponential factor
\begin{eqnarray}
\exp\left\{\int^\mu_{{\cal Q}}\frac{d\mu^\prime}{\mu^\prime}
{\gamma_j}(\alpha_s(\mu^\prime)) \right\} = \left(
\frac{\mu^2}{{\cal Q}^2}\right)^{\gamma_j(\alpha_s(\mu_r))/2}
\left[1-\frac{\alpha_s^2(\mu_r)}{(2\pi)^2} \frac{\beta_0}{8}
\gamma_j^{(0)} \ln\frac{{\cal Q}^2}{\mu^2} \ln\frac{\mu^2 {\cal
Q}^2}{\mu_r^4} + {\cal O }(\alpha_s^3) \right].
\label{eq:expCOPE}
\end{eqnarray}
Furthermore, we use the expansion
\begin{eqnarray}
\frac{2^{j+1+\gamma_j}
\Gamma\left(\frac{5}{2}+j+\frac{1}{2}\gamma_j\right)}{\Gamma\!\left(\frac{3}{2}\right)\Gamma\!\left(3+j+\frac{1}{2}
\gamma_j\right)} \left( \frac{\mu^2}{{\cal
Q}^2}\right)^{\gamma_j/2} = \frac{2^{j+1}
\Gamma(5/2+j)}{\Gamma(3/2)\Gamma(3+j)}\;\sum_{m=0}^\infty
 \frac{s^{(m)}_j({\cal Q}^2/\mu^2)}{2^m\, m!}
\left[ \gamma_j \right]^m, \label{eq:copeCexp1}
\end{eqnarray}
where the so-called shift coefficients, which also include the
factorization logs, are defined as
\begin{eqnarray}
s^{(m)}_j({\cal Q}^2/\mu^2)= \frac{d^m}{d\rho^m} \left(
\frac{\mu^2}{{\cal Q}^2} \right)^\rho \frac{4^\rho\Gamma(3+j)
\Gamma(5/2+j+\rho)}{\Gamma(5/2+j)\Gamma(3+j+\rho)}
\Big|_{\rho=0}\,. \label{eq:sQ2}
\end{eqnarray}
The first two shift coefficients read in terms of harmonic sums as
\begin{subequations}
\label{eq:s12Q2}
\begin{eqnarray}
s_j^{(1)}({\cal Q}^2/\mu^2)&= &
  S_1(j+3/2)-S_1(j+2) +2\ln(2)-\ln\frac{{\cal Q}^2}{\mu^2}\,, \quad
 \\
s_j^{(2)}({\cal Q}^2/\mu^2)&=& \left(s_j^{(1)}({\cal
Q}^2/\mu^2)\right)^2
   -S_2(j+3/2)+ S_2(j+2) \,,
\end{eqnarray}
\end{subequations}
where the analytical continuation of these sums are defined by
\begin{eqnarray}
S_1(z)= \frac{d}{dz} \ln \Gamma(z+1)+\gamma_E \quad  \mbox{and}\quad S_2(z)= -\frac{d^2}{dz^2}\ln
\Gamma(z+1) + \frac{\pi^2}{6}
\end{eqnarray}
with $\gamma_E=0.5772\dots$ being the Euler constant.
Taking into account the perturbative expansion of anomalous dimensions and DIS
Wilson coefficients, written as
\begin{eqnarray}
\gamma_j(\alpha_s)&\!\!\!=\!\!\!& \frac{\alpha_s}{2\pi}
\,\gamma_j^{(0)} + \frac{\alpha_s^2}{(2\pi)^2}\,   \gamma_j^{(1)}
+ \frac{\alpha_s^3}{(2 \pi)^3} \,   \gamma_j^{(2)}+ {\cal
O}(\alpha_s^4) \, ,
\nonumber \\
 c_{j}(\alpha_s)&\!\!\!=\!\!\!&  c_{j}^{(0)}
+\frac{\alpha_s}{2 \pi}\,  c_{j}^{(1)} + \frac{\alpha_s ^2}{(2
\pi)^2}\, c_{j}^{(2)} + {\cal O}(\alpha_s^3) \, ,
\end{eqnarray}
the DVCS Wilson coefficients (\ref{Def-WilCoeDVCScope}) take the
form
\begin{subequations}
\label{eq:copeCexp2all}
\begin{eqnarray}
\label{eq:copeCexp2} C_{j}
= \frac{2^{j+1}
\Gamma(j+5/2)}{\Gamma(3/2)\Gamma(j+3)} \left[ c_{j}^{(0)}
+\frac{\alpha_s(\mu_r)}{2 \pi}\, C_{j}^{(1)}({\cal Q}^2/\mu^2)
+ \frac{\alpha_s^2(\mu_r)}{(2 \pi)^2} C_{j}^{(2)}({\cal
Q}^2/\mu^2, {\cal Q}^2/\mu_r^2)
+ {\cal O}(\alpha_s^3) \right],
\end{eqnarray}
where
\begin{eqnarray}
 C_{j}^{(1)}
 &\!\!\!=\!\!\!& { c_{j}^{(1)}}
+ \frac{s^{(1)}_j({\cal Q}^2/\mu^2)}{2} \; { c_{j}^{(0)}}
   \; \gamma_j^{(0)}
\,,
\\
\label{eq:copeCexp2c}
C_{j}^{(2)}
&\!\!\!=\!\!\!&  c_{j}^{(2)} + \frac{s^{(1)}_j({\cal
Q}^2/\mu^2)}{2} \;
 \left[ c_{j}^{(0)} \; \gamma_j^{(1)}
 + c_{j}^{(1)} \;  \gamma_j^{(0)}\right]
+ \frac{s^{(2)}_j({\cal Q}^2/\mu^2)}{8} \; c_{j}^{(0)}
   \; \left( \gamma_j^{(0)}\right)^2
\\ &&\!\!\!+
   \frac{\beta_0}{2}\left(
    C_{j}^{(1)}({\cal
   Q}^2/\mu^2) \ln\frac{{\cal Q}^2}{\mu_r^2}
   + \frac{1}{4} c_{j}^{(0)} \gamma_j^{(0)} \ln^2\frac{{\cal Q}^2}{\mu^2}\right) \, . \nonumber
\end{eqnarray}
\end{subequations}
Notice that the renormalization group logs $\ln\!\left({\cal
Q}^2/\mu_r^2\right)$, caused by the running of $\alpha_s$, and
proportional to $\beta_0$, appear for the first time in the NNLO
correction $C_j^{(2)}$.

\subsubsection{Flavor nonsinglet Wilson coefficients}
\label{SubSec-PerExpWC}

The flavor nonsinglet Wilson coefficients have the form as
indicated in Eqs.\ (\ref{eq:copeCexp2})--(\ref{eq:copeCexp2c}). We
only have to decorate them  and the anomalous dimensions with the
corresponding superscripts. It has been already taken into account
that the DIS Wilson coefficients are normalized as
\begin{eqnarray}
\label{Def-DIS-WC-LO}
{^{\rm NS}\!c}_j^{{\rm I}(0)}=1 \,.
\end{eqnarray}
The
radiative corrections to NLO involve the
Wilson coefficients of the DIS unpolarized structure function
$F_1$,
\begin{eqnarray}
\label{eq:NScV1}
 {^{\rm NS}\!c}_j^{{\rm V}(1)}
=
     C_F \left[S^2_{1}(1 + j) + \frac{3}{2} S_{1}(j + 2)
      - \frac{9}{2}  + \frac{5-2S_{1}(j)}{2(j + 1)(j + 2)}
      - S_{2}(j + 1)\right]\, ,
\end{eqnarray}
for the vector case and those of the polarized structure function $g_1$,
\begin{eqnarray}
 {^{\rm NS}\!c}_j^{{\rm A}(1)}
=
 C_F \left[S^2_{1}(1 + j) + \frac{3}{2} S_{1}(j + 2)
  - \frac{9}{2}  + \frac{3-2S_{1}(j)}{2(j + 1)(j + 2)}
  - S_{2}(j + 1)\right]\,,
\label{eq:NScA1}
\end{eqnarray}
for the axial-vector one. The LO anomalous dimensions read in
both cases
\begin{eqnarray}
\label{eq:NSgamma0}
{^{\rm NS}\!\gamma}_{j}^{{\rm I}(0)} = - C_F \left( 3 + \frac{2}{(
j + 1 )( j + 2 )} - 4 S_{1}(j + 1) \right) \,,
\end{eqnarray}
 where $C_F=(N_c^2-1)/(2 N_c)$ with $N_c$ being the number of
colors.

The expressions for two-loop quantities are
lengthy. The nonsinglet anomalous dimensions $\gamma_j^{(1)}$ for the vector
and axial-vector case are given, e.g.,\ in Ref.\
\cite{CurFurPet80}, by analytic continuation of $j$-odd and $j$-even
expressions, respectively. The NNLO Wilson coefficients $c_j^{(2)}$ can be
read off for the vector and axial-vector cases from Refs.\
\cite{ZijNee92} and \cite{ZijNee94}, respectively. Implementation
of evolution to NNLO accuracy requires also the anomalous
dimensions to three-loop order \cite{MocVerVog04}. Instead of the
exact expressions one might alternatively use an analytic
approximation, given for the vector quantities in Refs.\
\cite{NeeVog99,MocVerVog04}.

\subsubsection{Flavor singlet Wilson coefficients}

So far the  $\{+,-\}$ basis, which resolves the mixing problem in
the flavor singlet sector, has been used for presenting the
conformal predictions.  However, the Wilson coefficients are in
the literature usually given in the $\{Q\equiv \Sigma,G\}$ basis.
For convenience, we might express  the flavor singlet results in
the basis of quark and gluon degrees of freedom.   We write the
Wilson coefficients as a two-dimensional row vector
\begin{eqnarray}
\label{Def-WilCoeDVCS} {\mbox{\boldmath $C$}_{j}^{\rm I}}({\cal
Q}^2/\mu^2,\alpha_s(\mu)) &\!\!\!=\!\!\!& \left({{^{\Sigma}
C_{j}^{\rm I}} \;,\; {^G C_{j}^{\rm I}}} \right)({\cal
Q}^2/\mu^2,\alpha_s(\mu))\quad\mbox{for}\quad {\rm I}=\{ {\rm V},
{\rm A}\}\,.
\end{eqnarray}
In the following we drop the superscript ${\rm I}$. The
rotation between the two bases is given by the
transformation matrix $\mbox{\boldmath $U$}_j$, defined in Eq.\
(\ref{Def-Bas-Tra}). Hence, for the Wilson coefficients we have
\begin{eqnarray}
\label{Def-WilCoeDVCSpm} {\mbox{\boldmath $C$}_{j}}({\cal
Q}^2/\mu^2,\alpha_s(\mu)) &\!\!\!=\!\!\!& \left({{^+ C_{j}}  \;,\;
{^- C_{j}}} \right)({\cal Q}^2/\mu^2,\alpha_s(\mu))\;\mbox{\boldmath
$U$}_j(\mu,{\cal Q}) \, ,
\end{eqnarray}
where the $\{+,-\}$ entries are given in Eq.\
(\ref{Def-WilCoeDVCScope}). Using the property (\ref{Def-U-ope})
of the transformation matrix within $\mu_0={\cal Q}$, derived in
Appendix \ref{App-BasFlaSinSec}, allows us to perform the
transformation at the normalization point $\mu={\cal Q}$ and
afterwards restore the scale dependence by an backward evolution:
\begin{eqnarray}
\label{Def-WilCoeDVCSpm1}
{\mbox{\boldmath $C$}_{j}}({\cal
Q}^2/\mu^2,\alpha_s(\mu)) = \left({{^+ C_{j}}  \;,\; {^- C_{j}}} \right)({\cal
Q}^2/\mu^2=1,\alpha_s({\cal Q}))\;\mbox{\boldmath $U$}_j({\cal
Q},{\cal Q})\, \mbox{\boldmath ${\cal E}$}_j^{-1}(\mu,\cal Q) \,.
\end{eqnarray}
Here $\mbox{\boldmath ${\cal E}$}_j^{-1}$ is the inverse
of the evolution operator in the $\{\Sigma,G\}$ basis, defined below in
Eq.\ (\ref{Def-EvoOpe}). Since $\mbox{\boldmath $U$}_j(\cal
Q,\cal Q)$ rotates the anomalous dimensions [depending on
$\alpha_s({\cal Q})$], we can express for $\mu={\cal Q}$ the
flavor singlet contributions in the form
\begin{eqnarray}
\label{eq:sinCexp1}
{ \mbox{\boldmath $C$}_{j}}({\cal Q }^2/\mu^2=1,\alpha_s({\cal Q})) = \frac{2^{j+1}
\Gamma(5/2+j)}{\Gamma(3/2)\Gamma(3+j)}\;\sum_{m=0}^\infty
 \frac{s^{(m)}_j({\cal Q}^2/\mu^2=1) }{2^m\, m!}\,
{\mbox{\boldmath $c$}_{j}} \left[{\mbox{\boldmath$\gamma$}_j}
\right]^m\,,\quad {\mbox{\boldmath $c$}_{j}}=
 \left({{^\Sigma\! c_{j}} \;,\; {^G\! c_{j}}} \right)
\, ,
\end{eqnarray}
where the anomalous dimension matrix is defined as
\begin{eqnarray}
\mbox{\boldmath$\gamma$}_j=\frac{\alpha_s}{2\pi}\left(
\begin{array}{cc}
{^{\Sigma \Sigma}\!\gamma}_{j}^{(0)} & {^{\Sigma G}\!\gamma}_{j}^{(0)}\\
{^{G \Sigma}\!\gamma}_{j}^{(0)} & {^{GG}\!\gamma}_{j}^{(0)}
\end{array}\right)+ {\cal O}(\alpha_s^2)\,.
\end{eqnarray}
This matrix valued expansion, although resulting from
somewhat more involved derivation, is in complete analogy to
the flavor nonsinglet result (\ref{eq:copeCexp1}).

Acting with the evolution operator $\mbox{\boldmath ${\cal
E}$}_j^{-1}(\mu,\cal Q)$ on Eq.\ (\ref{eq:sinCexp1}) removes the
constraint ${\cal Q}^2/\mu^2=1$ of the shift functions argument.
As in the preceding section, we  consequently expand with respect
to $\alpha_s(\mu_r)$ and finally write the Wilson coefficients
\req{Def-WilCoeDVCS} in complete analogy  to Eq.\
\req{eq:copeCexp2all} as
\begin{subequations}
\label{eq:copeCexp2allSI}
\begin{equation}
\label{Res-WilCoe-Exp-CS-SI}
 \mbox{\boldmath $C$}_{j}^{}
= \frac{2^{j+1} \Gamma(j+5/2)}{\Gamma(3/2)\Gamma(j+3)}
\left[{\mbox{\boldmath $c$}_{j}^{(0)}}  +
\frac{\alpha_s(\mu_r)}{2\pi} \mbox{\boldmath $C$}_j^{ (1)}({\cal
Q}^2/\mu^2) + \frac{\alpha^2_s(\mu_r)}{(2\pi)^2} \mbox{\boldmath
$C$}_j^{(2)}({\cal Q}^2/\mu^2, {\cal Q}^2/\mu_r^2) + {\cal
O}(\alpha_s^3) \right],
\end{equation}
where
\begin{eqnarray}
\label{Res-WilCoe-CS-NLO} \mbox{\boldmath $C$}_j^{(1)}({\cal
Q}^2/\mu^2) &\!\!\! =\!\!\!  & \mbox{\boldmath ${c}$}_j^{(1)}+
\frac{s^{(1)}_j({\cal Q}^2/\mu^2)}{2} \; \mbox{\boldmath
${c}$}_j^{(0)} \mbox{\boldmath ${\gamma}$}_j^{(0)}\,,
\\
\label{Res-WilCoe-CS-NNLO} \mbox{\boldmath $C$}_j^{(2)}({\cal
Q}^2/\mu^2) &\!\!\! =\!\!\!  & \mbox{\boldmath ${c}$}_j^{(2)} +
\frac{s^{(1)}_j({\cal Q}^2/\mu^2)}{2} \left[ \mbox{\boldmath
${c}$}_j^{(0)}\mbox{\boldmath ${\gamma}$}_j^{(1)} +
\mbox{\boldmath ${c}$}_j^{(1)} \mbox{\boldmath
${\gamma}$}_j^{(0)}\right] + \frac{s^{(2)}_j({\cal Q}^2/\mu^2)}{8}
\; \mbox{\boldmath ${c}$}_j^{(0)} \left(\mbox{\boldmath
${\gamma}$}_j^{(0)}\right)^2 \qquad
\\
&& +\frac{\beta_0}{2}
 \left[ \mbox{\boldmath ${C}$}_j^{(1)}({\cal Q}^2/\mu^2)\ln\frac{{\cal Q}^2}{\mu_r^2} +
\frac{1}{4} \mbox{\boldmath ${c}$}_j^{(0)} \mbox{\boldmath
${\gamma}$}_j^{(0)} \ln^2\frac{{\cal Q}^2}{\mu^2} \right]\,.
 \nonumber
\end{eqnarray}
\end{subequations}
To LO the  DIS Wilson coefficients are normalized as
\begin{eqnarray}
\label{Def-WilCoe-DisSLO}
{\mbox{\boldmath $c$}_{j}^{{\rm V}(0)}}  = {\mbox{\boldmath
$c$}_{j}^{{\rm A}(0)}} = (1\;,\;0) \,.
\end{eqnarray}
In NLO  the quark Wilson coefficients correspond to the non-singlet ones given
in Eqs.\ (\ref{eq:NScV1}) and (\ref{eq:NScA1}),
\begin{eqnarray}
 {^\Sigma\!c}_j^{{\rm V}(1)}\!\!\!&=&\!\!\! {^{\rm NS}\!c}_j^{{\rm V}(1)}
\, ,
\\
 {^\Sigma\!c}_j^{{\rm A}(1)}\!\!\!&=&\!\!\! {^{\rm NS}\!c}_j^{{\rm A}(1)}
\, ,
\end{eqnarray}
while the gluon ones read
\begin{eqnarray}
 {^G\!c}_j^{{\rm V}(1)}\!\!\!&=&\!\!\!
- n_f \frac{(4 + 3j + j^2)  S_{1}(j)  +2 + 3j + j^2}{( 1 + j)( 2 +
j)( 3 + j) } \label{Def-Coe-NLO-G-V}
\\
{^G\!c}_j^{A(1)}\!\!\!&=&\!\!\! - n_f
 \frac{j}{(1 + j)(2 + j)}\left[1 + S_{1}(j) \right]
\,. \label{Def-Coe-NLO-G-A}
\end{eqnarray}
The entries of the anomalous dimension matrices read to LO in the vector
case:
\begin{eqnarray}
{^{\Sigma G}\!\gamma}_{j}^{{\rm V}(0)} &\!\!\!=&\!\!\! -4n_f
T_F\frac{4 + 3\,j + j^2 }{( j + 1 )( j + 2 )( j + 3)}\,,
\\
\label{Def-LO-AnoDim-GQ-V}
{^{G \Sigma}\!\gamma}_{j}^{{\rm V}(0)} &\!\!\!=&\!\!\!
-2C_F\frac{4 + 3\,j + j^2 }{j( j + 1 )( j + 2 )}\,,
\\
\label{Def-LO-AnoDim-GG-V} {^{GG}\!\gamma}_{j}^{{\rm V}(0)}
&\!\!\!=&\!\!\! - C_A \left(-\frac{4}{( j + 1 )( j + 2
)}+\frac{12}{j( j + 3)} - 4S_1( j + 1 )  \right)+ \beta_0\,,
\end{eqnarray}
and in the axial-vector case:
\begin{eqnarray}
{^{\Sigma G}\!\gamma}_{j}^{{\rm A}(0)} &\!\!\!=&\!\!\! -4n_f
T_F\frac{j }{( j + 1 )( j + 2 )}\,,
\\
{^{G \Sigma}\!\gamma}_{j}^{{\rm A}(0)} &\!\!\!=&\!\!\! -2C_F\frac{
( j + 3 )}{( j + 1 )( j + 2 )}\,,
\\
\label{Def-LO-AnoDim-GG-A} {^{GG}\!\gamma}_{j}^{{\rm A}(0)}
&\!\!\!=&\!\!\! - C_A \left(\frac{8}{( j + 1 )( j + 2 )} - 4 S_1(
j + 1 ) \right)+ \beta_0\,,
\end{eqnarray}
where $C_A=N_c$ and $T_F=1/2$. At this order the anomalous
dimensions in the quark--quark channels are identical to the
flavor nonsinglet ones
\begin{equation}
{^{\Sigma \Sigma}\!\gamma}_{j}^{{\rm V}(0)} ={^{\Sigma
\Sigma}\!\gamma}_{j}^{{\rm A}(0)} ={^{\rm NS}\!\gamma}_{j}^{(0)}\,,
\end{equation}
given in Eq.\ \req{eq:NSgamma0}.  The NNLO DIS Wilson coefficients for the
structure functions $F_1$ and $g_1$ are given in Refs.\
\cite{ZijNee92,ZijNee94}.  The NLO singlet anomalous dimensions for
the vector and axial-vector cases can be found in Refs.\
\cite{CurFurPet80,GluReyVog90} and \cite{Vog96,MerNee96},
respectively. To treat the evolution to NNLO approximation we need
also the NNLO flavor singlet anomalous dimension matrix. It has been
calculated for the vector case in Ref.\ \cite{VogMocVer04}. We
remark that all these quantities are expressed in terms of rational
functions and harmonic sums. Several numerical routines for them
are available, e.g., in Refs.\
\cite{BluKur98,Ver98,BluMoc05,Mai05}. Here again one might rely
on analytic approximations for the quantities in question, see
Refs.\ \cite{VogNee00,VogMocVer04}.

\subsubsection{Expansion of the evolution operator}
\label{SubSec-ExpEvoOpe}

The evolution of the flavor nonsinglet (integer) conformal
moments in this $\overline{\rm CS}$ scheme is  governed by
\begin{eqnarray}
\label{Def-RGE-1} \mu\frac{d}{d\mu} F_{j} (\xi, \Delta^2,\mu^2)
&\!\!\!=\!\!\!& -\Bigg[ \frac{\alpha_s(\mu)}{2\pi} \gamma_j^{(0)}
+ \frac{\alpha_s^2(\mu)}{(2\pi)^2} \gamma_j^{(1)}+
\frac{\alpha_s^3(\mu)}{(2\pi)^3} \gamma_j^{(2)} +{\cal
O}(\alpha_s^4) \Bigg] \; F_{j}(\xi, \Delta^2,\mu^2)
\nonumber\\
&&\hspace{0.5cm} -\frac{\beta_0}{2}
\frac{\alpha_s^3(\mu)}{(2\pi)^3}\sum_{k=0}^{j-2}
\left[\Delta_{jk}^{\overline{{\rm CS}}}+{\cal O}(\alpha_s) \right]
\; F_{k}(\xi, \Delta^2,\mu^2)\,.
\end{eqnarray}
The solution of the renormalization group equation is given by the
path-ordered exponential, appearing in Eq.\
(\ref{Def-ResCOPEMomSpaTru}). Unfortunately, the mixing matrix
$\Delta_{jk}^{\overline{{\rm CS}}}$ is not completely known, and so
we can here only deal with the solution  for $\Delta_{jk}=0$:
\begin{equation}
F_{j} (\xi, \Delta^2,\mu^2) = {\cal E}_j(\mu,\mu_0)  F_{j} (\xi, \Delta^2,\mu_0^2)\,, \quad\mbox{where}\quad
{\cal E}_j(\mu,\mu_0) = \exp \left\{
    - \int_{\mu_0}^\mu \frac{d\mu^\prime}{\mu^\prime}\gamma_j(\mu^\prime)
\right\}\,.
\label{eq:evNS}
\end{equation}
In the numerical analysis we will resum only the leading
logarithms and expand the non-leading ones. The result, see Ref.\
\cite{MelMuePas02}, reads
\begin{eqnarray}
\label{Def-EvoOpeNS}
 {\cal E}_j(\mu,\mu_0) =  \left[
    1 + \frac{\alpha_s(\mu)}{2\pi} {\cal A}_j^{(1)}(\mu,\mu_0)
      + \frac{\alpha_s^2(\mu)}{(2\pi)^2} {\cal A}_j^{(2)}(\mu,\mu_0)
+O(\alpha_s^3) \right] \left[
\frac{\alpha_s(\mu)}{\alpha_s(\mu_0)}
\right]^{-\frac{\gamma^{(0)}_j}{\beta_0}}\, ,
\end{eqnarray}
where
\begin{eqnarray}
{\cal A}_j^{(1)}(\mu,\mu_0)\!\!\! &=&\!\!\!
 \left[1-\frac{\alpha_s(\mu_0)}{\alpha_s(\mu)}\right]
\left[ \frac{\beta_1}{2\beta_0} \frac{\gamma_j^{(0)}}{\beta_0}
-\frac{\gamma_j^{(1)}}{\beta_0} \right]\, ,
\\
{\cal A}_j^{(2)}(\mu,\mu_0) \!\!\! &=&\!\!\! \frac{1}{2}
\left[{\cal A}_j^{(1)}(\mu,\mu_0)\right]^2 -
\left[1-\frac{\alpha_s^2(\mu_0)}{\alpha_s^2(\mu)}\right]
\left[\frac{\beta_1^2-\beta_2 \beta_0}{8\beta_0^2}
\frac{\gamma_j^{(0)}}{\beta_0} -
\frac{\beta_1}{4\beta_0}\frac{\gamma_j^{(1)}}{\beta_0} +
\frac{\gamma_j^{(2)}}{2\beta_0}\right]\, . \nonumber
\end{eqnarray}
The expansion coefficients of the $\beta$ function are
\begin{eqnarray}
&&\!\!\!\!\! \frac{\beta}{g} = \frac{\alpha_s(\mu)}{4 \pi} \beta_0
+ \frac{\alpha_s^2(\mu)}{(4 \pi)^2} \beta_1
+\frac{\alpha_s^3(\mu)}{(4 \pi)^3} \beta_2+ O(\alpha_s^4),
\\
&&\!\!\!\!\! \beta_0 = \frac{2}{3} n_f -11,\quad \beta_1 =
\frac{38}{3} n_f - 102, \quad \beta_2 = -\frac{325}{54} n_f^2 +
\frac{5033}{18} n_f - \frac{2857}{2}\, . \nonumber
\end{eqnarray}

The flavor singlet case can be treated in the diagonal $\{+,-\}$
basis in the analogous way as the nonsinglet one. Here we
present the solution in the  $\{Q,G\}$ basis:
\begin{equation}
\left( \begin{array}{c}
       {^\Sigma\! F_j} \\
       {^G\! F_j} \\
      \end{array}
\right) (\xi, \Delta^2,\mu^2) =
 \mbox{\boldmath ${\cal E}$}_j(\mu,\mu_0)
\; \left( \begin{array}{c}
       {^\Sigma\! F_j} \\
       {^G\! F_j} \\
      \end{array}
\right) (\xi, \Delta^2,\mu_0^2) \, .
\label{eq:evS}
\end{equation}
We will expand the
evolution operator%
\footnote{Here the path ordering ${\cal P}$ is related to the
non-diagonality of the anomalous matrix in $\{Q,G\}$ basis.}
\begin{eqnarray}
\label{Def-EvoOpe}
 \mbox{\boldmath ${\cal E}$}_j(\mu,\mu_0) &\!\!\! =\!\!\! &   {\cal P}
 \exp{\left\{-\int_{\mu_0}^{\mu} \frac{d\mu^\prime}{\mu^\prime}
\mbox{\boldmath ${\gamma}_j$} (\alpha_s(\mu^\prime))\right\}}
\end{eqnarray}
up to order $\alpha_s^2$, and while leading logs remain resummed
the non-leading ones are expanded. To have a condensed notation, we
perturbatively expand the result in terms of the leading order
$\{+,-\}$ modes:
\begin{eqnarray}
\label{Exp--EvoOpe} \mbox{\boldmath ${\cal E}$}_j(\mu,\mu_0) =
\sum_{a,b=\pm}\left[
    \delta_{ab}\, {^{a}\!\mbox{\boldmath $P$}}_j +
      \frac{\alpha_s(\mu)}{2\pi}\, {^{ab}\!\!\mbox{\boldmath ${\cal A}$}}_j^{(1)}(\mu,\mu_0)
      + \frac{\alpha_s^2(\mu)}{(2\pi)^2}\,  {^{ab}\!\!\mbox{\boldmath ${\cal A}$}}_j^{(2)}(\mu,\mu_0)
+O(\alpha_s^3) \right]\left[ \frac{\alpha_s(\mu)}{\alpha_s(\mu_0)}
\right]^{-\frac{{^b\! \lambda}_j}{\beta_0}}  ,\nonumber\\
\end{eqnarray}
Here the projectors on the $\{+,-\}$ modes are
\begin{eqnarray}
\label{Def-ProP}
{^{\pm}\!\mbox{\boldmath $P$}}_j = \frac{\pm 1}{{^{+}\!
\lambda}_j-{^{-}\! \lambda}_j} \left(\mbox{\boldmath
$\gamma$}_j^{(0)}-{^{\mp}\! \lambda}_j \mbox{\boldmath
$1$}\right)\,,
\end{eqnarray}
where the eigenvalues of the LO anomalous dimension matrix (i.e.,
${^\pm \! \gamma}_j^{(0)}$ from Eq.\ \req{eq:gammapm}) are
\begin{eqnarray}
\label{Def-EigVal}
{^{\pm}\! \lambda}_j
 =\frac{1}{2}
\left({^{\rm \Sigma \Sigma}\!  \gamma}_j^{(0)} + {^{\rm GG}\!
\gamma}^{(0)}_j \mp
 \left( {^{\rm \Sigma \Sigma}\! \gamma}^{(0)}_j
 - {^{\rm GG}\!  \gamma}^{(0)}_j \right)
\sqrt{1 + \frac{4\, {^{\rm \Sigma G}\! \gamma}^{(0)}_j {^{\rm G
\Sigma}\! \gamma}^{(0)}_j}{ \left({^{\rm \Sigma \Sigma}\!
\gamma}^{(0)}_j - {^{\rm GG}\!  \gamma}^{(0)}_j\right)^2 }}
\right)\,,
\end{eqnarray}
A straightforward calculation leads to the matrix valued
coefficients
\begin{eqnarray}
\label{Def-A1} {^{ab}\!\!\mbox{\boldmath ${\cal A}$}}_j^{(1)}
&\!\!\!=\!\!\!& {^{ab}\!R}_j(\mu,\mu_0|1)\, {^{a}\!\mbox{\boldmath
$P$}}_j \left[ \frac{\beta_1}{2\beta_0}  \mbox{\boldmath
$\gamma$}_j^{(0)} -\mbox{\boldmath $\gamma$}_j^{(1)} \right]
{^{b}\!\mbox{\boldmath $P$}}_j
\\
{^{ab}\!\!\mbox{\boldmath ${\cal A}$}}_j^{(2)} &\!\!\!=\!\!\!&
\sum_{c=\pm} \frac{1}{\beta_0+{^{c}\! \lambda}_j-{^{b}\!
\lambda}_j}\left[{^{ab}\!R}_j(\mu,\mu_0|2)-{^{ac}\!R}_j(\mu,\mu_0|1)
\left(\frac{\alpha_s(\mu_0)}{\alpha_s(\mu)}\right)^{\frac{\beta_0+{^{c}\!
\lambda}_j-{^{b}\! \lambda}_j}{\beta_0}} \right]
{^{a}\!\mbox{\boldmath $P$}}_j \left[ \frac{\beta_1}{2\beta_0}
\mbox{\boldmath $\gamma$}_j^{(0)} -\mbox{\boldmath
$\gamma$}_j^{(1)} \right]
\nonumber\\
&&\!\!\!\!\times\, {^{c}\!\mbox{\boldmath $P$}}_j\left[
\frac{\beta_1}{2\beta_0}  \mbox{\boldmath $\gamma$}_j^{(0)}
-\mbox{\boldmath $\gamma$}_j^{(1)} \right] {^{b}\!\mbox{\boldmath
$P$}}_j - {^{ab}\!R}_j(\mu,\mu_0|2)\, {^{a}\!\mbox{\boldmath
$P$}}_j\left[\frac{\beta_1^2-\beta_2 \beta_0}{4\beta_0^2}
 \mbox{\boldmath $\gamma$}_j^{(0)} -
\frac{\beta_1}{2\beta_0} \mbox{\boldmath $\gamma$}_j^{(1)} +
 \mbox{\boldmath $\gamma$}_j^{(2)}\right]{^{b}\!\mbox{\boldmath $P$}}_j\,,
\label{Def-A2}
\end{eqnarray}
where the $\mu$ dependence is accumulated in the functions
${^{ab}\!R}_{j}(\mu,\mu_0|n)\equiv {^{ab}\!R}_{jj}(\mu,\mu_0|n)$
defined by
\begin{eqnarray}
{^{ab}\!R}_{jk}(\mu,\mu_0|n)
= \frac{1}{ n \beta_0+{^{a}\! \lambda}_j-{^{b}\! \lambda}_k}\left[
1- \left(\frac{\alpha_s(\mu_0)}{\alpha_s(\mu)}\right)^{\frac{n
\beta_0+{^{a}\! \lambda}_j-{^{b}\! \lambda}_k}{\beta_0}} \right]
\, .
\end{eqnarray}
The expansion of the evolution operator (\ref{Def-EvoOpeNS}) or (\ref{Exp--EvoOpe}) will
then be consistently combined with the Wilson coefficients (\ref{eq:copeCexp2all})
or (\ref{eq:copeCexp2allSI}), respectively, see for instance Ref.\
\cite{MelMuePas02}.

\subsection{$\overline{\rm MS}$ results up to NLO order}
\label{SubSec-MS}

The DVCS radiative corrections in the $\overline{\rm MS}$ scheme
are known only to NLO. The NLO corrections to the hard-scattering
amplitude have been obtained by rotation from the conformal
prediction \cite{Mue97a,BelMue97a} and agree with the
diagrammatical evaluation of Refs.\
\cite{ManPilSteVanWei97,JiOsb97,JiOsb98}. Employing Eq.\
\req{Def-HarSca2ConMom}, the corresponding Wilson coefficients can
be obtained by determining the conformal moments of the
hard-scattering amplitude, e.g., summarized in Ref.\
\cite{BelMueNieSch99}.

The integrals which are needed for the quark part are given in
Appendix C of Ref.\ \cite{MelMuePas02} for integer conformal spin
(see also the last paragraph in Appendix \ref{App-EvaConMom}).
The analytical continuation is straightforward and so in the
$\overline{\rm MS}$ scheme we have
\begin{eqnarray}
^{\Sigma}\!C_j^{{\rm V}(1)}({\cal Q}/\mu^2)&=& C_F \left[ 2
S^2_{1}(1 + j)- \frac{9}{2}  + \frac{5-4S_{1}(j+1)}{2(j + 1)(j +
2)} + \frac{1}{(j+1)^2(j+2)^2}\right] + \frac{^{\Sigma
\Sigma}\!\gamma_j^{(0)}}{2} \ln\frac{\mu^2}{{\cal Q}^2} \, ,
\nonumber \\
\label{Res-WilCoe-MS-NLO-V}
\\
^{\Sigma}\!C_j^{{\rm A}(1)}({\cal Q}/\mu^2)&=& C_F \left[ 2
S^2_{1}(1 + j)- \frac{9}{2}  + \frac{3-4S_{1}(j+1)}{2(j + 1)(j +
2)} + \frac{1}{(j+1)^2(j+2)^2}\right] + \frac{^{\Sigma
\Sigma}\!\gamma_j^{(0)}}{2} \ln\frac{\mu^2}{{\cal Q}^2} \, ,
\nonumber \\
\label{Res-WilCoe-MS-NLO-A}
\end{eqnarray}
in vector and axial-vector case, respectively, while $^{{\rm
NS}}\!C_j^{{\rm I}(1)}=^\Sigma\!\!C_j^{{\rm I}(1)}$.

The integral expressions needed for the determination of the
singlet gluon contributions we list in App.\ \ref{App-ConfMom-G}.
For the gluon conformal moments we obtain
\begin{eqnarray}
\label{Res-WilCoe-MS-NLO-Vg} ^{G}\!C_j^{{\rm V}(1)}({\cal
Q}/\mu^2)&=& -n_f\frac{(4 + 3j + j^2)
\left[S_{1}(j)+S_{1}(j+2)\right]  +2 + 3j + j^2}{
                                       ( 1 + j)( 2 + j)( 3 + j) }
+ \frac{^{\Sigma G}\!\gamma_j^{{\rm V}(0)}}{2}
\ln\frac{\mu^2}{{\cal Q}^2} \, , \quad
\\
\label{Res-WilCoe-MS-NLO-Ag} ^{G}\!C_j^{{\rm A}(1)}({\cal
Q}/\mu^2)&=& -n_f
 \frac{j}{(1 + j)(2 + j)}\left[1 + S_{1}(j)+ S_{1}(j+2) \right]
+ \frac{^{\Sigma G}\!\gamma_j^{{\rm A}(0)}}{2}
\ln\frac{\mu^2}{{\cal Q}^2} \, .
\nonumber \\
\end{eqnarray}

The complete anomalous dimension matrix in $\overline{\rm MS}$
scheme is known to two-loop accuracy \cite{Mue94,
BelMue98a,BelMue98c}. We present here the more involved case of
the evolution in singlet sector and will just state the
substitutions needed to obtain nonsinglet case results; for the
extensive account of the singlet case evolution see also Ref.\
\cite{BelMueNieSch98a}. Note that our normalization, adopted from
DIS, differs from the original one. Anomalous dimensions used in
this work  are related to those in Ref.\ \cite{BelMueNieSch98a},
denoted as
\begin{equation}
 \mbox{\boldmath $\gamma$}_{jk} =
 \mbox{\boldmath $N$}_{j}
 \mbox{\boldmath $\gamma$}_{jk} |_{\cite{BelMueNieSch98a}}
 \mbox{\boldmath $N$}_{k}^{-1}
  \, .
\label{eq:Utransf}
\end{equation}
The transformation matrix is
\begin{equation}
 \mbox{\boldmath $N$}_{j}
= N(j) \left(
\begin{array}{cc}
1 & 0
\\
0 & \frac{6}{j}
\end{array}
\right) \, ,
\end{equation}
with the normalization factor
\begin{equation}
N(j)= \frac{\Gamma(3/2) \Gamma (j+1)}{2^j\Gamma
\left(j+3/2\right)} \, ,
\end{equation}
originating from the altered definitions (\ref{Def-ConOpe-Q}) and
(\ref{Def-ConOpe-G}) of conformal operators. Explicitly, we have
for non-diagonal entries
\begin{eqnarray}
\mbox{\boldmath $\gamma$}_{jk}  =
\frac{2^k \Gamma(j+1) \Gamma(k+3/2)}{2^j \Gamma(k+1) \Gamma(j+3/2)}
 \left(
\begin{array}{cc}
{^{QQ}\!\gamma}_{jk}|_{\cite{BelMueNieSch98a}} & \frac{k}{6}
{^{QG}\!\gamma}_{jk}|_{\cite{BelMueNieSch98a}}
\\
\frac{6}{j} {^{GQ}\!\gamma}_{jk}|_{\cite{BelMueNieSch98a}} &
\frac{k}{j} {^{GG}\!\gamma}_{jk}|_{\cite{BelMueNieSch98a}}
\end{array}
\right) \, , \label{eq:Utrans2}
\end{eqnarray}
where the diagonal ones,
\begin{eqnarray}
\mbox{\boldmath $\gamma$}_{j}  \equiv \mbox{\boldmath $\gamma$}_{jj} =
 \left(
\begin{array}{cc}
{^{QQ}\!\gamma}_{j}|_{\cite{BelMueNieSch98a}} & \frac{j}{6}
{^{QG}\!\gamma}_{j}|_{\cite{BelMueNieSch98a}}
\\
\frac{6}{j} {^{GQ}\!\gamma}_{j}|_{\cite{BelMueNieSch98a}} &
{^{GG}\!\gamma}_{j}|_{\cite{BelMueNieSch98a}}
\end{array}
\right)\,,\label{eq:Utrans3}
\end{eqnarray}
coincide with the DIS anomalous dimensions.

The evolution operator in the $\overline{\rm MS}$-scheme leads
already at NLO to a mixing of conformal GPD moments. For integer conformal
spin these moments evolve as
\begin{eqnarray}
\mbox{\boldmath $F$}_{j}(\eta,\Delta^2,\mu)= \sum_{k=0}^{j}
\frac{1\mp(-1)^k}{2} \mbox{\boldmath ${\cal
E}$}_{jk}(\mu,\mu_0;\eta) \mbox{\boldmath
$F$}_{k}(\eta,\Delta^2,\mu_0) \,.
\end{eqnarray}
To NLO accuracy the evolution operator is expanded as%
\footnote{If one would be interested to resum all logs rather
than only the leading ones, one certainly has to deal also with the
resummation of the non-diagonal entries. This problem has been
solved in the flavor nonsinglet sector for $\eta=1$ in Ref.\
\cite{BakSte05}. }
\begin{eqnarray}
\label{Exp--EvoOpeSin} \lefteqn{ \mbox{\boldmath ${\cal
E}$}_{jk}(\mu,\mu_0;\eta)
} \nonumber \\
&=&
 \sum_{a,b=\pm}\left[
    \delta_{ab}\, {^{a}\!\mbox{\boldmath $P$}}_j \delta_{jk} +
      \frac{\alpha_s(\mu)}{2\pi}
     \left( {^{ab}\!\!\mbox{\boldmath ${\cal A}$}}_j^{(1)}(\mu,\mu_0)
     \delta_{jk} +
      {^{ab}\!\mbox{\boldmath ${\cal B}$}}_{jk}^{(1)}(\mu,\mu_0)
     \, \eta^{j-k}\right)
+O(\alpha_s^2) \right]\left[ \frac{\alpha_s(\mu)}{\alpha_s(\mu_0)}
\right]^{-\frac{{^b\! \lambda}_k}{\beta_0}}
\, . \nonumber\\
\end{eqnarray}
The diagonal term $\mbox{\boldmath ${\cal
E}$}_{jj}(\mu,\mu_0;\eta)$ coincides with the evolution operator
$\mbox{\boldmath ${\cal E}$}_{j}(\mu,\mu_0)$ of the preceding
section, where the diagonal anomalous dimensions $\mbox{\boldmath
$\gamma$}_{j}$ are the same in $\overline{\rm MS}$ and
$\overline{\rm CS}$ scheme. Also the projector
${^{a}\!\mbox{\boldmath $P$}}_j$ is the same as before and is
defined in Eq.\ \req{Def-ProP}. The matrix
${^{ab}\!\!\mbox{\boldmath ${\cal B}$}}_{jk}^{(1)}(\mu,\mu_0)$,
defined for $j-k=\{2,4,\cdots\}$ (while 0 otherwise),
assumes the form analogous to (\ref{Def-A1})
\begin{eqnarray}
{}^{ab}\!\mbox{\boldmath ${\cal B}$}_{jk}^{(1)} = -
{}^{ab}\!R_{jk}(\mu,\mu_0|1) \; {}^{a}\!\mbox{\boldmath $P$}_j \;
\mbox{\boldmath $\gamma$}_{jk}^{(1)} \; {}^{b}\!\mbox{\boldmath
$P$}_k \, , \qquad \mbox{for}\, j-k\, \mbox{even} \, ,
\end{eqnarray}
while $\mbox{\boldmath $\gamma$}_{jk}^{(1)}$ can be expressed as
\begin{equation}
\label{eq:gammajk} \mbox{\boldmath $\gamma$}_{jk}^{(1)} = \left[
\mbox{\boldmath $\gamma$}^{(0)}, \mbox{\boldmath $g$} +
\mbox{\boldmath $d$} (\beta_0- \mbox{\boldmath $\gamma$}^{(0)})
\right]_{jk} \, .
\end{equation}
This equation holds for parity even and odd anomalous dimension
matrices, where $\mbox{\boldmath $g$}_{jk}$ and $\mbox{\boldmath
$d$}_{jk}$ transform according to Eq.\ \req{eq:Utransf}. In
analogy to Eq.\  \req{eq:Utrans2}, the matrices $\mbox{\boldmath
$g$}_{jk}$, and $\mbox{\boldmath $d$}_{jk}$ can be read off from
the results given in Ref.\ \cite{BelMueNieSch98a}.
Taking into account the usual properties of projectors,
one can conveniently rewrite ${^{ab}\!\!\mbox{\boldmath ${\cal
B}$}}_{jk}^{(1)}(\mu,\mu_0)$ as
\begin{eqnarray}
{^{ab}\!\mbox{\boldmath ${\cal B}$}}_{jk}^{(1)}(\mu,\mu_0)= -
{}^{ab}R_{jk}(\mu,\mu_0|1) \; \left(
{}^a\!\lambda_j-{}^b\!\lambda_k \right) \left[ \left(
\beta_0-{}^b\!\lambda_k \right) {}^{a}\!\mbox{\boldmath $P$}_j
\;\mbox{\boldmath $d$}_{jk} {}^{b}\!\mbox{\boldmath $P$}_k +
{}^{a}\!\mbox{\boldmath $P$}_j \; \mbox{\boldmath $g$}_{jk}
{}^{b}\!\mbox{\boldmath $P$}_k \right] \, .
\end{eqnarray}

The NLO solution of the evolution operator convoluted with the
Wilson coefficients can be more easily treated as the evolution of
the conformal GPD moments itself. Namely, the convolution,
represented as an infinite series and  understood as an analytic
function of $j$, can be expressed in two equivalent forms
\begin{eqnarray}
\label{Res-EvoOpeMS}
\lefteqn{ \sum_{j=0}^{\infty} \left[1\mp(-1)^j\right] \xi^{-j-1}
\mbox{\boldmath $C$}_j({\cal Q}^2/\mu^2,\alpha_s(\mu))
\left(\sum_{k=0}^{j} \frac{1\mp(-1)^k}{2} \mbox{\boldmath ${\cal
E}$}_{jk}(\mu,\mu_0;\xi) \mbox{\boldmath
$F$}_k(\xi,\Delta^2,\mu_0) \right)} \nonumber \\ &=&
\sum_{j=0}^{\infty} \left[1\mp(-1)^j\right] \xi^{-j-1}
\left(\sum_{k=j}^\infty \frac{1\mp(-1)^k}{2} \mbox{\boldmath $C$}_k({\cal
Q}^2/\mu^2,\alpha_s(\mu)) \mbox{\boldmath ${\cal
E}$}_{kj}(\mu,\mu_0;1) \right) \mbox{\boldmath
$F$}_j(\xi,\Delta^2,\mu_0) \, .
\nonumber \\
\end{eqnarray}
The form on the r.h.s.\ corresponds to ``shifting'' the evolution
to Wilson coefficient which is more convenient for numerical
treatment of non-diagonal terms. In this representation it is also
obvious that in DVCS kinematics the mixing under evolution is {\em
not} suppressed by powers of $\xi^2$. The evolved Wilson
coefficients can furthermore be decomposed as
\begin{eqnarray}
\label{Cal-OffDigEvo0}
\mbox{\boldmath $C$}_{j}({\cal Q}^2/\mu_0^2,\mu,\mu_0)&\!\!=\!\!&
\mbox{\boldmath $C$}_{j}({\cal Q}^2/\mu^2,\alpha_s(\mu))\,
\mbox{\boldmath ${\cal E}$}_{jj}(\mu,\mu_0;1)+ \sum_{k=0\atop {\rm
even}}^\infty \mbox{\boldmath $C$}_{j+k+2}({\cal
Q}^2/\mu^2,\alpha_s(\mu))\, \mbox{\boldmath ${\cal
E}$}_{j+k+2,j}(\mu,\mu_0;1)\,.
\nonumber \\
\end{eqnarray}
Of course, we understand that a consequent expansion in $\alpha_s$
to NLO will be performed. To have a numerically more efficient
treatment this sum can be transformed into a Mellin--Barnes
integral. Employing the fact that the only residue of $\cot(\pi
k/2)$ is $2/\pi$ for even integer values of $k$, we
straightforwardly arrive at
\begin{eqnarray}
\label{Cal-OffDigEvo}
\mbox{\boldmath $C$}_{j}({\cal
Q}^2/\mu^2,\mu,\mu_0)&\!\!\!=\!\!\!& \mbox{\boldmath
$C$}_{j}({\cal Q}^2/\mu^2,\alpha_s(\mu))\, \mbox{\boldmath ${\cal
E}$}_{jj}(\mu,\mu_0;1)
\\
&& -\frac{1}{4i} \int_{c-i\infty}^{c+i\infty}\! dk\,
\cot\left(\frac{\pi k}{2}\right) \mbox{\boldmath
$C$}_{j+k+2}({\cal Q}^2/\mu^2,\alpha_s(\mu))\, \mbox{\boldmath
${\cal E}$}_{j+k+2,j}(\mu,\mu_0;1)\,,
\nonumber
\end{eqnarray}
where $-2<c<0$.

The formulae for the flavor nonsinglet sector (see, for example,
Ref.\ \cite{MelMuePas02}) are simply obtained by  reducing the matrix valued quantities
to real (or complex) valued ones, by performing the replacements
\begin{eqnarray}
&&\!\!\!\mbox{\boldmath $\gamma$}_{j}\to {^{\rm NS}\!\gamma}_{j}\,,
\hspace{2.3cm}
{^{a}\!\lambda}_j\to {^{\rm NS}\!\gamma}_{j\phantom{k}}^{(0)}={^{\rm \Sigma
\Sigma}\!\gamma}_{j\phantom{k}}^{(0)} \,,
\quad\, {^{a}\!\mbox{\boldmath $P$}}_j\to 1 \, ,
\nonumber \\
&&\!\!\!\mbox{\boldmath
$\gamma$}_{jk}\to {^{\rm NS}\!\gamma}_{jk} = {^{\rm \Sigma
\Sigma}\!\gamma}_{jk} \,,
\quad \mbox{\boldmath $g$}_{jk}\to
{^{\rm NS}\!g}_{jk}={^{\rm \Sigma \Sigma}\!g}_{jk} \,,
\qquad
\mbox{\boldmath $d$}_{jk}\to {^{\rm NS}\!d}_{jk}={^{\rm \Sigma
\Sigma}\!d}_{jk} \,.
\end{eqnarray}
We recall that in our convention the off-diagonal entries ($j>k$) are
dressed with a factor $N_j/N_k$, e.g.,
${^{\rm NS}\!\gamma}_{jk}=(N_j/N_k) {^{\rm QQ}\!\gamma}_{jk}|_{\cite{BelMueNieSch98a}}$,
and that only to NLO accuracy, considered here,
${^{\rm NS}\!\gamma}_{jk}$ coincides with the quark--quark
entry ${^{\rm \Sigma\Sigma}\!\gamma}_{jk}$.

\section{Parameterization of GPDs}
\label{Sec-GPDpar}

For the phenomenology of GPDs it is crucial that one has a
realistic ansatz at hand. At present, it is
popular to use a GPD ansatz \cite{Rad96a} that is based on the
spectral representation of GPDs \cite{MueRobGeyDitHor94, Rad96},
for a review on this subject see \cite{GoePolVan01}. We think that
our formalism offers the possibility to approach this problem from
a new perspective and that it finally leads to a more flexible
parameterization. It is also remarkable that the task of modelling
the $x$ dependence of GPDs by using the constraints of the lowest
moment \cite{DieFelJakKro04,GuiPolRadVan04,PerPunVan06}, i.e.,
$j=0$ yields elastic form factors, turns in our approach
into the inverse problem.
Namely, it is rather the (conformal) spin
dependence of the form factors that determines the
$x$ dependence, where $j=0$ serves as an `boundary' condition.
Since it is timely to find a parameterization that is the most
appropriate for a fitting procedure, we here only briefly outline
our considerations.

We recall that the GPD support can be separated into two different
momentum fraction regions, both having a dual partonic
interpretation, namely, as the exchange of partons in the
$s$-channel and as mesonlike configurations in the $t$-channel.
Moreover, these regions are tied owing to the Poincar\'e
invariance, see, e.g., Ref.\ \cite{MueSch05}. This  well-understood
support property and its partonic interpretation might be
considered a reflection of the conjectured duality between
$t$- and $s$-channel processes; for a review of its
phenomenological application, see Ref. \cite{MelEntKep05}. This
point of view opens the door to employ strong interaction
phenomenology (combined with some basic principles) in finding a
realistic model ansatz for GPDs and partial extraction of its
parameters. We believe that it is a promising starting point to
think in terms of $t$-channel exchanges combined with SO(3)
partial wave analysis and Regge phenomenology and then to perform
the crossing to the DVCS process. This approach is complementary
to those which are based on modelling GPDs in terms of partonic
degrees of freedom, see, for instance,
\cite{JiMelSon97,DieFelJakKro00,BroDieHwa00,ScoVen02,BofPasTra02,ChoJiKis01,TibMil02,NogTheVen02,AhmHonLiuTan06}.
In the next two sections we outline the SO(3) partial wave
analysis of DVCS and, based on an intuitive picture,
propose an ansatz for the partial wave amplitudes, appearing in
conformal GPD moments. In the third section we try to get
insight into the properties of this ansatz, and, finally,
in Eq.  \req{GPD--ModAns-Exp1} we
present its approximated version
convenient for numerical studies and used in subsequent sections.
The reader more interested in the applications
might just skip to that point.

\subsection{SO(3) partial wave decomposition of conformal GPD moments}
\label{SubSubSec-ParWav}

\begin{figure}[t]
\begin{center}
\mbox{
\begin{picture}(500,130)(0,0)
\put(-2,-4){\insertfig{6.7}{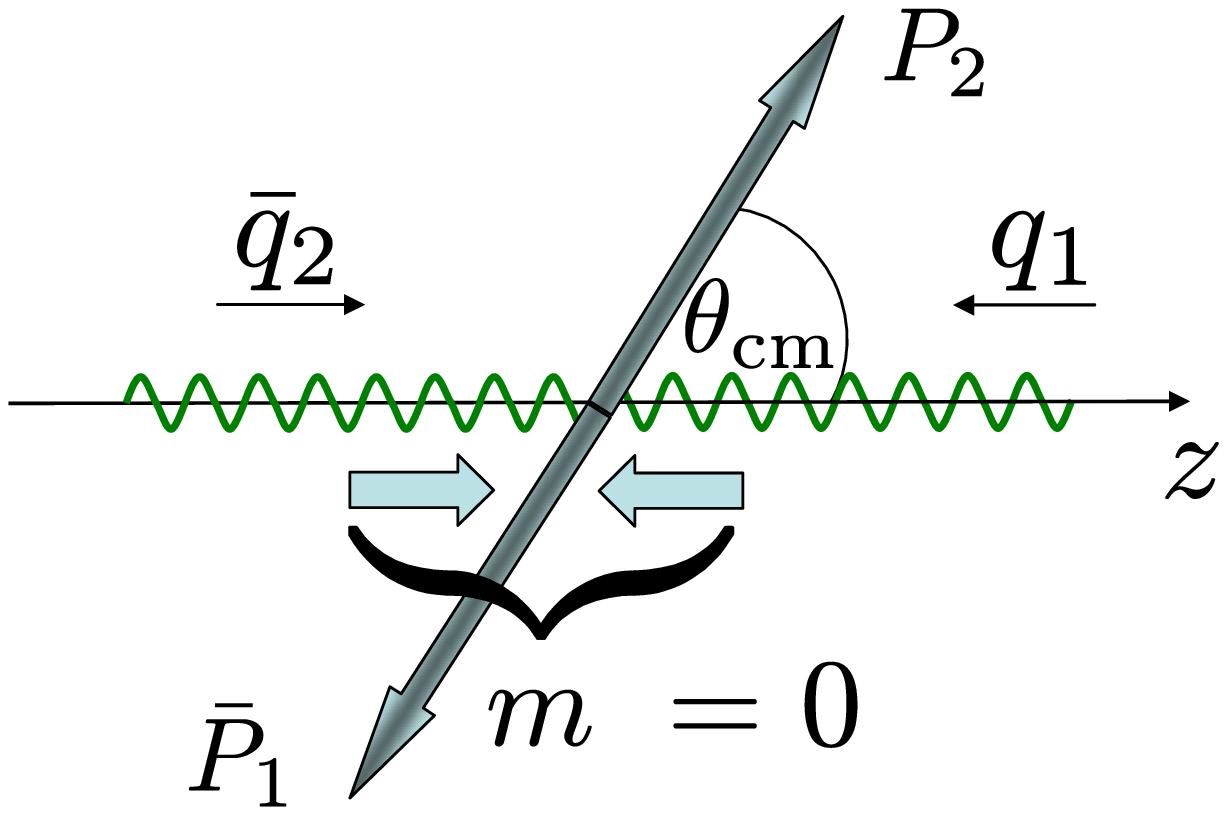}}
\put(200,4){\insertfig{6}{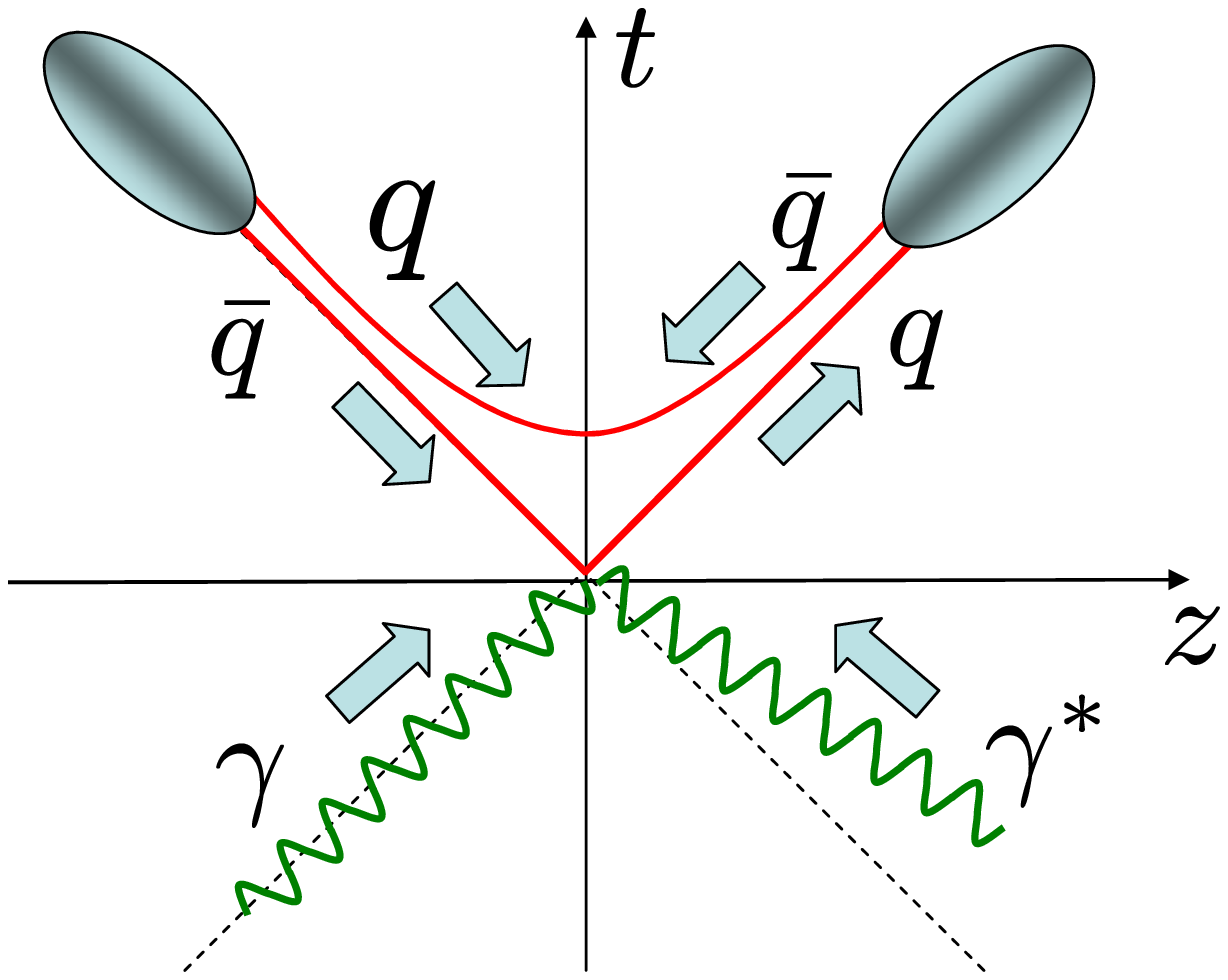}}
\put(400,10){\insertfig{3.2}{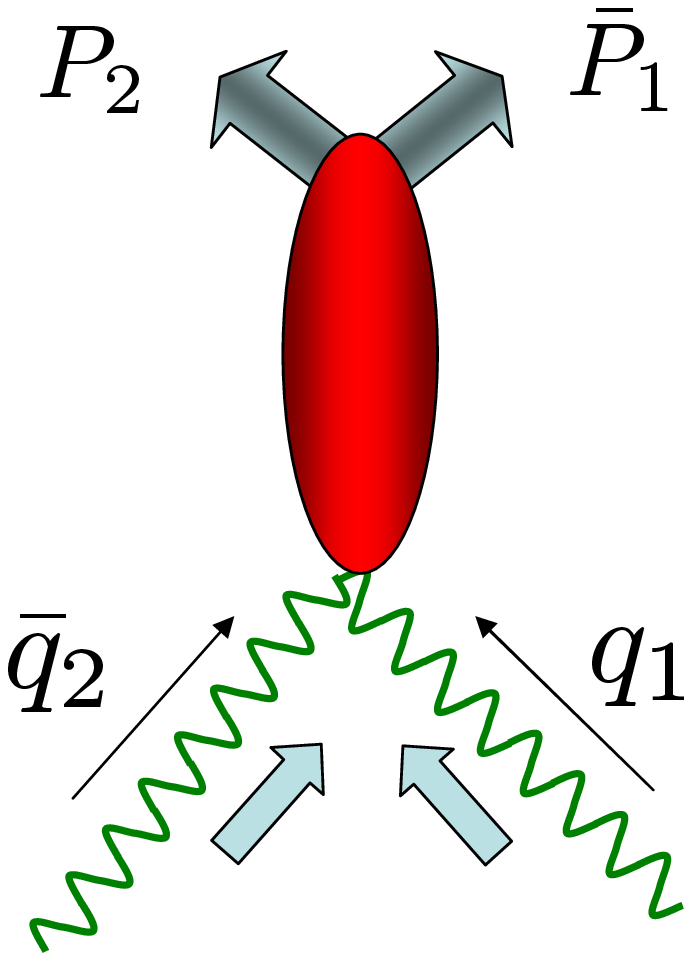}}
\put(75,-15){(a)}
\put(274,-15){(b)}
\put(440,-15){(c)}
\end{picture}
}
\end{center}
\caption{ \label{Fig-Cro} The fusion of two photons into a pair
of (massless) mesons  in the center--of--mass frame (a), partonic
space-time picture (b), and hadronic point of view (c). The thick
arrows indicate the spin projection of the particles.}
\end{figure}

It has already been proposed in Ref.\ \cite{PolShu02} to decompose
conformal moments of \emph{meson} GPDs into irreducible SO(3)
representations, namely, in terms of Legendre polynomials labelled
by the orbital angular momentum quantum number. These functions
enter the partial wave decomposition of the $t$-channel
scattering amplitude and depend on the scattering angle
$\theta\equiv\theta_{\rm cm}$, defined in the center--of--mass frame, see
Fig.\ \ref{Fig-Cro}(a). After crossing, $\cos\theta$ becomes%
\footnote{For the sake of simplicity, here we neglect
$\sqrt{1+M^2/\vec{P}^2}$ proportional corrections and those which
die out in the Bjorken limit.} the inverse of the
scaling variable $-\eta$, i.e., $\cos\theta=-1/\eta$
\cite{MueRobGeyDitHor94,PolShu02}. The basis of Legendre
polynomials has later been adopted for the conformal moments of
nucleon GPDs, too \cite{GuzTec06}. Let us here start with the
analysis and discussion which basis of polynomials one should
employ.

To avoid cumbersome technical details, we restrict
ourselves here to massless hadrons, so that the
crossing relations between helicity amplitudes become
simple: the signs of both four-momenta and helicities
of the crossed particles are to be changed. However,
one has to be careful with the virtual photon%
\footnote{We are
indebted to M.~Diehl for critique and discussion on this
subtle point.},
which we will not cross, but whose virtuality
causes a rotation of helicity amplitudes, i.e.,
when going from $t$- to $s$-channel amplitudes all
three helicities of the virtual photon contribute
whether it is crossed or not. Interestingly, the
dominant contribution for the kinematics we are
interested arises if its helicity will flip, too.
Hence, either in the $s$- or in the $t$-channel
one can take that both photons have the {\em same}
helicity in the center--of--mass frame. In the
real world, one has to take more care of the
rotation of helicity amplitudes, namely, a given
$s$-channel helicity amplitude is expressed as a
linear combination of $t$-channel ones, and for
massive and/or virtual particles all helicities
contribute.

The partial wave expansion of $t$-channel helicity amplitudes is
given in terms of Wigner $d$-matrices $d^J_{\mu,\nu}(\theta)$,
which are essentially expressed by Jacobi polynomials \cite{Wig59}.
Here the quantum number $J$ refers to the total angular momentum, and
$\mu$ ($\nu$) is given by the helicity difference of initial
(final) state particles.  In  the DVCS process, $\gamma^\ast h \to
h \gamma^{(\ast)}$,  the twist-two CFFs, considered in this
paper, appear in the helicity amplitudes in which both photons are
transversely polarized and have the same helicity
\cite{MueRobGeyDitHor94,DieGouPirRal97}.
As discussed above, since one photon is virtual,
one can take that the photons in the $t$-channel process
$\gamma^\ast \gamma^{(\ast)} \to h \bar{h}$ also have the same helicities.
For the rest of this section, we take them to be $-1$, denoted as
$\gamma^{\ast\downarrow}\gamma^{\downarrow}$  (the amplitude for
$\gamma^{\ast\uparrow}\gamma^{\uparrow}$ follows from reflection). Hence, for
scalar ($h \overline{h}$) and spin-$1/2$
particles (e.g., $h^{\uparrow} \bar{h}^{\uparrow}$)  the Wigner
$d$-matrices with $\mu=0$ (photons) and $\nu=\{+1,0,-1\}$ appear in the
partial wave expansion:
\begin{eqnarray}
 \{ d^J_{0,1}\; d^J_{0,-1}\} \quad  \mbox{for}\quad
\left\{ h^{\uparrow}\bar{h}^{\downarrow} \atop h^{\downarrow}\bar{h}^{\uparrow} \right\}
\qquad
{\rm and}\quad
d^J_{0,0}\quad \mbox{for}\quad
\left\{h \bar{h} \atop  h^{\uparrow} \bar{h}^{\uparrow}\right\} \,.
\end{eqnarray}

The SO(3) partial wave expansion of the $t$-channel helicity
amplitudes yields those for the crossed CFFs ${\cal H}^{(t)}$ and
$\widetilde {\cal H}^{(t)}$, defined via Eqs.\ (\ref{dec-FF-V})
and (\ref{dec-FF-A}) by the replacement $P_1 \to - \bar{P}_1$ and
$q_2\to -\bar{q}_2$. The  hadronic tensor in the $t$-channel is
obtained by the same substitutions from Eq.\ (\ref{decom-T}). The
evaluation of the helicity amplitudes in the center--of--mass
frame is straightforward. Taking into account the angular
dependence arising from the spinor bilinears,
\begin{eqnarray}
\label{ExaBiSp}
\frac{q_1^\alpha-\bar{q}_2^\alpha}{(q_1-\bar{q}_2)\cdot (P_2-\bar{P}_1) }
\overline{U}(P_2) \gamma_\alpha \frac{1\mp \gamma_5}{2}V(\bar{P}_1) =
\pm \frac{{\sqrt{1-\cos^2 \theta}}}{\cos\theta} \,,
\end{eqnarray}
we can read off the `effective' partial waves in the SO(3) expansion of the $t$-channel CFFs
${\cal H}^{(t)}$ and $\widetilde {\cal H}^{(t)}$ for massless hadrons:
\begin{eqnarray}
\label{ParWavExp}
{\cal F}^{(t)}(\cos\theta,\vartheta,s^{(t)},Q^2) =
\sum_{J=1}^\infty (2J+1) \frac{1\pm (-1)^J}{2} f_J(s^{(t)},\vartheta,Q^2)
\frac{\cos\theta}{\sqrt{1-\cos^2 \theta}}
 d^J_{0,1}(\theta)\,.
\end{eqnarray}
Here the upper (lower) sign refers to the parity even (odd)
case, in which the total angular momentum is even (odd).
The property of the Wigner matrices ensures then
that the CFFs as scalar valued functions are invariant under
reflection.
The lowest partial wave that appears is for a
total angular momentum $J=2\, (1)$.  Up to a phase factor,
which is not indicated, the crossing relation for helicity
amplitudes simply reads
\begin{eqnarray}
T_{1,\pm 1/2;1,\pm 1/2}(s,t,Q^2) =
T^{(t)}_{\mp 1/2,\pm 1/2, -1,-1}(t^{(t)},s^{(t)},Q^2)\,,
\qquad  t^{(t)} = s \,, \quad  s^{(t)} = t
\end{eqnarray}
and leads with $\cos\theta=-1/\eta$ to those  for the  CFFs:
\begin{eqnarray}
{\cal F}(\xi, \vartheta,\Delta^2,Q^2) =
{\cal F}^{(t)}(\cos\theta=-1/\eta, \vartheta,
s^{(t)}=\Delta^2,Q^2)
\end{eqnarray}
where the photon virtuality asymmetry $\vartheta= \eta/\xi$, cf.\
Eq.\ (\ref{Def-vartheta}), is invariant under crossing. Combining
this relation with the general form of the OPE
(\ref{Def-ConParDecInt}), where the Wilson coefficients, as
functions of $\vartheta$, are invariant under crossing, we can read
off within our conventions the crossing relation for conformal
moments \cite{Ter01,MueSch05}
\begin{eqnarray}
F_j(\eta, \Delta^2) =
\eta^{j+1}  F^{({t})}_j(\cos\theta=-1/\eta,
s^{(t)}=\Delta^2)\,.
\end{eqnarray}
Here $F^{({t})}_j$ are the conformal moments of generalized distribution amplitudes
\cite{MueRobGeyDitHor94,DieGouPirTer98}.
Finally, a formal comparison of the crossed version of the partial
wave expansion (\ref{ParWavExp}) with the OPE
(\ref{Def-ConParDecInt}) allows us to identify the partial waves
\begin{eqnarray}
\label{Def-ParWavV}
\hat d^J_{H}(\eta) &\!\!\!\propto\!\!\!&
\frac{\eta^J}{\sqrt{\eta^2-1}} d^J_{0,1}(\theta)|_{\cos\theta=-1/\eta}, \;\;
2 \le J \le j+1\;\;\mbox{for}\;   j=\{1,3,5,\cdots\}\,, \\
\label{Def-ParWavA}
\hat d^J_{\widetilde H}(\eta)
&\!\!\!\propto\!\!\!& \frac{\eta^{J}}{\sqrt{\eta^2-1}} d^J_{0,1}(\theta)|_{\cos\theta=-1/\eta}, \;\;
1 \le J \le j+1\;\;\mbox{for}\;  j=\{0,2,4,\cdots\}\,,
\end{eqnarray}
in the SO(3) expansion of conformal GPD moments $H_j$ and
$\widetilde H_j$, where $J$ is an even and odd number, respectively.  The
symmetry of the Wigner $d$-matrix ensures that these partial waves
are always even polynomials in $\eta$ of order $J-2$ or $J-1$.
Consequently, within odd (even) $j$ we see that $H_j$ ($\widetilde
H_j$) is an even polynomial in $\eta$, as it is required from time
reversal invariance. The order of these polynomials is $j-1$
and $j$ for odd  and even values of $j$, respectively.
Here we omit further technical details and now discuss the partial
wave amplitudes.

\subsection{Ansatz for  SO(3) partial wave amplitudes}

First, we would like to illustrate with two examples the potential
power of crossing, which allows us to consider the $t$-channel
process rather than the $s$-channel one. For DVCS off  a
(pseudo)scalar particle, in which the photon helicity is
conserved, we have one twist-two CFF $\cal H$, which belongs to
the parity even sector \cite{BelMueKirSch00}. The expansion of the
corresponding conformal GPD moments is done in terms of the Wigner
$d$-matrix
\begin{eqnarray}
\label{ParWavCon-j}
\eta^{j+1} d^J_{0,0}(\theta)|_{\cos\theta=-1/\eta} = \eta^{j+1} P_J(-1/\eta) \quad\mbox{with}\quad
0 \le J \le j+1,\; \mbox{and}\; J\,,(j+1)-\mbox{even},
\end{eqnarray}
which is a polynomial in $\eta$ of order $j+1$ expressed in terms of
a Legendre polynomial $P_J(\cos\theta) \equiv  d^J_{0,0}(\theta)$.  The meaning of the
quantum numbers in the crossed channel is obvious: two photons,
e.g., travelling oppositely along the $z$ axes, form a two-particle
state with zero magnetic quantum number, projected on the $z$ axes, and producing a pair of
(pseudo)scalar hadrons, cf.\ Fig.\ \ref{Fig-Cro}(a). The total
angular momentum of the initial state entirely transfer to the orbital angular
momentum of the hadrons and so the latter is equal to $J$. The
minimal value of the orbital angular momentum $J$,
determined from the magnetic quantum number of the photons, is zero.
Let us have a closer look at the partonic subprocess,
which is thought of as the production of a quark-antiquark pair
state, labelled by the conformal spin $j+2$, and travelling close
to the light cone in different directions.
Since at leading twist-two, only chirally even operators appear,
helicity conservation holds at short-distances. Namely, the quark
and  antiquark have {\em opposite} helicities and, obviously,
their spins point in the same direction, see Fig.\
\ref{Fig-Cro}(b).    The conformal spin $j+2$ must be larger than
$J+1$ and is an odd number in the parity even sector. Hence, all
odd conformal partial waves with $ J-1\le j$ contribute to the
SO(3) partial wave amplitude with the total orbital angular
momentum $J$. However, the quark-antiquark pair is also bound by
confinement and so they also represent  an intermediate mesonic
state with spin $J$, which consists of $J-1$ units of the orbital
angular momentum and one unit of the total angular momentum
arising from the aligned spins of the quark and the antiquark.
This mesonic resonance then decays into two scalar hadrons. The
quark (antiquark) must be combined with an {\em antiquark} ({\em
quark}), which is picked up from the vacuum. To not alter the
quantum numbers of the vacuum, however, a  {\em quark}-{\em
antiquark} pair must be picked up with opposite spin (magnetic
quantum number), as shown in Fig.\ \ref{Fig-Cro}(b). Suppose that
in one hadron the spins of the quark and {\em antiquark}  are
opposite, then in the other ones they are aligned. This would be
in contradiction with a naive quark model picture and it can be
resolved only if one of the produced quarks flips his helicity due
to nonperturbative effects.

Let us come back to DVCS off a nucleon, where again
$s$-channel helicity conservation for the photons is required.
Then the four CFFs ${\cal H},\cdots, \widetilde {\cal E}$ can be
expanded with respect to the $t$-channel partial waves
$d^J_{0,\pm 1}$ and $d^J_{0,0}$. However, in our fictional world
of massless particles, the helicity nonconserving quantities
${\cal E}$ and $\widetilde {\cal E}$ should vanish and so
$d^J_{0,0}$ is absent. Hence, we are left only with final states
which  have opposite helicities, i.e., $\cal H$ and
$\widetilde{\cal H}$, which are expanded with respect to the
partial waves (\ref{Def-ParWavV},\ref{Def-ParWavA}). We can again
employ our intuitive picture, and as in the intermediate resonance
before, the quark and the antiquark form a spin-one state with the orbital
angular momentum $J-1$. By picking up two quarks and antiquarks
from the vacuum the  final spin-1/2 nucleon and antinucleon are
formed and one would expect that they carry the helicity of the
produced quark and antiquark, respectively. The corresponding
conformal GPD moment for odd conformal spin $j+2$ is now a
polynomial of order $j-1$. In the case that the helicity of the
produced quark or antiquark is flipped owing to nonperturbative
effects, the helicities of the produced nucleons will be the same.
The partial wave $d^J_{0,0}$ appears in the conformal GPD moments
$E_j$ and also $H_j$, which are polynomials of order $j+1$.

Let us emphasise that we have therefore observed a relation between
the helicity flip and the order of conformal GPD moments. In general,
the conformal GPD moment $H_j$ is an even polynomial in $\eta$,
which is of order $j+1$ for odd $j$. The minimal value of $J$ is now given
by the magnetic quantum number of the final state and so it is one. We remind ourselves that in the partial
waves (\ref{Def-ParWavV}) a term of the order $j+1$ in $\eta$ is
absent and so it cannot appear in $H_j$. However, if we would
allow for helicity nonconserved quantities, such a term arises,
owing to the partial wave $d^J_{0,0}$, which will then appear in
$E_j$ and $H_j$, see the helicity representation for GPDs in Ref.\
\cite{Die01}. We conclude again that the restoration of full
polynomiality, namely, up to order $(j+1)$ for odd $j$, appears
owing to the nonperturbative interaction in which the helicity of
the parton is reversed. In other words, the breaking of chirality
is encoded in the conformal GPD moments, which for given odd
conformal spin $j+2$ are polynomials of order $j+1$. Certainly,
the highest possible term in $\eta$ naturally arises as a part
of the SO(3) partial waves. At this stage we see no reason to
treat such terms in a special way and to collect them into a
separate so-called $D$-term \cite{PolWei99}, which was introduced
to complete the common spectral representation of GPDs. As
mentioned in Sect.\ \ref{SubSubSec-subtraction-cte}, we suggest to
cure the spectral representation, e.g., as proposed in Ref.\
\cite{BelMueKirSch00}, rather than to add a $D$-term.

The partial wave expansion we wrote down for GPDs was borrowed
from the two-photon fusion into two hadrons. In the case of hard
vector meson production, the leading twist-two contributions arise
from a longitudinally polarized photon and vector meson.  It
is easy to name the Wigner matrices, which are the same as for
DVCS, namely, we have  $d^J_{0,\pm 1}$ and $d^J_{0,0}$ matrices
in the SO(3) expansion of the conformal GPD moments $H_j$ and
$E_j$. The former ones, given by $\sin(\theta)
C^{3/2}_{J-1}(\cos\theta)/\sqrt{J(J+1)}$, are expressed by the
Gegenbauer polynomials with index $\nu=3/2$ and the latter ones
are the Legendre polynomials, or, if one likes, Gegenbauer
polynomials with index $\nu=1/2$. Again, the helicity nonconserved
quantity $E_j$ is formed from $d^J_{0,0}$, while $H_j$ contains
$d^J_{0,1}$  and an admixture of the $d^J_{0,0}$ waves. If we now
replace the vector meson by a pseudoscalar ones, the same partial
waves appear for the parity odd quantities $\widetilde H_j$ and
$\widetilde E_j$.

We realize that the SO(3) partial wave expansions of GPDs are
universal, i.e., are the same for the considered processes. Such
an expansion has several advantages, e.g., for the analytic
continuation of even to odd $j$ values and allows for a simple
implementation of the normalization at $j=0$. It also leads in a
natural way to a term of order $\eta^{j+1}$ in the polynomial
$E_j$ for odd values of $j$. Moreover, it is convenient to have
for helicity conserved quantities the same functions that also
appear in the conformal SO(2,1) representation. In a conformally
invariant world there would appear only one SO(3) partial wave
with $J= j+1$.

Inspired by the hadronic view on the $t$-channel scattering
process, see Fig.\ \ref{Fig-Cro}(c), we now propose an ansatz for
the partonic partial wave amplitudes that appear in the SO(3)
expansion of the conformal GPD moments. Thereby, we  rely on the
Regge description of high-energy processes, which states that the
high-energy behavior of the $s$-channel process $\gamma^{(\ast)} h
\to h \gamma $ is dominated by linear Regge trajectories
$\alpha(t)$ of pomeron and meson exchanges in the $t$-channel. The
strength of the photon--photon--to--meson (pomeron) coupling is
contained in a vertex factor $f_{Jj}$ that depends on the
conformal spin, too. The conformal spin $j+2$ is considered as a
variable conjugated to the partonic momentum fraction $x$. Thus,
as in the case of mesonic distribution amplitudes that are
expanded with respect to the Gegenbauer polynomials $C^{3/2}_j(x)$,
$f_{Jj}$ can be viewed in the partonic language as a probability
amplitude for finding a quark-antiquark pair state with conformal
spin $j+2$ inside of a meson with given spin $J$. Furthermore, the
partial wave amplitudes also contain the propagator $1/(m^2(J)-t)
\propto 1/(J-\alpha(t))$ of the exchanged particles, as well as the impact
form factor, describing the interaction with the target. These form factors
will be modelled by a $p$-pole ansatz, i.e, monopole ($p=1$),
dipole ($p=2$), and so on, with a $J$-dependent cutoff mass
squared $M^2(J)$.

All together, we propose the following ansatz for the helicity
nonflip conformal GPD moments in terms of partonic partial waves
amplitudes, which we write down as a sum over the angular
momentum (we employ the fact that the polynomials are even):
\begin{eqnarray}
\label{GPD--ModAns-gen}
\left\{{H}_j  \atop \widetilde H_j\right\} (\eta,\Delta^2,\mu_0^2)=  \sum_{{J=2,{\rm even} \atop\phantom{J=} 1, {\rm odd}}}^{j+1}
\frac{f_{Jj}}{J-\alpha(\Delta^2)} \frac{\eta^{j+1-J} }{(1-\frac{\Delta^2}{M^2(J)})^p}
   \left\{  { {\hat d}_{0,1}^{J} (\eta) \atop    {\hat d}_{0,1}^{J+1} (\eta) }\right\}
  \quad \mbox{for} \; j= \left\{ { {\rm odd} \atop {\rm even} } \right. \,,
\end{eqnarray}
where for odd (even)   $j$ the sum runs over even (odd) $J$.
Here $ {\hat d}_{0,1}^{J}(\eta) \propto \eta^{J} d_{0,1}^{J}(1/\eta)/\sqrt(\eta^2-1)$ is the
`crossed version' (\ref{Def-ParWavV},\ref{Def-ParWavA}) of the Wigner matrix,
where the rotation matrix of the spinor bilinears
is taken off. It is simply  our partonic toy model
(\ref{Def-h-TreLev}), defined for complex valued $j$ in Eq.\
(\ref{DefAnaConC}):
\begin{eqnarray}
\label{Def-CroWigNonFli}
{\hat d}_{0,1}^{J=j+1}(\eta) =
\frac{\Gamma(3/2)\Gamma(3 +j)}{2^{1+j}\Gamma(3/2 + j) } \eta^j\;
{_2\!F_1}\left( { -j,j+3\atop 2
}\Big|\frac{\eta-1}{2\eta}\right)\,.
\end{eqnarray}
For our later convenience, the crossed $d$-matrices are normalized in the forward limit to one:
\begin{eqnarray}
\lim_{\eta\to 0}  \hat d^J_{\mu,\nu} (\eta) = 1.
\end{eqnarray}
The conformal GPD moments (\ref{GPD--ModAns-gen}) are even polynomials in
$\eta$ of order $j-1$ or $j$, as required.  We also note that they are build  from
even polynomials $d_{0,1}^{l+1}(\cos\theta) \propto \cos\theta\, C_{l}^{3/2}(\cos\theta)$, where
$l=\{0,2,\cdots, j-1\}$ or $\{0,2,\cdots, j\}$,
with  eigenvalue $+1$ under parity transformation. For
helicity nonconserved quantities $E_j$ and $\widetilde E_j$,
analogous ans\"atze can be written down in terms of Legendre
functions. Thereby $H_j$  will get an
admixture from $d_{0,0}^J$ partial waves, too.

\subsection{Modelling of conformal GPD moments}

It is beyond the scope of this paper to present a thorough study of realistic ans\"atze
for GPD moments. We would rather like to convince the reader that the
proposed parameterization  generically works and then use some
simplified version for our numerical studies. One
important aspect is  the skewness dependence and its approximations.
Another, new one, is the implementation of lattice results.

The problem of how different approximations of the skewness
dependence will affect the size of the corresponding CFF will be
investigated within a toy model. It is similar to Eq.\ (\ref{Def-h-TreLev}), which
describes the Compton scattering process at tree level or, in other
words, within a noninteracting parton picture. To make it somewhat more
realistic, we multiply it with the generically valid Mellin moments of
an unpolarized valencelike parton density $35x^{-1/2}(1-x)^3/32$
(here normalized to one):
\begin{eqnarray}
\label{toy-exa}
H^{{\rm toy}}_j(\xi) =
\frac{\Gamma(1/2+j)\Gamma(9/2)}{\Gamma(9/2+j)\Gamma(1/2)}\, {\hat d}_{H}^{j+1}(\eta=\xi) \,.
\end{eqnarray}
Here the partial wave ${\hat d}_{H}^{j+1}(\eta)$, given in Eq.\ (\ref{Def-CroWigNonFli}), is
normalized to one for vanishing skewness $\eta$.
For the sake of illustration, we now expand $H^{\rm toy}_j(\xi)$
with respect to Legendre polynomials:
\begin{eqnarray}
\label{ParWav1}
{\hat d}_{H}^{j+1}(\eta) = \hat{d}^j_{0,0}(\eta) + \frac{\eta^2}{4}\sum_{l=0}^{j-2} \frac{1 + (-1)^{j-l}}{2} \left(\frac{\eta}{2}\right)^{j-l-2}
 \frac{\Gamma(l+3/2)\Gamma(j+1) }{\Gamma(j+3/2)\Gamma(l+1)}\hat{d}^l_{0,0}(\eta)
 \,,
\end{eqnarray}
where $\hat{d}^l_{0,0}(\eta) =  (\eta/2)^l \sqrt{\pi} \Gamma(l+1)
P_l(1/\eta)/\Gamma(l+1/2)$. The first term on the r.h.s.\ is the
leading partial wave, while  all other partial waves, contained in
the remaining sum, are suppressed by powers of  $\eta^2$. We will
now study the numerical deviation in the CFF that is induced by an
approximation of the partial wave ${\hat d}_{H}^{j+1}(\eta)$. For
this purpose we make three choices: we drop  the skewness
dependence altogether, take  the leading partial wave, and include
also the next-to-leading ones:
\begin{eqnarray}
\label{ParWav-2}
{\hat d}_{H-0}^{j+1} = 1\,, \qquad
{\hat d}_{H-{\rm LO}}^{j+1}  = {\hat d}_{0,0}^{j}(\eta)\,,
\qquad
{\hat d}_{H-{\rm NLO}}^{j+1} =
 {\hat d}^{j}_{0,0}(\eta) + \frac{(j-1)j}{(4j^2-1)} \eta^2  {\hat d}^{j-2}_{0,0}(\eta)\,.
\end{eqnarray}
Note that some care is needed in the truncation of
exact partial waves. The approximated partial wave ${\hat
d}_{H-{\rm NLO}}^{j+1}$ differs for $j=1$ (and also $j=0$)  from
the exact one by $\eta/27$ ($3/\eta$). This is induced by a pole in
${\hat d}^{-1}_{0,0}(\eta)$ (${\hat d}^{-2}_{0,0}(\eta)$) that
cancels the zero at $j=1$ (or  $j=0$) in the expansion
coefficient. Finally, this leads to an addenda in the real part of
the CFFs that must be subtracted. In our toy example the
subtraction term for the next-to-leading approximation reads
$-5/27 \eta$.

\begin{figure}[t]
\begin{center}
\mbox{
\begin{picture}(450,150)(0,0)
\put(180,125){(a)}
\put(0,10){\insertfig{7.4}{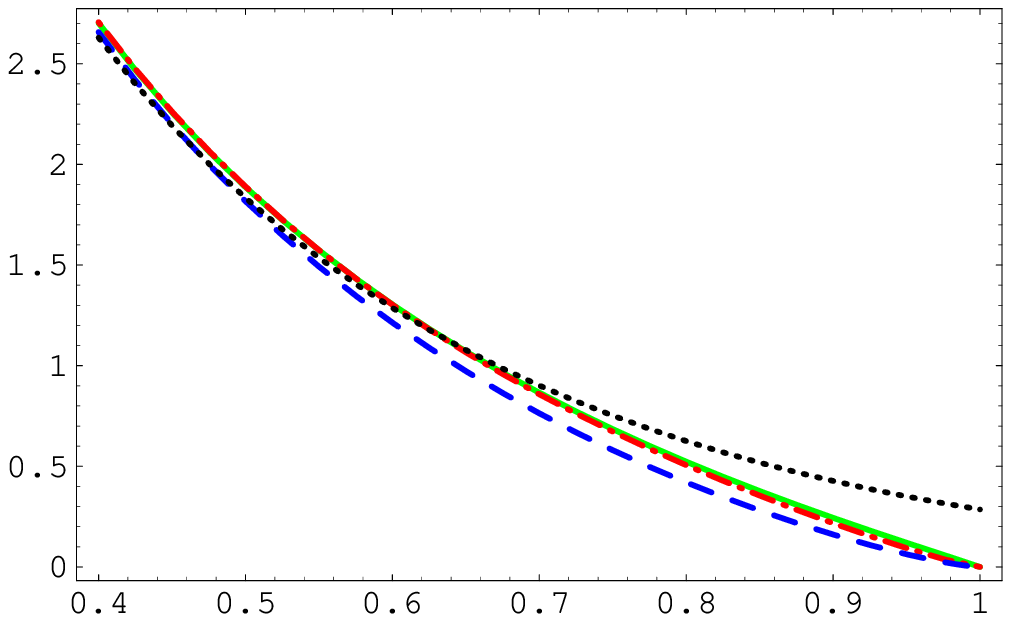}}
\put(200,0){$\xi$}
\put(420,125){(b)}
\put(250,10){\insertfig{7.4}{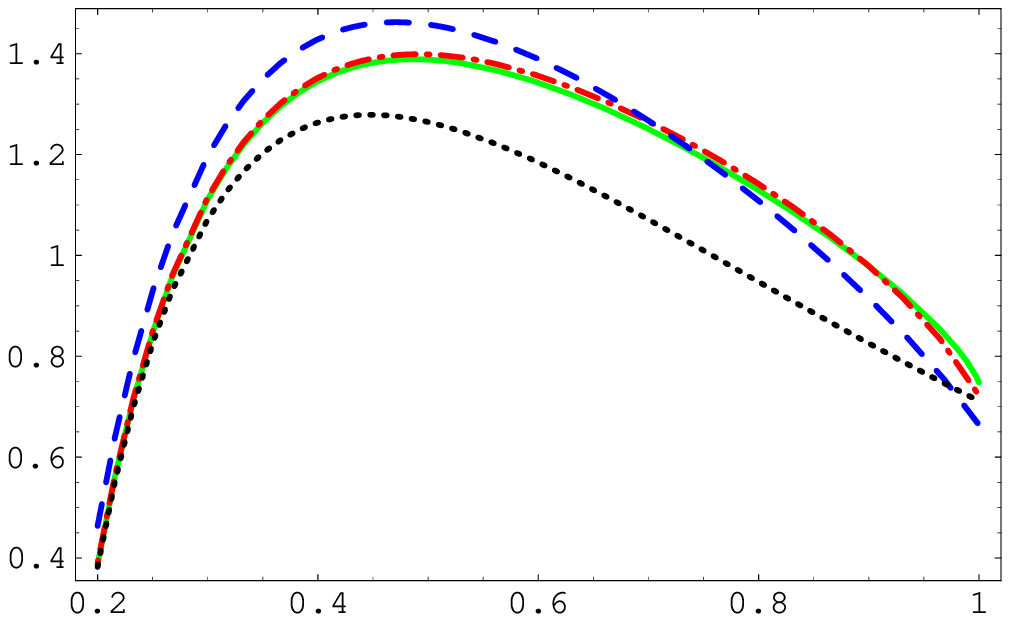}}
\put(-20,50){\rotatebox{90}{$\Im{\rm m}\, {\cal H}^{\rm toy}_j(\xi)$}}
\put(230,50){\rotatebox{90}{$\Re{\rm e}\,  {\cal H}^{\rm toy}_j(\xi)$}}
\put(450,0){$\xi$}
\end{picture}
}
\end{center}
\caption{ \label{Fig-Appr} The imaginary (left) and real (right)
part of the CFF ${\cal H}(\xi,\Delta^2=0)$ arising from the toy
ansatz (\ref{toy-exa})  for conformal GPD moments  within
different approximations (\ref{ParWav-2}): exact (solid), next-to-leading
(dash-dotted), leading (dashed) SO(3) partial waves and
$\xi=0$ (dotted) term.}
\end{figure}
The outcome for the imaginary and real part of the CFF ${\cal
H}^{\rm toy}_j(\xi)$ is displayed  in the left and right panel of
Fig.\ \ref{Fig-Appr}, respectively. In the next-to-leading
approximation (dash-dotted) one can hardly see a
difference to the exact CFF (solid). If we take only the leading
SO(3) partial wave (dashed), we realize that the deviation
from the exact CFF (solid) is small over the whole $\xi$
region, in particular for the imaginary part. Furthermore, making
the expansion in powers of $\xi^2$ and retaining only the leading
$\xi = 0$ term (dotted) yields the imaginary part that approaches
the exact result already for $\xi \lesssim 0.7$. However, it has an
unrealistic feature that it does not vanish in the limit $\xi\to
1$, as it should. Concerning the real part, both of these
approximations start to digress from the exact result for $ \xi \gtrsim
0.3 $, where the deviation is larger for the expansion in $\xi$.

To summarize, the expansion with respect to the angular momentum
looks promising and works rather well in the case that the leading
pole factorizes. We have also found in the expansion with respect to
Legendre polynomials, which was used in Ref.\ \cite{GuzTec06},
that the minimal version of the dual model, i.e., taking leading
and next-to-leading partial waves, yields also an astonishing agreement
with the exact result. We have also observed that in the experimentally
accessible kinematical region the expansion in $\xi$ practically
coincides with the exact result, in particular for the imaginary
part. Nevertheless, there could be a drawback if the (leading)
Regge poles are nonfactorizable as it is the case in our model
ansatz (\ref{GPD--ModAns-gen}). The next-to-leading term in the
partial wave expansion generates an `artificial' pole, e.g.,
$1/(j-1-\alpha(\Delta^2))$, that is situated on the r.h.s.\ of the
leading Regge pole $1/(j+1-\alpha(\Delta^2))$. Such a pole leads
to an addendum in the Mellin--Barnes representation for CFFs with
the same small $\xi$-behavior as the leading pole. Thus, we expect
that the normalization of the CFFs, e.g., for $\Delta^2=0$, is
governed by all terms in the partial wave expansion. A closer look
at this potential problem should be given somewhere else. For the
time being, we rely on the leading term in the expansion with
respect to $\eta$,
\begin{eqnarray}
\label{GPD--ModAns-Exp}
{\!F}_j(\eta,\Delta^2,\mu_0^2)= \frac{f_{j+1,j}}{j+1-\alpha(\Delta^2)}
\frac{1}{(1-\frac{\Delta^2}{M^2_j})^p}  +{\cal O}(\eta^2)\,,
\end{eqnarray}
which we will use in our numerical studies.

Next we fix the normalization of conformal GPD moments at $\Delta=0$:
\begin{eqnarray}
\label{ForKin}
{\!F}_j(\eta=0,\Delta^2=0,\mu_0^2)=  \frac{f_{j+1,j}}{1+j-\alpha(0)}\,.
\end{eqnarray}
The conformal moments of helicity nonflip GPDs  $H_{j}$ and
$\widetilde H_{j}$ are then reduced to the Mellin moments of
unpolarized ($q$)  and polarized ($\Delta q$) parton densities,
\begin{eqnarray}
\label{Con-ForParDen}
\frac{h_{j+1,j}}{1+j-\alpha(0)} =  \int_0^1\! dx\,  x^j  q(x,\mu_0) \qquad \text{and}\qquad
\frac{\widetilde h_{j+1,j}}{1+j-\alpha(0)} =  \int_0^1\! dx\,  x^j  \Delta q(x,\mu_0)\,,
\end{eqnarray}
respectively.
They are parameterized with the guidance of
Regge phenomenology, determining its small  $x$ behavior to be
$x^{-\alpha(0)}$, whereas counting rules suggest their large $x$ behavior,
parameterized as $(1-x)^\beta$ for $x\to 1$. Corresponding to Eq.\
(\ref{Con-ForParDen}), the generic ansatz for parton densities in
the $x$ space yields the following Mellin moments, e.g.,
\begin{eqnarray}
\label{Cal-Nor}
q(x,\mu_0^2) = N x^{-\alpha_0} (1-x)^\beta
\quad \Rightarrow  \quad
h_{j+1,j} = N (1+j-\alpha(0)) B(1-\alpha_0+j,\beta+1)\,,
\end{eqnarray}
where $B(a,b)=\Gamma(a)\Gamma(b)/\Gamma(a+b)$ is the Euler beta
function. Note that the intercept $\alpha_0$ of the
Regge trajectory occurring in our ansatz and that obtained from a
given fit of parton densities must agree. Hence, the leading pole
in the Mellin moments (\ref{Cal-Nor}) is cancelled and replaced
by the corresponding
trajectory%
\footnote{Remaining nonleading poles at
$j=\{-2+\alpha_0,-3+\alpha_0,\cdots\}$ are considered an artifact
of the parameterization and appear as subleading contributions in
the CFFs.}. Certainly, a more realistic ansatz for parton
densities, to be used in global fits, could be obtained by a
linear combination of such building blocks. In practice, either one can
take  the Mellin moments of one of the standard
parameterizations of parton densities
\cite{MarStiTho06,Ale05,Pumetal02,GluReyVog98,LeaSidSta05,
BluBot02,GluReyStrVog00,GehSti95} or, if they are plagued by
larger errors or theoretical uncertainties, one can perform a
simultaneous fit of exclusive (e.g., DVCS) and inclusive data.
Here one has to bear in mind that parton densities are
scheme-dependent quantities.

Further constraints for the conformal moments arise at  $j=0$.
There they are  given by the so-called partonic form factor
\begin{eqnarray}
\label{Con-FF}
{\!F}_{j=0}(\eta,\Delta^2)=
\frac{N_0}{1-\alpha(\Delta^2)} \frac{1}{(1-\frac{\Delta^2}{M^2_0})^p}\,.
\end{eqnarray}
After adjusting the flavor quantum numbers, the partonic form
factors coincide with the measured elastic form factors of the
proton. The free parameters left, i.e., the cut-off mass $M^2_0$
and the power behavior $p$ at large $\Delta^2$ can be taken from a
fit to experimental data, where a refinement of our
parameterization might be necessary. We note that form factors are
scheme independent, except the axial one in the flavor singlet
sector. Another advantage of the parameterization
(\ref{GPD--ModAns-gen}), which we will not use here, is that the
partonic partial wave amplitudes $f_{Jj}$ are related to physical
ones, denoted as $f_{J}(W^2,\vartheta,Q^2)$, in the $t$-channel.
The physical amplitudes $f_{J}$ are given as a series of partonic
ones, where the sum runs over the conformal spin $j+2$. Note that
such a relation depends on our scheme conventions, too.

Finally, plugging the normalization (\ref{Cal-Nor}) into the ansatz
(\ref{GPD--ModAns-Exp}), we end up with the following
simplified GPD parameterization:
\begin{eqnarray}
\label{GPD--ModAns-Exp1}
{\!F}_j(\eta,\Delta^2,\mu_0^2)=
{\!N} B(1-\alpha(0)+j,\beta+1)\frac{j+1-\alpha(0)}{j+1-\alpha(\Delta^2)}
\frac{1}{(1-\frac{t}{M^2_j})^p}  +{\cal O}(\eta^2)\,.
\end{eqnarray}
We remark that for large $|\Delta|^2$, the counting rules predict
a power-like falloff of form factors as ($1/|\Delta|^2)^{n_s}$,
where $n_s$ is the number of spectators, while the large $x$
behavior of parton densities is $(1-x)^{2n_s-1}$, i.e.,
\begin{eqnarray}
\label{CouRul}
p=n_s-1\,, \qquad \beta=2 n_s-1\,.
\end{eqnarray}
An inclusive--exclusive relation between the unpolarized DIS
structure function $W_2$ and the electromagnetic form factor $F_2$ has
also been derived by  Drell and Yan  within a field--theoretical
model that accounts for the dynamics of partons \cite{DreYan69}.
This result coincides with the counting rules for valence quarks.
However, one should be aware that for sea-quarks and gluons,
these counting rules might be modified; for a discussion see,
e.g., Ref.\ \cite{BroBurSch94}. For simplicity, we will not
account for that here.

\begin{figure}[t]
\begin{center}
\mbox{
\begin{picture}(450,150)(0,0)
\put(200,125){(a)}
\put(0,10){\insertfig{7.9}{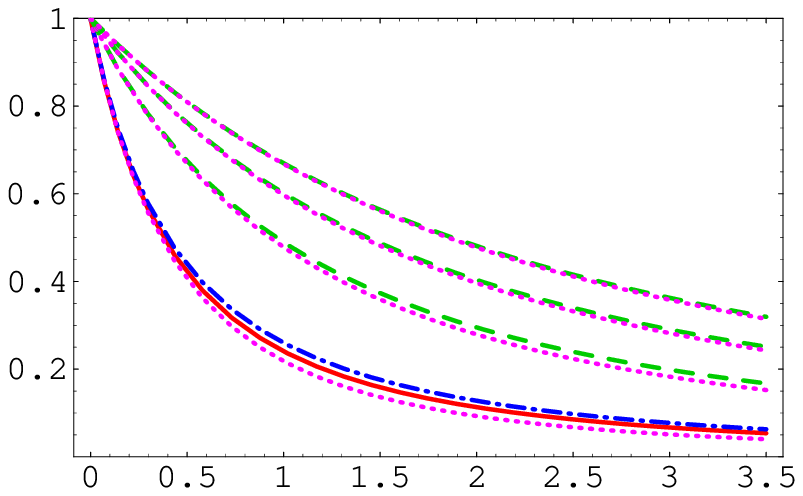}}
\put(160,0){$-{\Delta}^2\ [\GeV^2]$}
\put(180,34){\tiny \boldmath $j=0$}
\put(180,49){\tiny \boldmath $j=1$}
\put(180,59){\tiny \boldmath $j=2$}
\put(180,72){\tiny \boldmath $j=3$}
\put(270,125){(b)}
\put(250,10){\insertfig{7.5}{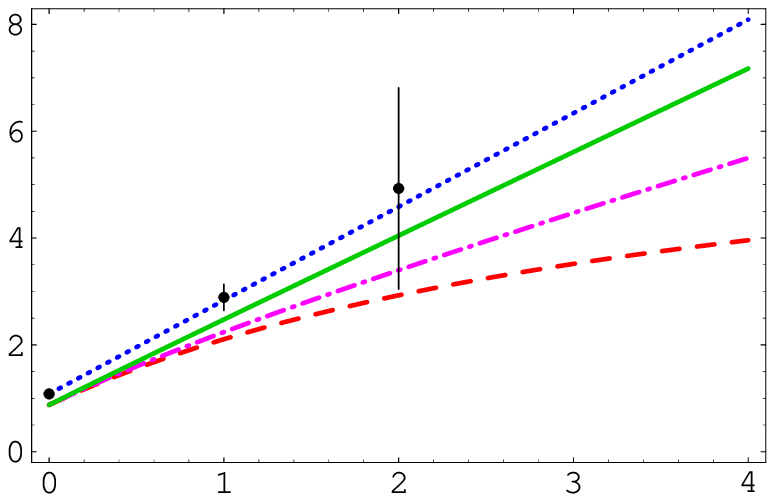}}
\put(-20,30){\rotatebox{90}{${\!H}^{\rm val}_j(0,\Delta^2)/{\!H}^{\rm val}_j(0,0)$}}
\put(240,40){\rotatebox{90}{$M^2_{\rm dipol}(j)\, [\GeV^2]$}}
\put(450,0){$j$}
\end{picture}
}
\end{center}
\caption{ \label{Fig-F1} The $\Delta^2$ dependence of the
conformal GPD moments (\ref{Ans-MomVal0}) for valence quarks
(left) and the spin dependence of the effective dipole mass
(\ref{Def-DipMass}) (right). In the left panel the solid line
displays the dipole fit (\ref{Def-DipFit}) to the experimental data
and the dash-dotted one is our ansatz (\ref{Ans-MomVal0}) with
$j=0$. The dashed and dotted lines show the changes with
$j=\{1,2,3\}$ for our ansatz and a dipole fit within the dipole
masses (\ref{Def-DipMass}), where $\Delta M^2 = (M_{\rm p})^2$. In
the right panel we show the effective dipole masses
(\ref{Def-DipMass}) as a function of spin within $\Delta M^2$:
$0$ (dashed), $M_{\rm p}^2$ (dash-dotted), $(2 M_{\rm p})^2$
(solid). The dotted line has a slope  that arises from $\Delta M^2
=2(2 M_{\rm p})^2$ and is compatible with lattice measurements
\cite{Gocetal04} for the flavor nonsinglet combination $u-d$
in the heavy-pion world,
which was linearly extrapolated to the physical pion mass.}
\end{figure}
Finally, we would like to demonstrate that our  ansatz
(\ref{GPD--ModAns-Exp1}) can be easily adjusted to the
experimental data on electromagnetic form factors and, moreover,
that it is well suited to include lattice data. In particular,
present lattice measurements
\cite{Hagetal03,Hagetal04,Hagetal04a,Gocetal03,Gocetal05,Edwetal06}
give insight into the functional change of the
$\Delta^2$ dependence with increasing conformal spin. This
dependence arises from two sources: the Regge trajectory and the
impact form factor. As an example, we consider the conformal GPD
moments of valence quarks, where the Regge trajectory is
generically correctly described by $\alpha(t)= \alpha(0) +
\alpha^\prime t$ with the intercept $\alpha(0)= 1/2$ and the slope
$\alpha^\prime =1\, \GeV^{-2} $. For two spectators the counting
rules state that the impact form factor is a  monopole, i.e.,
$p=1$. For its cutoff mass we naturally choose two times the
proton mass $M_0 \sim 2 M_{\rm p} = 1.88\ \GeV$. Hence, our
conformal GPD moments are written as
\begin{eqnarray}
\label{Ans-MomVal0}
{\!H}^{\rm val}_j(\eta,\Delta^2,\mu_0^2) \propto
\frac{1}{1-\frac{\Delta^2}{m_j^2}} \frac{1}{1-\frac{\Delta^2}{M^2_j}}
+{\cal O}(\eta^2)\,.
\end{eqnarray}
Here we have introduced the monopole mass squared $m_j^2 =(1-\alpha(0)
+j)/\alpha^\prime = (1/2+j)\, \GeV^2$, which arises from the
mesonic Regge trajectory.  We might also assume that the spin
dependence of the monopole mass squared $M_j^2$ is linear
\begin{eqnarray}
\label{CutofMasImpFF}
M^2_j = M^2_0 + \Delta M^2 j \,.
\end{eqnarray}

The generic ansatz (\ref{Ans-MomVal0}) leads to a satisfying
description of the electromagnetic proton form factor. In
Fig.\ \ref{Fig-F1} (a) we confront this ansatz
(dash-dotted) with the dipole fit (solid) of the electromagnetic
proton form factor $F_1$,
\begin{eqnarray}
\label{Def-DipFit}
F_1(Q^2=-\Delta^2)= \frac{1+Q^2/1.26\ \GeV^2}{\left(1+\frac{Q^2}{m^2_{\rm dipol}}\right)^2
\left(1 +\frac{Q^2}{M_{\rm p}^2}\right)}\,,\quad m^2_{\rm dipol} = 0.71\ \GeV^2\,,
\end{eqnarray}
and realize that they fairly agree. We also display the $\Delta^2$ dependence of the
conformal moments within the choice $\Delta M^2= (M_{\rm p})^2$. A larger value of
$\Delta M^2$ is compatible with the slope measured on lattice in the
heavy-pion
world, as shown in Fig.\ \ref{Fig-F1} (b) (dotted line).
However, the intercept
differs (in fact, the lowest moment does
not describe the $F_1(Q^2)$ data). The masses are extracted from a dipole fit to the
lattice data. To compare with our ansatz, we calculated the `effective' masses
\begin{eqnarray}
\label{Def-DipMass}
M^2_{\rm dipol}
=
2\left(\frac{\alpha^\prime}{1+j-\alpha(0)}+\frac{1}{M_0^2+  \Delta M^2 j} \right)^{-1}
\,,
\end{eqnarray}
appearing in a dipole fit.
For $|\Delta^2| < 1\, \GeV^2$ such a `refitting procedure'  only
weakly modifies the GPD moments, compare dashed and dotted
lines in Fig.\ \ref{Fig-F1} (a), and so this procedure is justified
to some extent.

\section{Perturbative corrections of the DVCS cross section}
\label{Sec-PerCorCrosSec}

This section is devoted to the numerical analysis of radiative
corrections to CFFs. We will concentrate on the CFF $\cal H$,
since it is the dominant contribution in most of the DVCS observables.
The three remaining twist-two CFFs are usually
suppressed by kinematical factors \cite{BelMueKir01}. To reveal
these CFFs from experimental data, it is therefore crucial to
understand the theoretical uncertainties of $\cal H$ . Our
findings can be qualitatively adapted for the helicity
nonconserved CFF $\cal E$, which enters the parity even sector,
too. In this  sector we face a peculiarity that is related to the
appearance of the pomeron trajectory in high-energy scattering.
Technically, it shows up as an essential singularity of
the evolution operator at $j=0$, see anomalous dimensions
(\ref{Def-LO-AnoDim-GQ-V}) and (\ref{Def-LO-AnoDim-GG-V}). Such a
singularity is absent from the parity odd sector. Hence, our results
in the small $\xi$ region are not directly applicable to the study of radiative
corrections of the two remaining parity odd CFFs $\widetilde {\cal
H}$ and $\widetilde {\cal E}$.

Based on our model (\ref{GPD--ModAns-gen}), we introduce in the
next section  a simplified generic ansatz for conformal GPD
moments,  which will serve us in our numerical studies of the
experimentally accessible kinematical regions. In Sect.\
\ref{SubSecEvoLO} we shortly discuss the features of flavor
singlet and nonsinglet CFFs to LO accuracy. In the following two
sections we then elaborately analyze  the size of radiative
corrections to NLO and, finally, to NNLO accuracy. Thereby,
independently of the considered order and scheme, we take the same
conformal GPD moments which thus leads to different CFFs. However, note
that CFFs are physical observables and do not depend on our
conventions. Therefore, conversely, the conformal GPD moments revealed from a
measurement of the physical  CFFs will depend on both our scheme
conventions and the approximation.  Nevertheless, the results of
our analysis give us a measure for both the
reparameterization of the GPD ansatz needed to compensate convention change
and for the convergency of the perturbation theory.

\subsection{A simplified generic ansatz for conformal GPD moments}
\label{SubSec-GenAns}

The kinematics of interest can be restricted to $\eta =\xi \le
0.5$, i.e., the Bjorken scaling variable $x_{\rm Bj} =
2\xi/(1+\xi)$ is bounded by $2/3$. For simplicity, we do not resum
the partial waves and rely on the leading term in
the $\eta$ expansion (\ref{GPD--ModAns-Exp1}).

Let us specify the other parameters in the ansatz
(\ref{GPD--ModAns-Exp1}) for a valence-like helicity non-flip GPD
$H$. The leading meson Regge trajectory is generically given by
$\alpha(t)= \alpha(0) + \alpha^\prime t$ with $\alpha(0)= 1/2$ and
$\alpha^\prime =1\, \GeV^{-2}$. For two spectators the counting
rules (\ref{CouRul}) state that $p=1$ and $\beta=3$. For the cut-off
mass of the impact form factor we  choose
(\ref{CutofMasImpFF}) with $M_0 = 2 M_{\rm
p} =2\Delta M= 1.88\ \GeV$. Hence, we have for our GPD moments:
\begin{eqnarray}
\label{Ans-MomVal}
{\!H}^{\rm val}_j(\eta,\Delta^2,\mu_0^2)= \frac{B(1/2+j,4)}{B(1/2,4)}
\frac{1}{1-\frac{2 \Delta^2}{(1+2 j)\, \GeV^2}} \frac{1}{1-\frac{\Delta^2}{M_{\rm p}^2 (4+j) }}
+{\cal O}(\eta^2)\,,
\end{eqnarray}
which is normalized to one for $j=0$ and $\Delta^2=0$.

The conformal GPD moments with the flavor nonsinglet combination
(\ref{Fla-Nf3}) or (\ref{Fla-Nf4}), which is also relevant for DVCS,
is built from valence and sea quarks. For four active
quarks and the SU(2) flavor symmetric sea, the sea quark part arises
from the difference of charm and strange (anti-)quarks, cf.\ Eq.\
(\ref{Fla-Nf4}). Suppose we are at the charm threshold and the
charm sea is generated dynamically. Then, essentially, only the
(anti-) strange quark counts. We might assume that the
$\bar{u},\bar{d},\bar{s}$ antiquarks have the same conformal GPD
moments and so the breaking of SU(3) flavor symmetry is described
by one single parameter $R_{\bar{s}/\bar{u}}$, defined as the
ratio of $\bar{s}$ to $\bar{u}$ antiquarks. For the purpose of
illustration we consider two alternative cases: one without and one with
sea quark admixture:
\begin{eqnarray}
\label{Ans-ConMomNS-pur}
{^{\rm NS}\! H}_j(\eta, \Delta^2,\mu_0^2) &\!\!\! = \!\!\!&
{\!H}^{\rm val}_j(\eta,\Delta^2,\mu_0^2)\,,
\\
\label{Ans-ConMomNS-pse}
{^{\rm NS}\! H}_j(\eta, \Delta^2,\mu_0^2) &\!\!\! = \!\!\!&
{\!H}^{\rm val}_j(\eta,\Delta^2,\mu_0^2)  -
\frac{R_{\bar{s}/\bar{u}}}{2+R_{\bar{s}/\bar{u}}} H_j^{\rm sea}(\eta, \Delta^2,\mu_0^2)\,.
\end{eqnarray}
For  the valence quark content we rely on Eq.\ (\ref{Ans-MomVal})
and the sea quark GPD moments are specified below in the ansatz
(\ref{Ans-Momsea}) with $N_{\rm sea}= 4/15$. For the SU(3)
flavor symmetry breaking parameter we choose the often used value
$R_{\bar{s}/\bar{u}}=1/2$.

In the flavor singlet sector we need an ansatz for both quark
and gluon conformal GPD moments:
\begin{eqnarray}
\label{Ans-ConMomS} \mbox{\boldmath $H$}_j(\eta, \Delta^2,\mu_0^2) = \left(
\begin{array}{c}
H_j^{\Sigma}  \\
H_j^{\rm G}
\end{array}
\right)(\eta, \Delta^2,\mu_0^2)\,,\quad
H_j^{\Sigma}= H_j^{\rm sea} +  H_j^{u_{\rm val}} +  H_j^{d_{\rm val}} = H_j^{\rm sea} +3 H_j^{{\rm val}} \,.
\end{eqnarray}
The singlet quark combination consists of sea and valence quarks,
cf.\ (\ref{Fla-Nf3}) and (\ref{Fla-Nf4}), and we employ isospin
symmetry to express the latter within the ansatz
(\ref{Ans-MomVal}). Moreover, we rely on  the counting rules
(\ref{CouRul}), i.e, $\beta_{\rm G}=5,p_{\rm G}=2, \beta_{\rm
sea}=7,p_{\rm sea}=3$, and take the same cut-off mass for the
impact form factors as for the valence quarks before. Hence, analogously to Eq.\ (\ref{Ans-MomVal}),
the generic ansatz (\ref{GPD--ModAns-Exp1}) then reads:
\begin{eqnarray}
\label{Ans-Momsea}
{\!H}^{\rm sea}_j(\eta,\Delta^2,\mu_0^2) &\!\!\! = \!\!\!&
N_{\rm sea}\frac{B(1-\alpha_{\rm sea}(0)+j,8)}{B(2-\alpha_{\rm sea}(0),8)}
\frac{1}{1-\frac{\Delta^2}{(m^{\rm sea}_j)^2}} \frac{1}{\left(1-\frac{\Delta^2}{(M^{\rm sea}_j)^2}\right)^3} +{\cal O}(\eta^2)\,,
\\
\label{Ans-MomG}
{\!H}^{\rm G}_j(\eta,\Delta^2,\mu_0^2) &\!\!\! = \!\!\!&  N_{\rm G}\frac{B(1-\alpha_{\rm G}(0)+j,6)}{B(2-\alpha_{\rm G}(0),6)}
\frac{1}{1-\frac{\Delta^2}{(m^{\rm G}_j)^2}} \frac{1}{\left(1-\frac{\Delta^2}{(M^{\rm G}_j)^2}\right)^2} +{\cal O}(\eta^2)\,,
\end{eqnarray}
where $M^{\rm G}_j = M^{\rm sea}_j = M_{\rm p}^2 (4+j)$ and $m_j^2 =(1-\alpha(0)
+j)/\alpha^\prime$ is expressed in terms of the intercept and slope of the Regge trajectories, specified below.

The leading trajectory arises now from the `pomeron'
exchange. We remind that in deeply inelastic scattering the
structure function $F_2 \sim (1/x_{\rm Bj})^{\lambda(Q^2)}$ grows
with increasing $Q^2$. Here the exponent is governed by the
intercept of the Regge trajectory $\lambda=\alpha(0)-1$, which is,
in the language of Regge phenomenology, that of the soft pomeron
(for $Q^2 \to 0$):
\begin{eqnarray}
 \alpha_{\mathbb{P}}(t) =   \alpha_{\mathbb{P}}(0) + \alpha_{\mathbb{P}}^\prime t\,\qquad
 \alpha_{\mathbb{P}}(0) \simeq 1\,, \quad \alpha_{\mathbb{P}}^\prime =0.25\,.
\end{eqnarray}
However, in hard processes the trajectory will effectively change
due to evolution, which differently effects the behavior of quark
and gluon parton densities. In particular, the value of
$\alpha(0)$ increases, while that of $\alpha^\prime$
decreases with growing resolution scale ${\cal Q}^2$, e.g.,
see  Ref.\  \cite{Mue06}.
In the flavor singlet sector the size of radiative corrections and
the strength of evolution crucially depends on the effective pomeron parameters, see for instance
the variation of NLO corrections obtained in \cite{FreMcD01b}.
In our numerical studies we shall consider two scenarios
in which the radiative corrections to NLO accuracy are respectively
small and large. It is known that the corrections to the quark
sector are rather stable, while the main uncertainty arises from
the gluons, which enter in the perturbative description of CFFs at
NLO. Small and large NLO corrections can be obtained by choosing a
softer and harder gluon, respectively:
\begin{eqnarray}
\label{AnsSinSof}
\hspace{1cm} \mbox{``soft'' gluon:}&& \hspace{1cm}
N_G=0.3,\hspace{1cm} \alpha_{\rm G}(0) =\alpha_{\rm sea}(0)-0.2\, ,\\
\label{AnsSinHar}
\hspace{1cm} \mbox{``hard'' gluon:} && \hspace{1cm} N_G=0.4,\hspace{1cm} \alpha_{\rm G}(0) =\alpha_{\rm sea}(0)+0.05\, .
\end{eqnarray}
Furthermore, we choose a realistic value for $\alpha_{\rm
sea}(0)=1.1$ and $\alpha^\prime_{\rm sea}=\alpha^\prime_{\rm
G}=0.15$, all at the input scale ${\cal Q}_0^2=2.5\, \GeV^2$.

We remark that we normalize the sea quark (\ref{Ans-Momsea}) and
gluon (\ref{Ans-MomG}) moments  at  $j=1$, so $N_{\rm sea}$ and
$N_{\rm G}$ give the amount of momentum fraction carried by the
considered parton species. Because of the momentum sum rule
(\ref{Def-SumRulMom-FK}), valid in the forward kinematics, we have
the constraint
\begin{eqnarray}
\quad N_{\rm G} + N_{\rm sea} + \int_{0}^1\!dx\;x \left[u_{\rm v}+
d_{\rm v}\right](x) \equiv 1\,.
\end{eqnarray}
Within our toy ansatz for valence quark moments (\ref{Ans-MomVal}) we
have for their momentum the generic value $1/3$ and so $N_{\rm
sea}= 2/3- N_{\rm G}$. We note, however, that the separate
contributions will change during the evolution. In the asymptotic
limit $\cal Q\rightarrow \infty$, the evolution equation tells us
that $ N_{\rm G} = 4 C_F/(4 C_F+n_f)$, i.e., that more than $50\%$
of the longitudinal proton momentum is carried by gluons. As it is
experimentally verified, at a scale of a few $\GeV^2$ the gluons
already carry about $40\,\%$ of the momentum.

\subsection{Compton form factors to LO accuracy}
\label{SubSecEvoLO}

The  CFFs ${^{\rm S}\!{\cal H}}$ and ${^{\rm NS}\!{\cal H}}$ are
evaluated from the specified conformal GPD moments by  the
Mellin--Barnes integral (\ref{Res-ImReCFF}). Essentially, we
employ here the same technique that is well established in deeply
inelastic scattering, only the integrand is now more intricate.
The integral does not depend on the integration path, going from
$c-i\infty$ to $c+i\infty$, as long as we do not cross any
singularities, see Fig.\ \ref{FigCou2} (a). In practice we use this
property to get closer to the leading singularity, lying left
of the integration path. Since the conformal GPD moments
rapidly decrease with growing conformal spin $j$, we can in practice
cut the integration path to a finite length. Moreover, the
convergency can be improved  by a separate rotation of the
integration path in the upper and lower half-plane in such a way that
along the path the real part decreases, as shown in
Fig.\ \ref{FigCou2} (b). Nevertheless, it is always a good
idea to exercise proper care in the choice of the integration path
and to check the numerical accuracy. Once this is done for given
conformal GPD moments, the numerical treatment is simple, fast,
and stable even in NNLO. We use two different codes for the
numerical evaluation, one specifically written in FORTRAN and,
alternatively, the integration routine from MATHEMATICA.

For the  evaluation of the CFFs to LO accuracy, we use the flavor
nonsinglet Wilson coefficients (\ref{eq:copeCexp2}) and
(\ref{Def-DIS-WC-LO}) as well as  the singlet ones
(\ref{Res-WilCoe-Exp-CS-SI}) and (\ref{Def-WilCoe-DisSLO}). The
${\cal Q}^2$ evolution is governed by the flavor nonsinglet
operator (\ref{Def-EvoOpeNS}) and singlet one
(\ref{Exp--EvoOpe})--(\ref{Def-EigVal}), approximated to LO. We
equate the factorization scale with the photon virtuality:
$\mu^2={\cal Q}^2$. The various ans\"atze, given in Eqs.\
(\ref{Ans-ConMomNS-pur}), (\ref{Ans-ConMomNS-pse}),
(\ref{Ans-Momsea}), and (\ref{Ans-MomG}),   are taken at the input
scale ${\cal Q}_0^2 =2.5\, {\GeV}^2$. The running coupling in LO
approximation is normalized to $\alpha_s({\cal Q}_0^2 =2.5\,
{\GeV}^2)/\pi =0.1$, where the number of active quarks is four.

We found it  useful to describe the complex valued CFFs
in polar coordinates,
\begin{eqnarray}
{\cal H}(\xi,\Delta^2,{\cal Q}^2) =
\left|{\cal H}(\xi,\Delta^2,{\cal Q}^2)\right|
\exp\left\{i \varphi(\xi,\Delta^2,{\cal Q}^2) \right\}\,,
\quad -\pi < \varphi={\rm arg} \left({\cal H}\right) \le \pi\,,
\end{eqnarray}
rather than Cartesian ones. This avoids a discussion of radiative
corrections in the vicinity of zeros, appearing, e.g., in the real part
of certain CFFs. Moreover, the polar coordinates reveal a simple
shape of CFFs in their functional dependence on $\xi$ and ${\cal
Q}^2$.
\begin{figure}[t]
\begin{center}
\mbox{
\begin{picture}(700,280)(0,0)
\psfrag{-1}[cc][cc]{\tiny $10^{-1}$}
\psfrag{-2}[cc][cc]{\tiny $10^{-2}$}
\psfrag{-3}[cc][cc]{\tiny $10^{-3}$}
\psfrag{-4}[cc][cc]{\tiny $10^{-4}$}
\psfrag{-5}[cc][cc]{\tiny $10^{-5}$}
\psfrag{-6}[cc][cc]{}
\psfrag{0}[cc][cc]{\tiny $0$}
\psfrag{20}[cc][cc]{\tiny $20$}
\psfrag{40}[cc][cc]{\tiny $40$}
\psfrag{60}[cc][cc]{\tiny $60$}
\psfrag{80}[cc][cc]{\tiny $80$}
\psfrag{100}[cc][cc]{\tiny $100$}
\psfrag{25}[cc][cc]{}
\psfrag{50}[cc][cc]{\tiny $50$}
\psfrag{75}[cc][cc]{}
\psfrag{2}[cc][cc]{\tiny $2$}
\psfrag{4}[cc][cc]{\tiny $4$}
\psfrag{6}[cc][cc]{\tiny $6$}
\psfrag{8}[cc][cc]{\tiny $8$}
\psfrag{10}[cc][cc]{\tiny $10$}
\psfrag{0.1}[cc][cc]{\tiny $.1$}
\psfrag{0.2}[cc][cc]{\tiny $.2$}
\psfrag{0.3}[cc][cc]{\tiny $.3$}
\psfrag{0.4}[cc][cc]{\tiny $.4$}
\psfrag{0.5}[cc][cc]{\tiny $.5$}
\psfrag{1}[cc][cc]{\tiny $1$}
\psfrag{1.2}[cc][cc]{\tiny $1.2$}
\psfrag{1.4}[cc][cc]{\tiny $1.4$}
\psfrag{1.5}[cc][cc]{\tiny $1.5$}
\psfrag{1.6}[cc][cc]{\tiny $1.6$}
\psfrag{2}[cc][cc]{\tiny $2$}
\put(165,100){\small (c)}
\put(30,0){\insertfig{6}{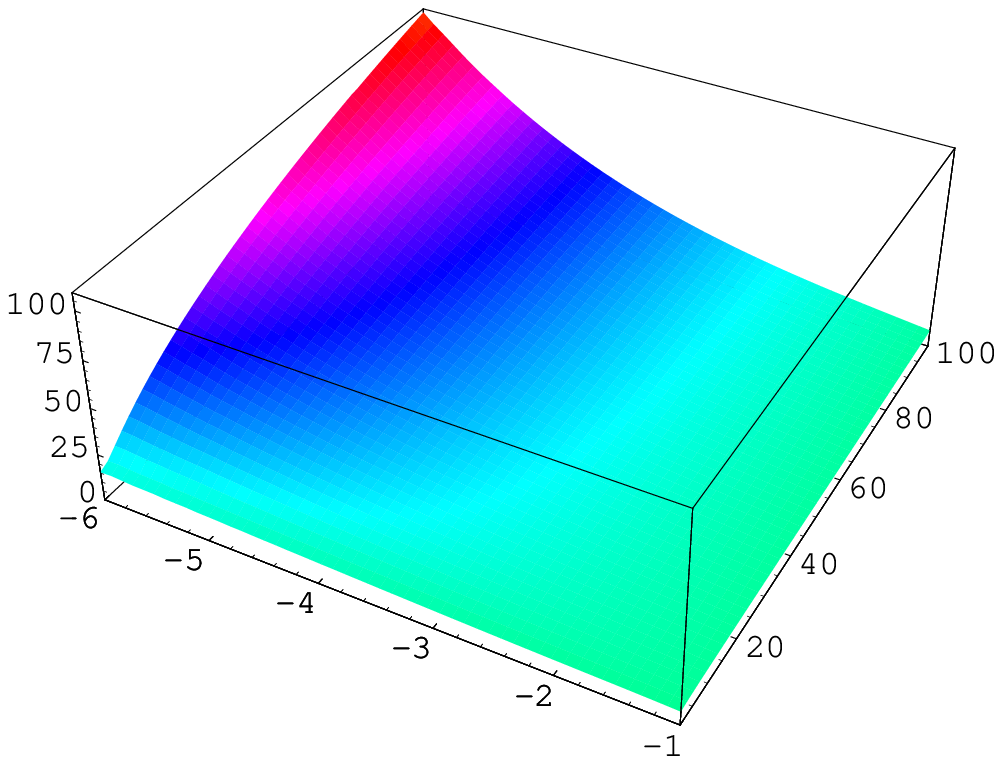}}
\put(80,15){\small $\xi$}
\put(168,13){\rotatebox{59}{\small ${\cal Q}^2\ [\GeV^2] $}}
\put(415,100){\small (d)}
\put(280,0){\insertfig{6}{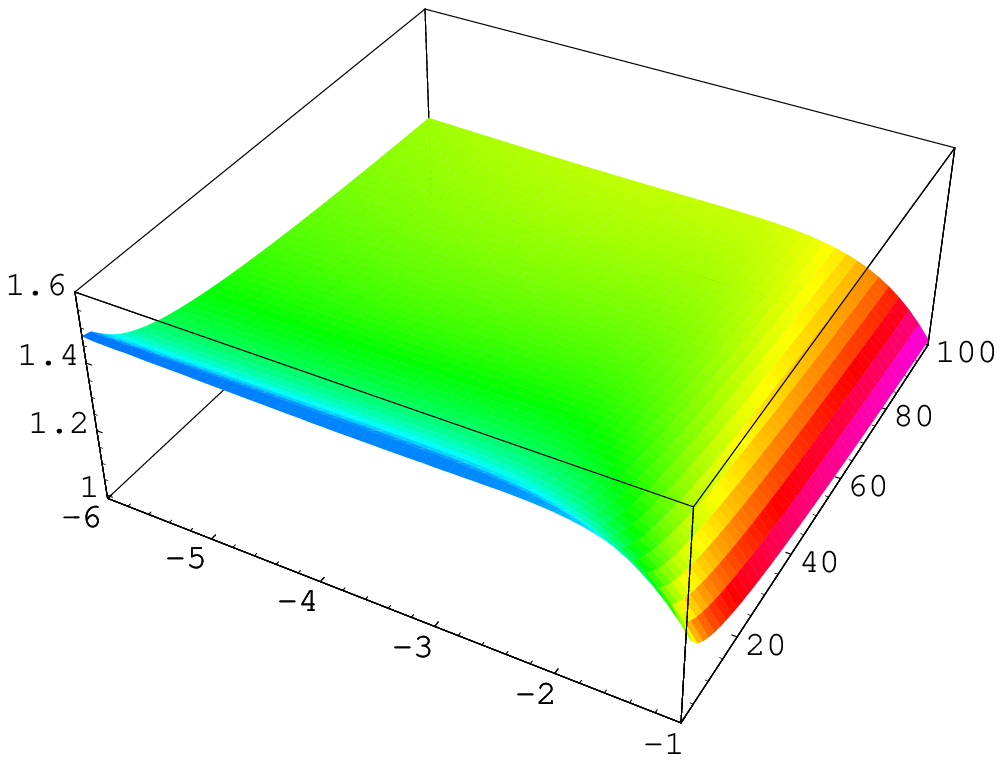}}
\put(10,50){\rotatebox{90}{\small $\xi\left|{^{\rm S}\!{\cal H}} \right|$}}
\put(260,50){\rotatebox{90}{\small ${\rm arg}\left( {^{\rm S}\!{\cal H}}\right)$}}
\put(330,15){\small $\xi$}
\put(415,160){\rotatebox{60} {\small ${\cal Q}^2\ [\GeV^2] $ }}
\put(165,250){\small (a)}
\put(30,150){\insertfig{6}{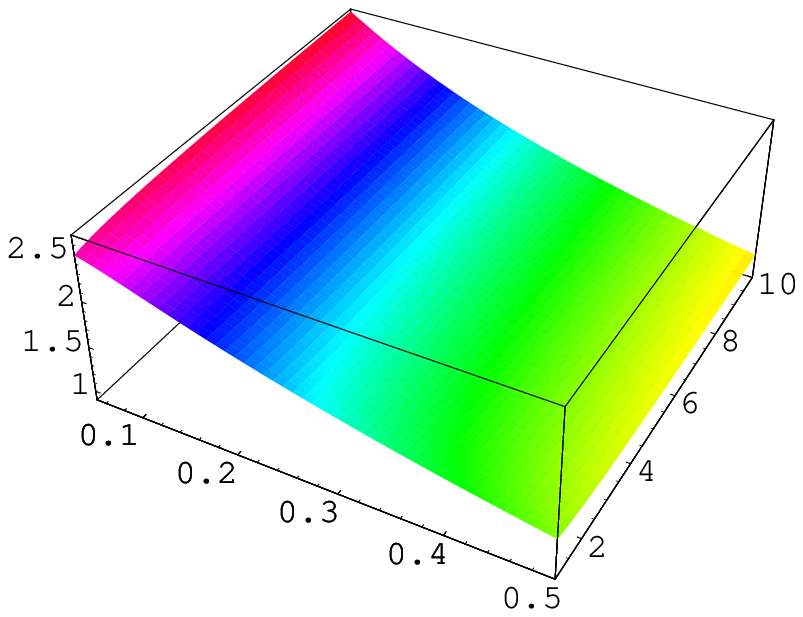}}
\put(80,160){\small $\xi $}
\put(168,160){\rotatebox{60}{\small ${\cal Q}^2\ [\GeV^2] $}}
\put(415,250){\small (b)}
\put(280,150){\insertfig{6}{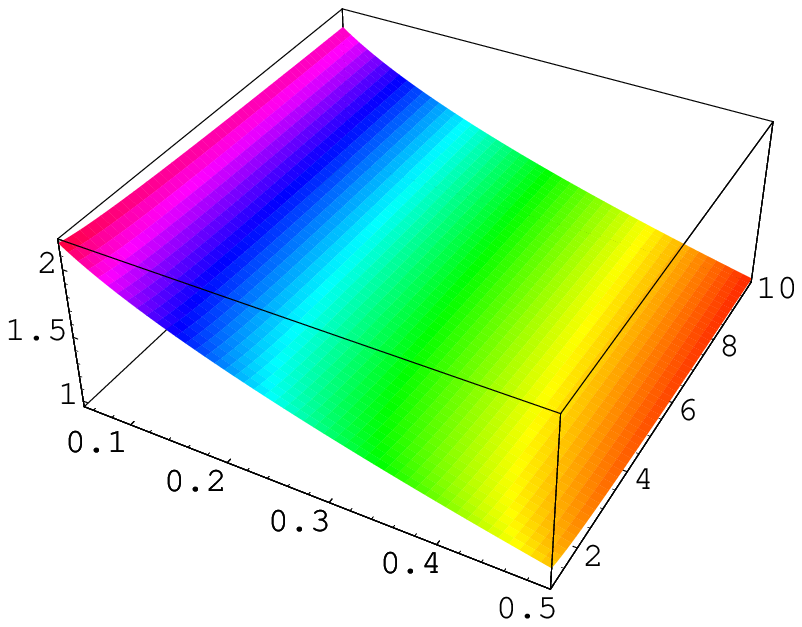}}
\put(10,200){\rotatebox{90}{\small $ \sqrt{\xi} \left|{^{\rm val}\!{\cal H}} \right|$}}
\put(260,200){\rotatebox{90}{\small ${\rm arg}\left( {^{\rm val}\!{\cal H}}\right)$}}
\put(340,160){\small $\xi $}
\put(420,13){\rotatebox{60}{\small ${\cal Q}^2\ [\GeV^2]  $}}
\end{picture}
}
\end{center}
\caption{ \label{Fig-EVLO} The  scaled moduli (a,c) and phases (b,d) of the
flavor nonsinglet (up) and singlet (down) CFF $\cal H$
are displayed to LO accuracy versus ${\cal Q}^2$ and $\xi$ for fixed
$\Delta^2=-0.25\, \GeV^2$. At the input scale ${\cal Q}_0^2=2.5\, \GeV^2$
we use the ans\"atze (\ref{Ans-ConMomNS-pur}) and (\ref{AnsSinHar})
for conformal GPD moments.  The running coupling is normalized to
$\alpha_s({\cal Q}_0^2)/\pi = 0.1$, where the number
of active quarks is set to four.}
\end{figure}
This is demonstrated in Fig.\ \ref{Fig-EVLO}, where we display
the modulus (left) and the phase (right) of the pure valence
nonsinglet (up) and the ``hard gluon'' singlet (down) CFF,
respectively. For the former one we choose the kinematical region
that covers the phase space of present fixed target experiments,
namely $0.05 \lesssim \xi \lesssim 0.5$ [$0.1 \lesssim x_{\rm Bj}
\lesssim 0.65$] and $1\, \GeV^2 \le {\cal Q}^2\le 10\, \GeV^2$.
For the latter one we also include the phase space that is
explored in collider experiments, i.e., $10^{-6} \lesssim \xi
\lesssim 0.5$ [$2\cdot 10^{-6} \lesssim x_{\rm Bj} \lesssim 0.65$]
and $1\, \GeV^2 \le {\cal Q}^2\le 100\, \GeV^2$. The simple shape
of both the modulus and the phase is perhaps a more general feature of
CFFs and should  not be considered as an artifact of our
approximation \cite{FreMcD01b}. It is for $1\, \GeV^2 \le {\cal
Q}^2\le 10\, \GeV^2$ almost independent on the photon virtuality.
In Sect.\ \ref{subsubsec-Num-FlaNS} we will have a closer look to
the evolution. Here we remark only that both the moduli and
phases of CFFs are rather ``planar'' in fixed target kinematics. With
increasing $1/\xi$ the evolution starts strongly to affect the
modulus in the singlet sector. It grows with increasing scale
${\cal Q}^2$, namely, in such a way that the resumed logarithmical
scaling violations lead to a power like change of its
$1/\xi$-dependence. This feature is well known from unpolarized
DIS and in agreement with experimental DVCS results
\cite{Adletal01,Chekanov:2003ya,Aktas:2005ty}. The phase is much
less affected by evolution and in the small $\xi$ region it is
nearly independent of $\xi$. It monotonously decreases from its
input value to be changed by less than $25\%$ at ${\cal Q}^2 =
100\ \GeV^2$.

\subsection{Size of NLO radiative corrections:
 $\overline{\rm CS}$ versus $\overline{\rm MS}$ scheme }
\label{SecCon2MSscheme}

Now we  explore  the radiative corrections to NLO accuracy and in
particular the differences between the $\overline{\rm CS}$ and
$\overline{\rm MS}$ schemes, separately for the flavor
nonsinglet and singlet sector.  Before we do so let us explain
the relation between these schemes and define the quantities
that serve us as measure for the size of perturbative corrections.

The perturbative expansion of a CFF in the $\overline{\rm CS}$
scheme, e.g., for the flavor nonsinglet case, might be
structurally written to NLO as
\begin{equation}
\label{GenFor-PerExpCS}
\mathcal{F} =   \left[ C^{(0)}
+ \frac{\alpha_{\rm s}}{2\pi}
\big(C^{(1)\overline{\rm CS}} +
C^{(0)} \mathcal{A}^{(1)} \big)\right]\mathcal{E}^{(0)}
\otimes F^{\overline{\rm CS}}+ {\cal O}(\alpha_s^2)\,.
\end{equation}
Here $\mathcal{E}^{(0)}$ denotes the evolution operator
(\ref{Def-EvoOpeNS}) in LO approximation and $\mathcal{A}^{(1)} $
is the NLO correction. The convolution symbol $\otimes$ indicates
the integration over the complex valued conformal spin $j$, which
is for shortness not particularized, where the measure includes
the appropriate normalization and $\xi$ dependence, see
(\ref{Res-ImReCFF}) and (\ref{eq:copeCexp2}). By construction, in
the forward kinematics this scheme is {\em identical} with the
$\overline{\rm MS}$ one, see normalization condition
(\ref{NorCon-c}). However, for DVCS kinematics the form of Wilson
coefficients and evolution operator in the $\overline{\rm MS}$
scheme are already at NLO modified by off-diagonal, i.e., $\eta$
proportional, terms, see Eqs.\ (\ref{Def-EvoEq-gen-Ope})  and
(\ref{Def-EvoEq-gen-WC}). Because of the particularity of the DVCS
process, where $\eta=\xi$, both Wilson coefficients and evolution
operator are finally modified by an infinite sum of terms, which
appears then in front of the conformal GPD moments, cf. Eq.\
(\ref{Res-EvoOpeMS}):
\begin{equation}
\label{GenFor-PerExpMS}
\mathcal{F} =  \left[ C^{(0)} + \frac{\alpha_{\rm s}}{2\pi}
\big(C^{(1)\overline{\rm MS}} + C^{(0)} \mathcal{A}^{(1)} +
C^{(0)}\oplus\mathcal{B}^{(1)} \big) \right]\mathcal{E}^{(0)}\otimes
F^{\overline{\rm MS}}+ {\cal O}(\alpha_s^2)\,.
\end{equation}
Here $C^{(0)}\oplus\mathcal{B}^{(1)}$ contains off-diagonal part of the evolution operator, where
$\oplus$ symbolizes the summation over the conformal
spin, cf.\ Eq.\ (\ref{Cal-OffDigEvo0}).
Since the physical observable $\mathcal{F}$ is independent of our
conventions, the conformal moments in both schemes are related to
each other by a scheme transformation. At the input scale ${\cal
Q}^2 = {\cal Q}_0^2$, where $\mathcal{E}^{(0)} =1$, $\mathcal{A}^{(1)}=0$, and
$\mathcal{B}^{(1)}=0$,  we might express this transformation by a
finite factorization (or renormalization) constant $z^{(1)}$:
\begin{eqnarray}
F^{\overline{\rm MS}} = F^{\overline{\rm CS}} + \frac{\alpha_s}{2 \pi}
z^{(1)}(\eta) \oplus F^{\overline{\rm CS}}\,,
\quad
C^{(1)\overline{\rm CS}} = C^{(1)\overline{\rm MS}} + C^{(0)} \oplus z^{(1)}\,.
\end{eqnarray}
For (positive) integer conformal spin the triangular matrix
$z^{(1)}_{jk}(\eta) = \eta^{j-k} z^{(1)}_{jk}$ contains {\em only}
off-diagonal entries, i.e., $j \ge k+2$. Note that the change of
Wilson coefficients is of course $\eta$-independent, while the
skewness dependence of the conformal GPD moments is altered, which
is at least suppressed by a factor $\eta^2$. In particular, the
two lowest GPD moments are untouched by the scheme transformation,
while the (positive non-vanishing integer) moments are modified by
$\alpha_s$ and $\eta^2$ suppressed contributions:
\begin{eqnarray}
F_0^{\overline{\rm MS}} =  F_0^{\overline{\rm CS}}
\,,\
F_1^{\overline{\rm MS}} =  F_1^{\overline{\rm CS}}\, ,
\quad\mbox{and}\quad F^{\overline{\rm MS}}_j =
F_j^{\overline{\rm CS}} +  \frac{\alpha_s}{2 \pi} {\cal
O}(\eta^2)\quad\mbox{for}\quad j=3,4,\cdots\,.
\end{eqnarray}

Strictly spoken, the truncation of the perturbative expansion also
induces a discrepancy in the CFFs between these two schemes, which
is beyond the approximation we are dealing with, i.e., of order
$\alpha_s^2$.  However, the conformal GPD moments revealed from a
given data set will differ to order $\alpha_s$. In the following
we study the NLO corrections in both schemes within the same
ansatz. The resulting  deviation in the CFFs can be viewed as a
measure for the needed reparameterization of conformal GPD moments
by altering their $\eta$ dependencies.

We introduce now  the quantities that we utilize as a measures of
the scheme dependence and, foremostly, as indicators of the
convergency of the perturbation series. It is natural to employ
for this purpose the ratio of the CFF at order ${\rm N}^{P}{\rm
LO}$ to the one at order ${\rm N}^{P-1}{\rm LO}$, where $P=\{0,1,2\}$ stands
for LO, NLO, and NNLO order, respectively:
\begin{eqnarray}
\label{Def-Krat}
\frac{{\cal H}^{P}}{{\cal H}^{P-1}}(\xi,\Delta^2,{\cal Q}^2|{\cal Q}_0^2) \equiv
K^{P}(\xi,\Delta^2,{\cal Q}^2|{\cal Q}_0^2)\exp\{i \delta^{P}\varphi(\xi,\Delta^2,{\cal Q}^2|{\cal Q}_0^2)\}\,.
\end{eqnarray}
The phase difference
\begin{eqnarray}
\label{Def-PhaDif}
\delta^P \varphi=
{\rm arg}\!\left( \frac{{\cal H}^{{\rm N}^P{\rm LO}}}{ {\cal H}^{{\rm N}^{P-1}{\rm LO}}}\right)
\sim {\cal O}\left(\alpha_s^P\right)
\end{eqnarray}
is formally of order $(\alpha_s/2\pi)^P$. If convergency holds, it
diminishes in higher orders. Under this circumstance the ratio of
moduli
\begin{eqnarray}
K^{P}=
\frac{\left|{{\cal H}}^{{\rm N}^P{\rm LO}}\right|}
{\left|{{\cal H}}^{{\rm N}^{P-1}{\rm LO}}\right|}
\end{eqnarray}
approaches one, i.e., its deviation from one vanishes, too:
\begin{eqnarray}
\label{Def-Del}
\delta^{P} K= K^{P}-1 \sim {\cal O}\left(\alpha_s^P\right)\,.
\end{eqnarray}
If we suppose that the perturbative expansion is not ill-behaved,
then  the radiative corrections should not overshoot the size
itself of the CFF in a given order, i.e.,
\begin{eqnarray}
\label{Con-ConPerSer}
\left|{{\cal H}}^{{\rm
N}^{P}{\rm LO}} -{{\cal H}}^{{\rm N}^{P-1}{\rm LO}} \right| = r
\left| {{\cal H}}^{{\rm N}^{P-1}{\rm LO}}\right| \quad \mbox{with} \quad  r < 1\,.
\end{eqnarray}
If this inequality holds true, the phase difference (\ref{Def-PhaDif}) is geometrically
constrained by the value of $r$, otherwise it is independent.
More precisely, we have the upper bound
\begin{eqnarray}
\label{Con-Pha-Mod}
\left|\sin \delta^P \varphi\right|\le   r
 \quad \Rightarrow\quad
 \left|\delta^P \varphi\right| \le  \frac{\pi}{2} r\quad  \mbox{for}\quad r \le 1 \,.
\end{eqnarray}
Moreover, the triangle inequality, applied to Eq.\ (\ref{Con-ConPerSer}),
constrains the variation of the modulus, namely, $r \ge
\left|\delta^P K\right|$. Employing the $\cos$-theorem, we can
appraise the radiative corrections (\ref{Con-ConPerSer}),
quantified by the ratio $r$, from the variation of the phase
and modulus. It will turn out that in our analysis the phase
difference $\left|\delta^P \varphi\right|$ is always small, i.e.,
$\left|\delta^P \varphi\right| \ll \pi/2$. Hence, we can rely on the
expanded version of this theorem,
\begin{eqnarray}
r^2 \approx (\delta^P K)^2 + (1+\delta^P K) (\delta^P \varphi)^2\,,
\end{eqnarray}
which gives us a simple form of the constraint among the variations
of the modulus and phase as well as the size of radiative
corrections. Note that in the case of a (very) small phase change the
variation of the modulus is roughly estimated to be $\left|\delta^P K\right| \sim r$.

\subsubsection{Flavor nonsinglet sector}
\label{subsubsec-Num-FlaNS}

The flavor nonsinglet CFF ${^{\rm NS}{\cal H}}$ is
straightforwardly evaluated from the two ans\"atze
(\ref{Ans-ConMomNS-pur}) and (\ref{Ans-ConMomNS-pse}) by means of
the Mellin--Barnes integral (\ref{Res-ImReCFF}).  The
Wilson coefficients, needed for our NLO ($P=1$) analysis,  are
listed for the $\overline{\rm CS}$ and $\overline{\rm MS}$ scheme
in Sect.\ \ref{SubSec-PerExpWC} and \ref{SubSec-MS}, respectively,
while the relevant expansion of the evolution operator  can be
read off from Sects. \ref{SubSec-ExpEvoOpe} and \ref{SubSec-MS}.
As said above, we combine the perturbative expansion of the
Wilson coefficients with that of the evolution operator in a
consistent manner, where the leading logs are resummed, cf.\ Eqs.\
(\ref{GenFor-PerExpCS}) and (\ref{GenFor-PerExpMS}). As long as it is
not stated otherwise we equate the factorization $\mu_f \equiv \mu$ and
renormalization $\mu_r$ scales with the photon virtuality $\cal
Q$. As input scale for the conformal GPD moments we use as before
${\cal Q}^2_0 =2.5\, \GeV^2$. This input scale serves us also to
normalize the coupling constant $\alpha_s({\cal Q}^2_0)/\pi =
0.1$, where its running is described by the {\em exact} numerical
solution of the NLO renormalization group equation.

\begin{figure}[t]
\begin{center}
\mbox{
\begin{picture}(450,300)(0,0)
\put(-20,0){\insertfig{17.5}{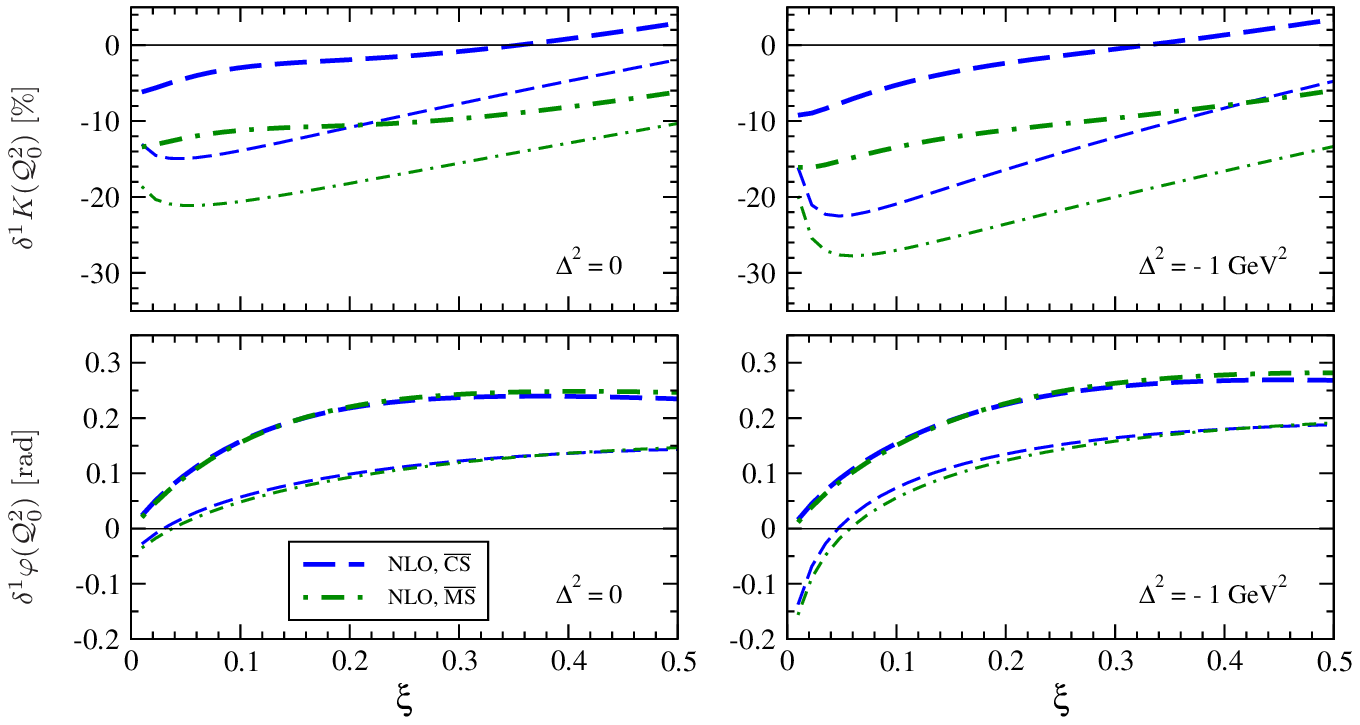}}
\end{picture}
}
\end{center}
 \caption{ \label{FigNLO-WilNS} The relative NLO radiative
corrections (\ref{Def-Del}) and (\ref{Def-PhaDif}) to ${^{\rm
NS}{\cal H}}$  for $\Delta^2=0$ (left) and $\Delta^2=
-1\,\mbox{GeV}^2$ (right) are plotted versus $\xi$ for the moduli
(up) and phases (down) within the $\overline{\rm CS}$ (dashed) and
$\overline{\rm MS}$ (dash-dotted) scheme. Thin and thick lines
refer to a  valence-like ansatz (\ref{Ans-ConMomNS-pur}) and one
with a sea-quark admixture (\ref{Ans-ConMomNS-pse}), given at the
input scale ${\cal Q}^2_0= 2.5\, \GeV^2$. We equated the scales
$\mu_f=\mu_r= {\cal Q}$ and take the normalization condition
$\alpha_s({\cal Q}^2_0)/\pi = 0.1$.
 }
\end{figure}
In Fig.\ \ref{FigNLO-WilNS} we display the relative NLO
corrections (\ref{Def-Del}) and (\ref{Def-PhaDif}) in the
$\overline{\rm CS}$ (dashed) and $\overline{\rm MS}$
(dash-dotted) scheme to the moduli (up) and phases (down) for the
phase space of present fixed target experiments. Thereby, we
employ the pure valence-like ansatz (thin lines), i.e., Eq.\
(\ref{Ans-ConMomNS-pur}), and the one with a sea quark admixture
(thick lines), given in Eq.\ (\ref{Ans-ConMomNS-pse}). For the
momentum transfer squared we take  two extreme values, namely,
$\Delta^2 =0$ (left) and $\Delta^2 =-1\, \GeV^2$ (right). Comparing the
resulting radiative corrections for both choices, one realizes
that the $\Delta^2$-dependencies only slightly influence their
size.

Due to the radiative corrections the phases (lower panels)
increase by a small amount, except for the valence-like ansatz
(thin lines) where we observe a decrease for smaller values of
$\xi$. In any case, the absolute value of the phase differences
$\left|\delta^P\varphi \right|$ does not exceed $0.1\pi$ rad. The
influence of the NLO corrections  on the moduli (upper panels) is
more pronounced.  Generally, they {\em moderately} reduce the
moduli, however, for the sea quark admixture ansatz  in the
$\overline{\rm CS}$ (thick dashed) there is a small increase at
larger values of $\xi$. The modulus is more affected
for the valence-like ansatz (thin) than for the one with a sea quark
admixture (thick), while in the case of the phase difference the situation is
reversed.

Comparing the  corrections in the $\overline{\rm CS}$ (dashed) and
$\overline{\rm MS}$ (dash-dotted) scheme,  one realizes that in
the former scheme they are smaller for the moduli, while the phase
differences are almost independent of the specific choice. This is
in agreement with the findings of Ref.\ \cite{Mue05a}, where a
slightly different ansatz has been chosen. As explained above, the
difference between the two schemes originates from the skewness
dependence. For positive integer conformal spin, we would count them
in the conformal GPD moments as $\eta^2$ effects, suppressed
by $\alpha_s/2\pi$,. However, as we also spelled out, in DVCS
kinematics  we should not use $\eta=\xi$ as an expansion
parameter, since the change of Wilson coefficients is
$\eta$-independent. Indeed, roughly spoken, the moduli differences
in both schemes are of the same order as the radiative corrections
themselves and nearly $\xi$ independent. This observation should
be understood as a warning that $\eta^2$ suppressed terms in
conformal GPD moments perhaps cannot be simply neglected by
formal $\eta^2$ counting.

We consider now the evolution to LO and NLO accuracy, where we compare the
modulus and phase at a given scale with those at the input scale,
quantified by the ratios:
\begin{eqnarray}
\label{Def-MesEvo}
\Delta K({\cal Q}^2,{\cal Q}^2_0)=
\frac{\left|{{\cal H}}({\cal Q}^2)\right|}
{\left|{{\cal H}}({\cal Q}^2_0)\right|}-1\, \qquad
\Delta \varphi({\cal Q}^2,{\cal Q}^2_0)=
{\rm arg}\!\left( \frac{{\cal H}({\cal Q}^2)}{ {\cal H}{(\cal Q}^2_0) }\right)\,.
\end{eqnarray}
Note that in contrast to the definitions
(\ref{Def-PhaDif}) and (\ref{Def-Del}), in which the variation is
denoted by $\delta$, here we  do not compare the CFFs in different order,
but rather measure the strength of the evolution within a given order.
The evolution in the $\overline{\rm MS}$ scheme is consistently
treated, i.e., the mixing of conformal GPD moments is taken into
account. We remark, however, that this effect is tiny and can be
safely neglected. For instance, for the quantity
\begin{eqnarray}
\frac{\left|{^{\rm NS}\!{\cal H}}({\cal Q}^2)\right|
- \left|{^{\rm NS}\!{\cal H}}^{\rm dia}({\cal Q}^2)\right|}
{\left|{^{\rm NS}\!{\cal H}}({\cal Q}^2)\right|}\,,
\nonumber
\end{eqnarray}
where the superscript `dia' stands for neglecting the non-diagonal
parts in the anomalous dimension $^{\rm NS}\!\!\ \gamma_{jk}$,
we find in a broad range of
$\xi$ and $\cal Q$ a value on the level of few per mil.
The phase differences are negligible, too.

\begin{figure}[t]
\begin{center}
\mbox{
\begin{picture}(450,300)(0,0)
\put(-20,0){\insertfig{17.3}{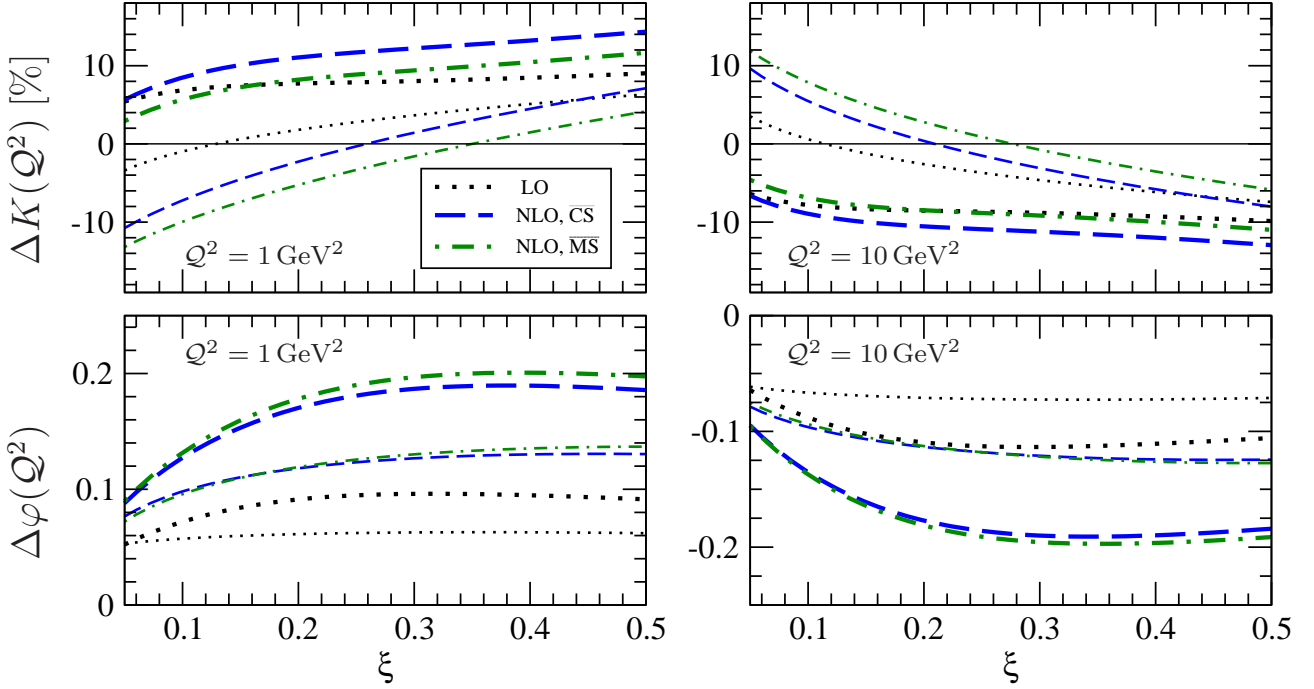}}
\end{picture}
}
\end{center}
\caption{ \label{FigNLO-EvoNS} The modulus (up) and phase (down)
changes of the evolved CFFs ${^{\rm NS}{\cal H}}$ are plotted for
fixed ${\cal Q}^2 = 1\,\GeV^2$ (left), ${\cal Q}^2 = 10\,\GeV^2$
(right), and $\Delta^2=-0.25\, \GeV^2$ versus $\xi$ in LO (dotted)
and NLO:  $\overline{\rm CS}$ (dashed) and $\overline{\rm MS}$
(dash-dotted) scheme. The ans\"atze and scale setting
prescriptions are the same as in Fig.\ \ref{FigNLO-WilNS}. }
\end{figure}
In Fig.\ \ref{FigNLO-EvoNS} we show the evolution effects for the
same ans\"atze as before, in LO (dotted) and NLO for both the
$\overline{\rm CS}$ (dashed) and $\overline{\rm MS}$ (dash-dotted)
schemes. In the left and right panels we plot the quantities
(\ref{Def-MesEvo}) at a scale ${\cal Q}^2= 1\, \GeV^2$ and ${\cal
Q}^2= 10\, \GeV^2$, respectively. For simplicity, we do not
perform any matching at the charm threshold. As already observed
for the radiative corrections at the input scale, see lower panels
in Fig.\ \ref{FigNLO-WilNS}, the phase differences in right [left]
panels in Fig.\ \ref{FigNLO-EvoNS} are again rather small and lie
for forward [backward] evolution in the interval $-0.06 \pi\cdots
-0.02\pi$ [$0.02 \pi \cdots 0.06\pi$] rad. The signs tell us that
the phases decrease during the evolution to a larger scale.
Radiative corrections amplify this effect, where the differences
between both schemes are again tiny. Comparing the upper left and
right panels, we see that for the valence-like ansatz  in LO (thin
dotted line) the evolution mildly affects the moduli, i.e., of the
order of about $\pm 5\%$ at the edges of the phase space%
\footnote{For $-\Delta^2 \sim 1\ \GeV^2$ this evolution effect
might increase to become of the order of $\pm 10\%$ for smaller
values of $\xi$. This is caused by the shift of the leading meson
Regge pole to the left, approaching the $j=-1$ pole of the
anomalous dimension, see Eq.\ (\ref{eq:NSgamma0}).}
explored in the present fixed target experiments.

Furthermore, the modulus crosses for low and high values of ${\cal
Q}^2$ (upper left and right panels) the zero line at almost the
same point $\xi \simeq 0.12$ to LO accuracy (thin dotted).  This
means that the CFF in the vicinity of this point is nearly scale
independent.  As in DIS, the evolution predicts a decreasing
modulus for $\xi > \xi_0$ and increasing one for $\xi < \xi_0$
with growing ${\cal Q}^2$. The value of $\xi_0$ is a function of
$\Delta^2$ and depends on the input.  In our example it is shifted
by radiative corrections to the right (thin dashed and dash-dotted
lines). This balance effect arises from the fact that forward
evolution suppresses (enhances) the large (small) momentum
fraction region.

The radiative corrections lead to an amplification of the
evolution effects. Note that the Wilson coefficients now also
depend on ${\cal Q}^2$ and the perturbative corrections are
getting smaller with increasing ${\cal Q}^2$. Now the variation of
the CFF modulus reaches for the valence-like ansatz (thin lines)
the $\mp 10\%$ [$\mp 15\%$] (for small $\xi$) to $\pm 8\%$ [$\pm
6\%$] (for large $\xi$) level in the $\overline{\rm CS}$ (dashed)
[$\overline{\rm MS}$ (dash-dotted)] scheme. For the ansatz with
sea quark admixture (thick lines) there is no crossing point with
the zero line anymore and so the `balance-point' is shifted
outside of the discussed kinematical region. Also here the
strength of evolution grows by radiative corrections and varies in
the range from $\pm 6\%$ (small $\xi$) to $\pm 15\%$ (large
$\xi$).

As we realized, in both ans\"atze the NLO corrections within the
$\overline{\rm MS}$ (dash-dotted lines)  and the $\overline{\rm
CS}$ (dashed lines) change the LO prediction (dotted line). The
differences, caused by the scheme dependence, mainly arises from
the different Wilson coefficients to NLO, which are multiplied
with the LO evolution operator. The contribution of the diagonal
part of the NLO anomalous dimensions is small, i.e, about one
percent. As we spelled out above, the influence of the
non-diagonal part, appearing in the $\overline{\rm MS}$ scheme in
the anomalous dimensions, is negligible.

We would like to briefly confront evolution effects with experimental
measurements from the Hall A experiment at Jefferson LAB
\cite{Cametal06}, where scaling was reported. Within the lever arm
$1.5\ \GeV^2\le {\cal Q}^2\le 2.5\ \GeV^2 $, we find  for the
valence-like ansatz that the scaling violation  due to the LO
evolution is small, namely, the modulus of the CFF varies by about
$1.2\%$ for $x_{\rm Bj}=0.36$ (i.e., $\xi=0.22$) and
$\Delta^2=-0.25\ \GeV^2$, while a sea quark admixture can lead to
a change of up to 4\%. In NLO the variation for the former
ansatz is $-0.5\%\, [-2\%]$ for the $\overline{\rm CS}$
[$\overline{\rm MS}$] scheme, while for the later one we find
$5.7\%\, [4.3\%]$. If one naturally assumes that for $x_{\rm
Bj}=0.36$ the valence components dominate, one might conclude that
the observed scaling in Ref.\ \cite{Cametal06} indicates the
smallness of higher twist contributions. However, in modelling of
GPDs one usually realizes that the role of sea quarks in the CFFs
is more pronounced than in DIS structure functions.
Interestingly, it has been argued that indeed the leading Regge
trajectory essentially contributes to DVCS even in the valence region,
in contrast to DIS \cite{SzcLon06}. We only like to point out
here that even for fixed target kinematics a detailed view on
scaling breaking effects for the net contribution to CFFs is {\em
necessary} and that it might be used to constrain the GPD ansatz.

So far we have considered only the scale setting prescription
$\mu_f=\mu_r = \cal Q$. Change obtained by choosing another
prescription within the same input scale ${\cal Q}_0$ is often
considered as an estimate for the possible size of higher order
corrections. Let us have a closer look at this point of view. A
change of $\mu_r$ modifies the size of the Wilson coefficients by
formally inducing a $\beta_0$ proportional contribution
$\alpha_s^2/(2\pi)^2$ that is multiplied with the NLO correction
to the Wilson coefficients themselves, see Eq.\
(\ref{eq:copeCexp2c}). A modification of $\mu_f$ essentially
corresponds to the difference between the expanded and
non-expanded version of the evolution operator which is formally
also of higher order in $\alpha_s$. However, besides non-leading
log terms, it contains also leading ones, e.g., proportional to
$\alpha_s^2/(2\pi)^2 \ln^2\left(\mu_f/{\cal Q}_0\right)$.
Certainly, whenever a new entry appears in the next order that is
completely independent of these quantities then the rough higher
order estimate can fail.

\begin{table}
\begin{center}
\begin{tabular}{|c|c|c|r|r|r|r|}
  \hline
 variation, order/ $\xi$ &                           0.05 & 0.1 & 0.25 & 0.5   \\ \hline\hline
 $\mu_f$, LO &                                     3.7 [-6.9] & 0.7 [-8.5] & -3.8 [-9.5] & -8.0 [-10.9]
 \\ \hline\hline
 $\mu_f$, NLO ($\overline{\rm CS}$)  &             0.5  [-0.8]& 0.7 [-0.7] & 0.8 [-0.3]& -0.1 [-0.8] \\
 \hline\hline
  $\mu_f$, NLO ($\overline{\rm MS}$) &              0.3 [-0.6]  & 0.7 [-0.3]  & 1.0 [0.3] & 0.4 [0.2]\\
 \hline\hline
  $\mu_r$, NLO ($\overline{\rm CS}$) &             5.5 [1.3] & 4.9 [0.6] & 2.9 [-0.4] & 0.3 [-1.6]  \\
  \hline\hline
   $\mu_r$, NLO  ($\overline{\rm MS}$) &             7.9 [3.9]& 7.6 [3.2] & 5.7 [2.2] & 2.9 [0.8]   \\
 \hline
\end{tabular}
\caption{ \label{Tab-ScaSetNS} Variation (\ref{Def-VarScaDep}) in
percent of the nonsinglet CFFs ${^{\rm NS}\! {\cal H}}$ induced
by separate factorization and renormalization scale changes from
${\cal Q}^2/2 \cdots 2{\cal Q}^2$. Here we used the valence-like
[with sea quark admixture] conformal GPD moments
(\ref{Ans-ConMomNS-pur}) [(\ref{Ans-ConMomNS-pse})] and set ${\cal
Q}^2 = 4\,\GeV^2$  and $\Delta^2 = -0.25\, \GeV^2$.}
\end{center}
\end{table}
Let us explore these estimates in more detail by employing the definitions
\begin{eqnarray}
\label{Def-VarScaDep}
\delta_i = \frac{\left|{{\cal H}}({\cal Q}^2|\mu^2_i =2 {\cal Q}^2)\right| -
\left|{{\cal H}}^{\rm}({\cal Q}^2|\mu^2_i ={\cal Q}^2/2)\right|}
{\left|{{\cal H}}({\cal Q}^2|\mu^2_i ={\cal Q}^2)\right|}\,,
\end{eqnarray}
where $\mu_i$ is the factorization ($i=f$) [renormalization
($i=r$)] scale and the renormalization [factorization] scale is
fixed to be ${\cal Q}^2$. At LO we can only change the
prescription for the ambiguous factorization scale setting.  As
one can read off from Tab.\ \ref{Tab-ScaSetNS}, the scale
uncertainty to LO goes from about $4\%$ $[-7\%]$ to $-8\%$
$[-11\%]$ with increasing $\xi$ for the valence-like [with sea
quark admixture] ansatz. Comparing with the upper right panel in
Fig. \ref{FigNLO-EvoNS}, we realize that this uncertainty reflects
nothing else but the LO evolution itself. As we have expected,
these numbers are not correlated to the perturbative corrections.
At NLO the factorization scale dependence is drastically reduced
and is now only about $\pm 1\%$ or even smaller in both schemes.
At this order, the renormalization scale
dependence arises and the modulus of the CFF can vary of up to
$6\%\, [8\%]$ by changing the scale in the $\overline{\rm CS}$
[$\overline{\rm MS}$] scheme. We will come back to these numbers
and compare them with the actual NNLO corrections in the $\overline{\rm
CS}$ scheme, evaluated in Sect.\ \ref{SubSecRadCorNNLO}. We should
stress here that the uncertainties with respect to the
factorization scale setting are `maximized' at LO. Or, in other
words, revealing GPDs from experimental data in this approximation
means that one does not know at which resolution scale
$\mu_f^2$ this information was extracted. Was it ${\cal Q}^2$,
${\cal Q}^2/2$, or \ldots ?

\subsubsection{Flavor singlet sector}

For the numerical studies in the flavor singlet sector we use the
same scale prescriptions and normalization conditions as in the
preceding section. As input we alternatively take the `soft' and
`hard' gluonic ans\"atze (\ref{AnsSinSof}) and (\ref{AnsSinHar}),
respectively. The Wilson coefficients for the $\overline{\rm CS}$
are listed in Eqs.\ (\ref{Res-WilCoe-Exp-CS-SI}) and
(\ref{Res-WilCoe-CS-NLO}) and the evolution operator can be read
off from Eq.\ (\ref{Exp--EvoOpe}). The results for the
$\overline{\rm MS}$ scheme are collected in Sect.\
\ref{SubSec-MS}.

\begin{figure}[t]
\begin{center}
\mbox{
\begin{picture}(600,270)(0,0)
\put(-12,0){
\insertfig{17.5}{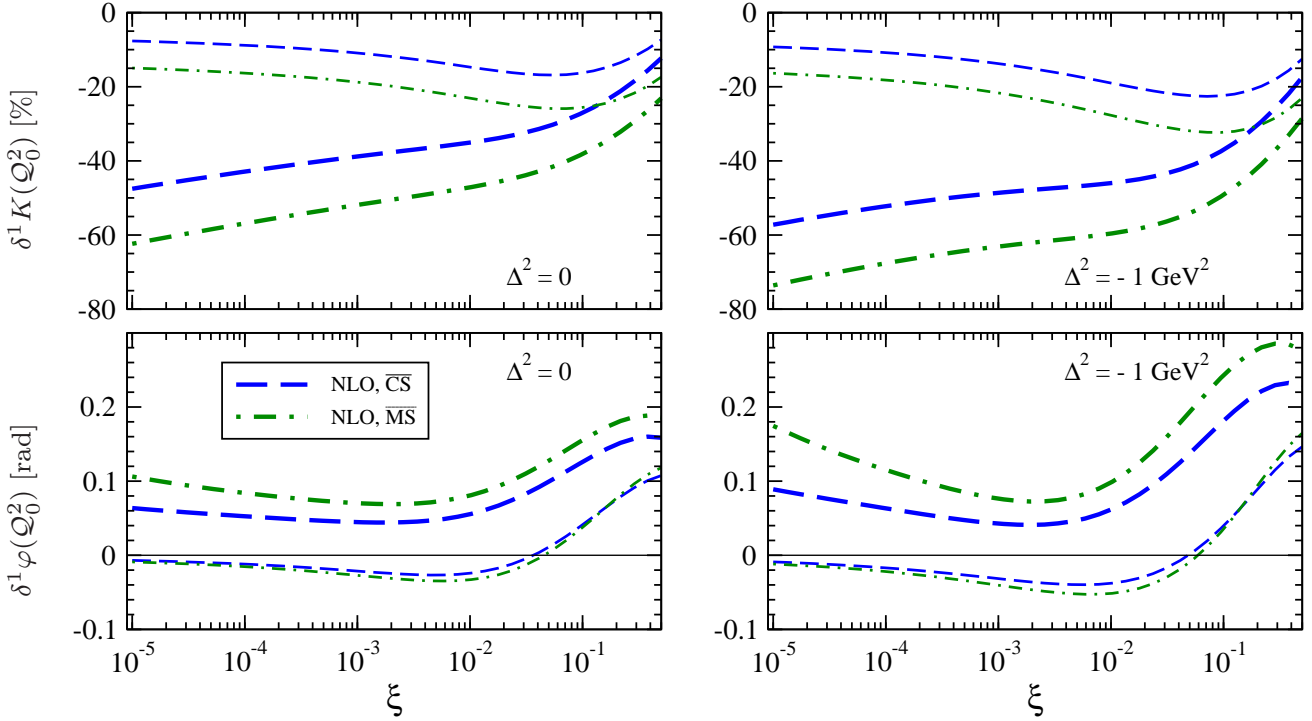}}
\end{picture}
}
\end{center}
\caption{ \label{FigNLO-Wil}
The relative NLO radiative corrections (\ref{Def-PhaDif}) and (\ref{Def-Del})
are plotted versus $\xi$ for the modulus (up) and phase (down) of
${^{\rm S}{\cal H}}$ for $\Delta^2=0$ (left) and $\Delta^2=
-1\,\mbox{GeV}^2$  (right): $\overline{\rm CS}$  (dashed) and
$\overline{\rm MS}$ (dash-dotted) scheme. Thick (thin) lines refer to the ``hard''
(``soft'') gluon parameterization, where the scale
setting prescriptions are the same as in Fig.\ \ref{FigNLO-WilNS}. }
\end{figure}
In Fig.\ \ref{FigNLO-Wil} we plot the relative NLO corrections
(\ref{Def-PhaDif}) and (\ref{Def-Del}) for the moduli (up) and
phase differences (down) at the input scale ${\cal Q}_0^2 = 2.5\,
\GeV^2$, again for the two extreme values of the momentum transfer squared:
$\Delta^2= 0$ (left) and $\Delta^2=-1\, \GeV^2$ (right).  As in
the nonsinglet case, the variation of the phase differences from
LO to NLO is not large and does not exceed $0.09 \pi$ rad for
$\Delta^2=-1\, \GeV^2$ and is even smaller for $\Delta^2= 0$.
However, the NLO corrections to the moduli have now a wider
variety. This is related to the fact that at NLO the gluons enter the
hard scattering part of the DVCS amplitude the first time.
For the `soft' gluon ansatz (thin lines) they lead to a small
decrease of the CFF of about $10-20\%$ and $15-30\%$ for the
$\overline{\rm CS}$ and $\overline{\rm MS}$ scheme, respectively.
In contrast, if the gluon is `hard' (thick lines) it cancels partly
the sea quark dominated LO contribution and reduces so drastically
the modulus of the CFF.
\begin{figure}[t]
\begin{center}
\mbox{
\begin{picture}(500,300)(0,0)
\put(-12,0){
\insertfig{17.5}{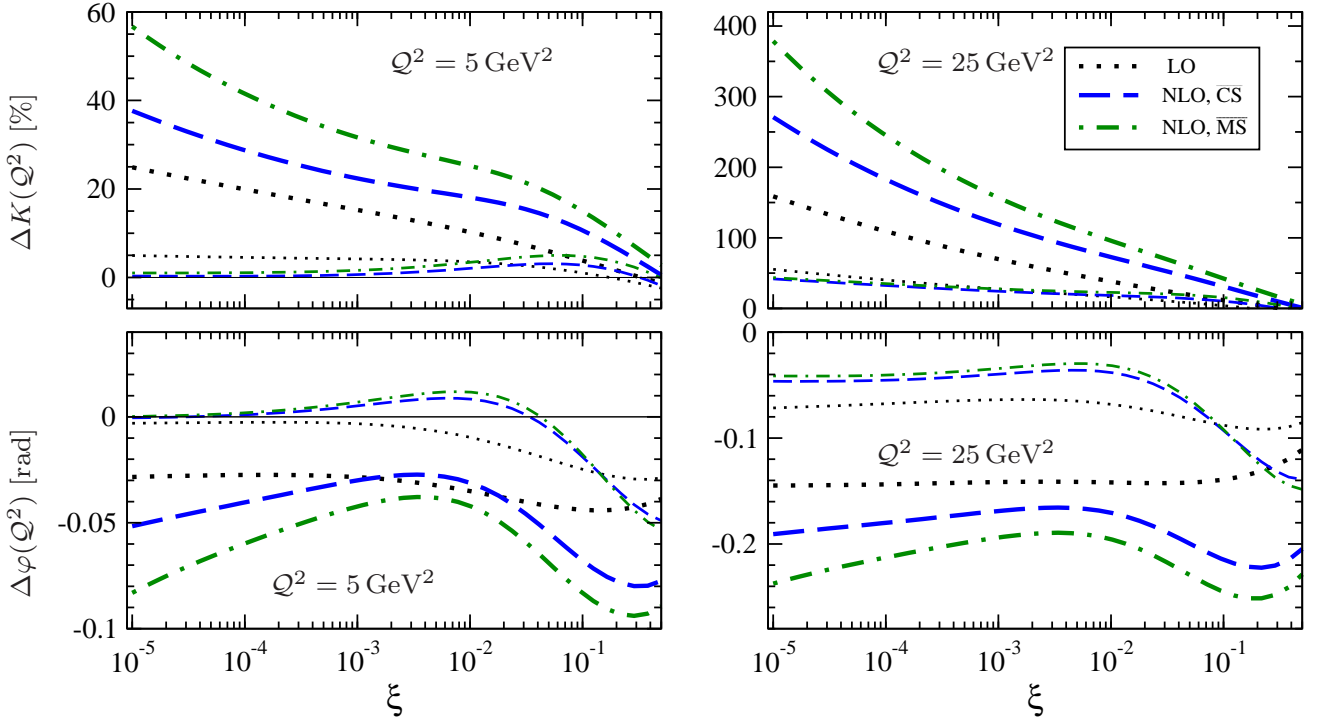}
}
\end{picture}
}
\end{center}
\caption{
\label{FigNLO-EvoS}
Evolution of the flavor singlet CFFs ${^{\rm S}{\cal
H}}$. The moduli (up) and phases (down) are
plotted for fixed ${\cal Q}^2 = 5\,\GeV^2$ (left), ${\cal Q}^2 =
25\,\GeV^2$ (right), and $\Delta^2=-0.25\, \GeV^2$ versus $\xi$ in
LO (dotted) and NLO for the $\overline{\rm CS}$ (dashed) and
$\overline{\rm MS}$ (dash-dotted) scheme. The ans\"atze and scale
setting prescriptions are the same as in Fig.\ \ref{FigNLO-Wil}.
}
\end{figure}
Since in this scenario the gluonic part grows with decreasing $\xi$
faster than the sea
quark one, the modulus of the CFF
monotonously decreases, too. For our ansatz the reduction reaches
$80\%$ at very small $\xi$ and large $\Delta^2$. As in the
nonsinglet case, we observe again that the
radiative corrections are a bit smaller at $\Delta^2=0$. Another similarity is
that in the $\overline{\rm CS}$ scheme they are up to $5-20\%$
smaller than in the $\overline{\rm MS}$ one. Although the size of
perturbative corrections can be very large, the phase differences
are still small. This is caused by the fact that the phase is
dominated by the leading pole of the ansatz. Let us stress that
the rather large corrections to the hard scattering part, induced
by gluons, should not be considered an argument against the
applicability of perturbation theory.

We study now the evolution effects to LO and NLO approximation.
The ratios (\ref{Def-MesEvo}) are plotted in Fig.\
\ref{FigNLO-EvoS} versus $\xi$ for two fixed values ${\cal Q}^2 =
5\, \GeV^2$ and ${\cal Q}^2 = 25\, \GeV^2$ , where $\Delta^2=
-0.25\, \GeV^2.$ The moduli (up) in the `soft' gluon scenario
(thin lines) are relatively mildly affected by evolution and they
grow with decreasing $\xi$. Both the modification of the LO
(dotted) prediction and the difference between the $\overline{\rm
CS}$ (dashed) and $\overline{\rm MS}$ (dash-dotted) schemes are
rather small. However, we remark that the NLO contributions to the
anomalous dimensions are getting large in the small $\xi$ region,
e.g., about 100\%, which is eventually compensated by the evolved
NLO Wilson coefficients. We also found that the off-diagonal
entries in the anomalous dimensions cannot be neglected anymore.
Their contribution to the net result can grow from a few percent
in the large $\xi$ region to over $25\%$ in the small one. This is
related to the fact that the off-diagonal entries
(\ref{eq:gammajk}) contain now $j=0$ poles, which arise from the
LO anomalous dimensions (\ref{Def-LO-AnoDim-GQ-V}) and
(\ref{Def-LO-AnoDim-GG-V}). In the ``hard'' gluon scenario the
evolution effects, the NLO corrections and the scheme dependence
are quite large. The NLO corrections to the evolution are
dominated by those to the anomalous dimensions. We again observe
that the NLO corrections are smaller in the $\overline{\rm CS}$
scheme. The scheme dependence partly arises from the NLO
Wilson coefficients, yielding in the former scheme smaller
corrections, which evolve with the LO evolution operator, however,
also due to the off-diagonal part in the anomalous dimensions.
Corresponding to the evolution effects that appear in the moduli,
the phase differences in the ``soft'' scenario are much smaller
than in the ``hard'' one. However, also in the latter case they cannot be
considered large. Again, we see that at least within our
ans\"atze the phase is protected from radiative
corrections, since their leading pole is in the vicinity of $j=0$.

\begin{table}
\begin{center}
\begin{tabular}{|c|c|c|r|r|r|r|}
  \hline
 variation, order/ $\xi$ &                      $10^{-5}$  &   $10^{-3}$ & $10^{-1}$ & $0.25$ & $0.5 $  \\ \hline\hline
 $\mu_f$, LO &                                  13.2 [49.5] &  9.4 [29.4] & 2.1 [7.3] & -1.5 [1.7] & -4.8 [-2.6]
 \\ \hline\hline
 $\mu_f$, NLO ($\overline{\rm CS}$)  &            -26.8[-67.6]  & -15.8 [-30.1]  & -2.8 [0.4]  & -0.5 [2.2] & -0.3 [1.4] \\
 \hline\hline
 $\mu_f$, NLO ($\overline{\rm MS}$) &            -32.0[-48.8]  & -19.1 [-40.5]  & -3.6 [0.2]  & -0.6 [2.6] & 0.0 [2.0]  \\
 \hline\hline
  $\mu_r$, NLO ($\overline{\rm CS}$) &            9.3 [46.8] & 7.8 [26.8] & 6.5 [11.1] & 4.6 [7.0] & 2.4 [3.8]  \\
  \hline\hline
   $\mu_r$, NLO  ($\overline{\rm MS}$) &             13.8 [76.5] & 12.1 [42.3] & 10.9 [17.7] & 8.7 [12.2] & 6.1 [8.0]   \\
 \hline
\end{tabular}
\caption{
\label{Tab-ScaSetS}
Relative changes (\ref{Def-VarScaDep}) in percent of the
singlet CFFs ${^{\rm S}\! {\cal H}}$ within the separate variation of
the factorization and renormalization scale from  ${\cal
Q}^2/2 \cdots 2{\cal Q}^2$. Here we used the soft [hard]
conformal GPD moments (\ref{AnsSinSof}) [(\ref{AnsSinHar})] and set ${\cal Q}^2 =
4\,\GeV^2$  and $\Delta^2 = -0.25\, \GeV^2$.}
\end{center}
\end{table}
Table \ref{Tab-ScaSetS} lists the changes of the CFF that come
from the variation of the factorization and renormalization
scales, see Eq.\ (\ref{Def-VarScaDep}). The first row demonstrates
that the factorization scale variation in LO  is correlated with
the evolution, compare with dotted lines in Fig.\
(\ref{FigNLO-EvoS}). In the fixed target region the evolution and
the associated variation is weak. However, approaching the small
$\xi$ region, its strength is growing, in particular for the
`hard' gluon ansatz. One would normally expect that the scale variation
is getting smaller at NLO. But this is not the case for the small
$\xi$ region, rather the sign is reversed and its magnitude
increases. This behavior completely differs from the one in the
fixed target kinematics and it tells us that the factorization
logs in the small $\xi$ region are enhanced. Hence, already from
these NLO findings one might wonder whether a perturbative
treatment of the evolution in the usual manner is justified. The
change of the renormalization scale yields variations of the few
to ten percent in the `soft' gluon scenario, and, with decreasing
$\xi$, to much larger ones in the `hard' gluon scenario. These numbers
reflect simply the size of the NLO corrections for the former
and latter scenario, respectively, and do not necessarily indicate that
the perturbative expansion of the Wilson coefficients is
ill-defined. Whether this is the case or not can be only  clarified by
a NNLO evaluation.

\subsection{Radiative corrections beyond NLO}
\label{SubSecRadCorNNLO}

\begin{figure}[t]
\begin{center}
\mbox{
\begin{picture}(600,200)(0,0)
\put(-5,0){\insertfig{17.5}{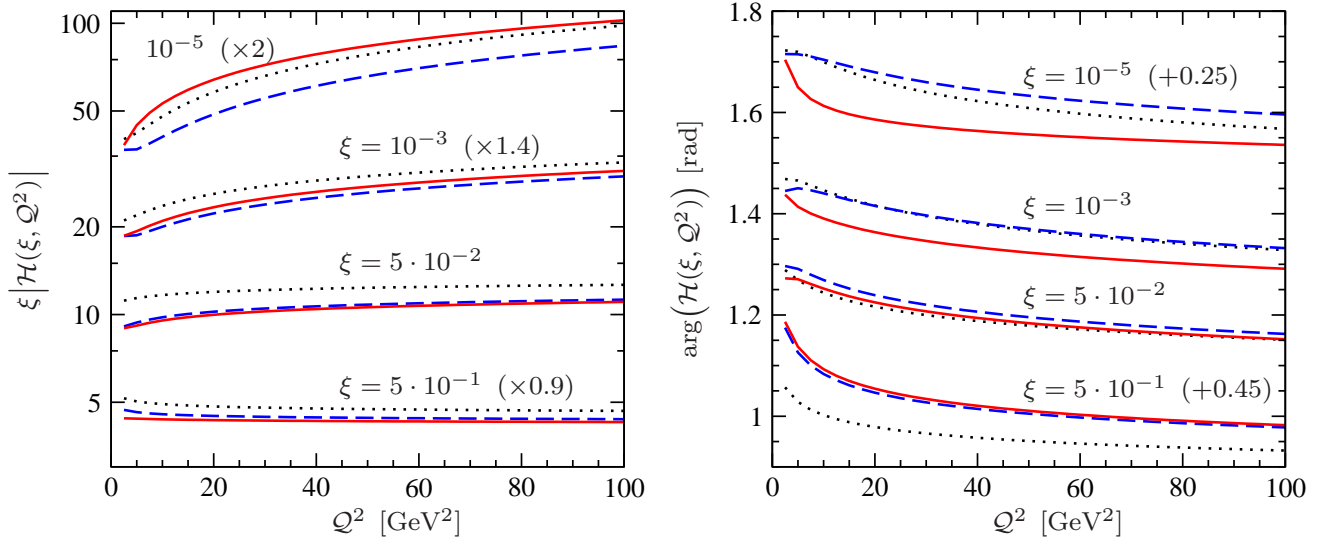}}
\end{picture}
}
\end{center}
\caption{\label{FigComRadCor} The modulus (left) and the phase
(right) of the rescaled singlet CFF $\xi\, {^S\! {\cal
H}}(\xi,{\cal Q}^2)$ versus ${\cal Q}^2$. Here the `soft' gluon
ansatz is used within $\Delta^2=-0.25\, \GeV^2$ to LO (dotted),
NLO (dash-dotted), and NNLO (solid).}
\end{figure}

To investigate the radiative corrections in NNLO, we use the
$\overline{\rm CS}$ scheme within the Wilson coefficients
(\ref{eq:copeCexp2})-(\ref{eq:copeCexp2c}) and
(\ref{Res-WilCoe-Exp-CS-SI})--(\ref{Res-WilCoe-CS-NNLO}), as well
as the evolution operators (\ref{Def-EvoOpeNS}) and
(\ref{Exp--EvoOpe}).  We stress again that the perturbative
expansion is consistently done as a power series in
$\alpha_s/2\pi$ to the order ${\rm N}^2{\rm LO}$, where the
running of the coupling is described to the same order. We again
calculate CFF $\cal H$, where the model ansatz for conformal GPD
moments, scale setting, and normalization of the running coupling
are spelled out above. We remind that the mixing term, appearing
at three-loop level in the anomalous dimensions is unknown.
Fortunately, we found that at NLO the mixing term in the
$\overline{\rm MS}$ scheme is small (tiny) for flavor
(non-)singlet CFFs in the fixed target kinematics. Therefore, we
expect that it is justified to neglect a NNLO mixing term for
these quantities in the $\overline{\rm CS}$ scheme. Unfortunately,
this is not true for the singlet part at smaller values of
$\xi$, where we observed at NLO about $30\%$ effect at $\xi=10^{-5}$
and ${\cal Q}^2 = 100\, \GeV^2$. Roughly speaking, we would presume that
the mixing at NNLO is given by the contribution of the diagonal
NLO anomalous dimensions times $(-\beta_0) \alpha_s/2\pi \sim
0.4$, which is additionally suppressed by the initial condition.
All together, we expect that for small $\xi$ the neglected mixing contributes to the net result
at the $10\%$ level.

In Fig.\ \ref{FigComRadCor} we visualize the general features of
the radiative corrections for fixed target and collider
experiments up to NNLO in the parity even sector. Thereby, we
employ the flavor singlet CFF ${^{\rm S}\!{\cal H}}$ within the
`soft' gluon ansatz (\ref{AnsSinSof})  at $\Delta^2=-0.25\, \GeV^2$. As pointed out
in the preceding section, in this ansatz both the modulus, scaled
with $\xi$, (left panel) and the phase (right panel) are mildly
affected by NLO corrections (dashed). As it can be seen, the NNLO
corrections (solid) are insignificant for both of them at the
input scale ${\cal Q}^2 = 2.5\, \GeV^2$ over the whole $\xi$
region. As long as we stay away from the very small $\xi$ region,
the perturbative prediction for the evolution is stable, starting at NLO, too.
But approaching the small $\xi$ region, NNLO corrections
are growing in size, which already shows up in a splitting
of the $\xi=10^{-3}$ NNLO and NLO trajectories, smaller for the modulus and
larger one for the phase. The $\xi=10^{-5}$ NNLO trajectory,
compared to the NLO one, is affected by rather large corrections,
which are of the same size as the NLO ones, but with a
competing overall sign. With increasing ${\cal Q}^2$ the NNLO
trajectory approaches the LO one. Such an ill behavior reflects
the competition of the Bjorken limit, i.e., ${\cal Q}^2\to
\infty$, and the high energy limit, i.e.,   $1/\xi \to \infty$.
Indeed, the expansion parameter is rather $\ln(1/\xi)
\alpha_s({\cal Q})/2\pi$. Since the slopes of the NLO and NNLO
trajectories are getting closer at large values of ${\cal Q}^2$,
one can easily imagine that using a larger input scale would
alleviate this problem. Indeed, it entirely originates from the
resummed poles in the anomalous dimensions at $j=0$ and so it is universal
(process independent) and the same one that appears for parton
densities. Below we will come back to this issue.
The changes of the phases with respect to ${\cal Q}^2$ is for any
given $\xi$ smaller than $0.03 \pi$ rad and according to Eq.\
(\ref{Con-Pha-Mod}) they are bound by the ratio of moduli. In the
small $\xi$ region  the phase approaches the value $\pi/2$ and so
the CFF is dominated by the imaginary part. This value is driven
by both the pole, which is in our ansatz (\ref{Ans-ConMomS}) in
the vicinity of $j=0$, and the essential singularity of the
evolution operator at $j=0$, resulting from poles in the anomalous
dimensions (\ref{Def-LO-AnoDim-GQ-V}) and
(\ref{Def-LO-AnoDim-GG-V}).

\begin{figure}[t]
\begin{center}
\mbox{
\begin{picture}(450,240)(0,0)
\put(-10,0){\insertfig{17.5}{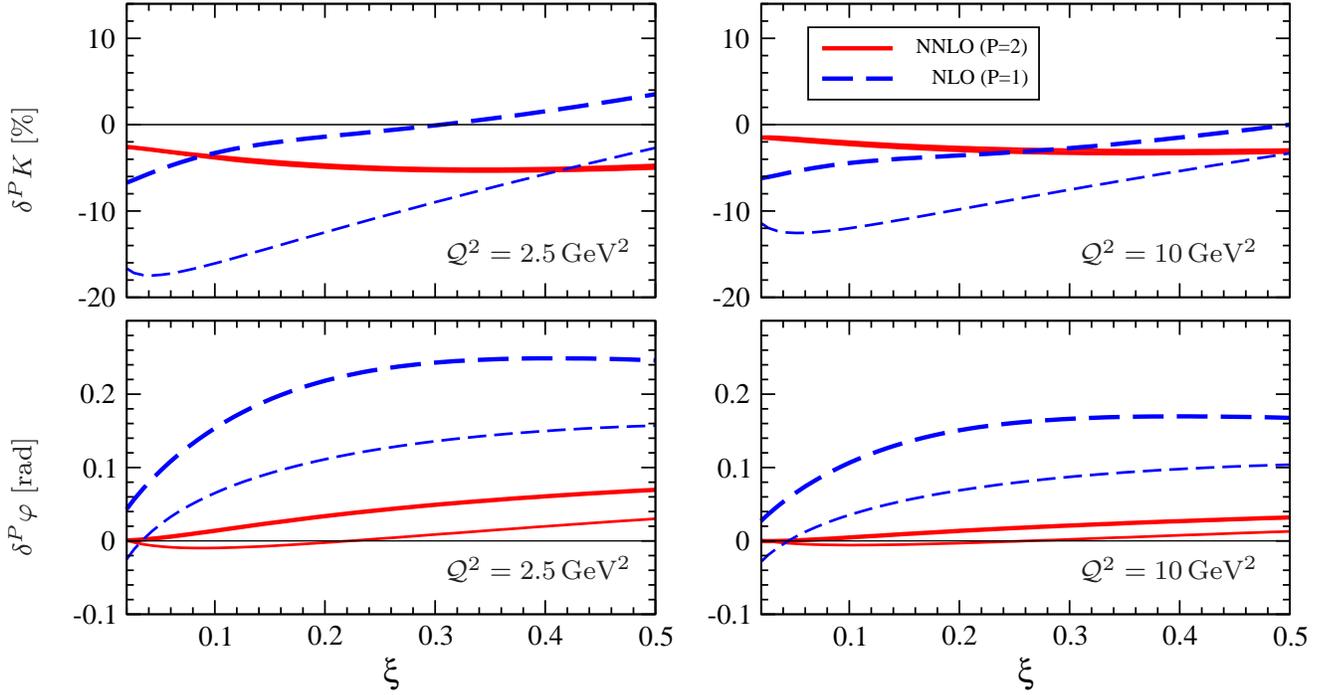}}
\end{picture}
}
\end{center}
\caption{ \label{FigNNLO-WilNS} The relative NLO (dashed)
and NNLO (solid) radiative corrections to the flavor nonsinglet
CFF ${^{\rm NS}{\cal H}}$ in the $\overline{\rm CS}$ scheme at the
input scale ${\cal Q}^2_0= 2.5\, \GeV^2$ (left) and ${\cal
Q}^2=10\, \GeV^2$ (right) and $\Delta^2=-0.25$. The moduli
(\ref{Def-Del})  and phase  differences (\ref{Def-PhaDif}) are
shown in the upper and lower panels, respectively. The results
without and with sea quarks are shown as thin and thick lines,
respectively, which are indistinguishable for NNLO moduli.}
\end{figure}
In Fig.\ \ref{FigNNLO-WilNS} we present a more detailed view on
the radiative corrections in the flavor nonsinglet sector for
the moduli (up) and the phase differences (down). The left
panels show the radiative corrections at the input scale ${\cal
Q}_0^2 =2.5\, \GeV^2$, while in the right panels we evolve the input
to the scale ${\cal Q}^2 =10\, \GeV^2$. This gives a measure of
the radiative corrections arising from the evolution operator.
Let us first discuss those for the moduli that arise at the
input scale, i.e., the left upper panel. The NLO corrections for
both a valence-like ansatz (thin) and one with a sea quark admixture
(thick) have for the relevant fixed target kinematics a variance
of about $20\%$. In NNLO this is reduced to the $5\%$ level. More
precisely, the radiative corrections reduce the CFF both without and
with sea quark ansatz by about $2-5 \%$. If we
evolve them to $10\, \GeV^2$ then the NNLO radiative corrections
further reduce by $1-2 \% $, while at NLO the variance of them is of
about $13\%$, cf.\ right upper panel. The radiative corrections to
the phases are already dictated by those to the moduli. Already to
NLO, they are smaller than $0.05 \pi$ [$0.08 \pi$] rad for the
ansatz without [with] sea quarks and shrink further with
evolution. They become tiny at NNLO, see lower panels.
Obviously, there is an improvement in the perturbative expansion,
which is roughly of the same order we estimated from the variation
of the scales, see Tab.\ \ref{Tab-ScaSetNS}. However, a closer
look to the separate contributions, arising from different color
factors, shows that there is a cancellation between the $C_F^2$
and the $\beta_0$ proportional terms \cite{Mue05a}. The latter is
negative and about two times larger than the former, positive one,
Hence both of them partly compensate each
other. Without this delicate cancellation the NNLO corrections
would be almost on the $10\%$ level.  Let us mention that the scale
dependencies are now almost $\xi$ independent and are of the order
of $-1.5\%$ and $3\%$ for the factorization and renormalization
scale, respectively.

\begin{figure}[t]
\begin{center}
\mbox{
\begin{picture}(600,270)(0,0)
\put(-10,0){\insertfig{17.5}{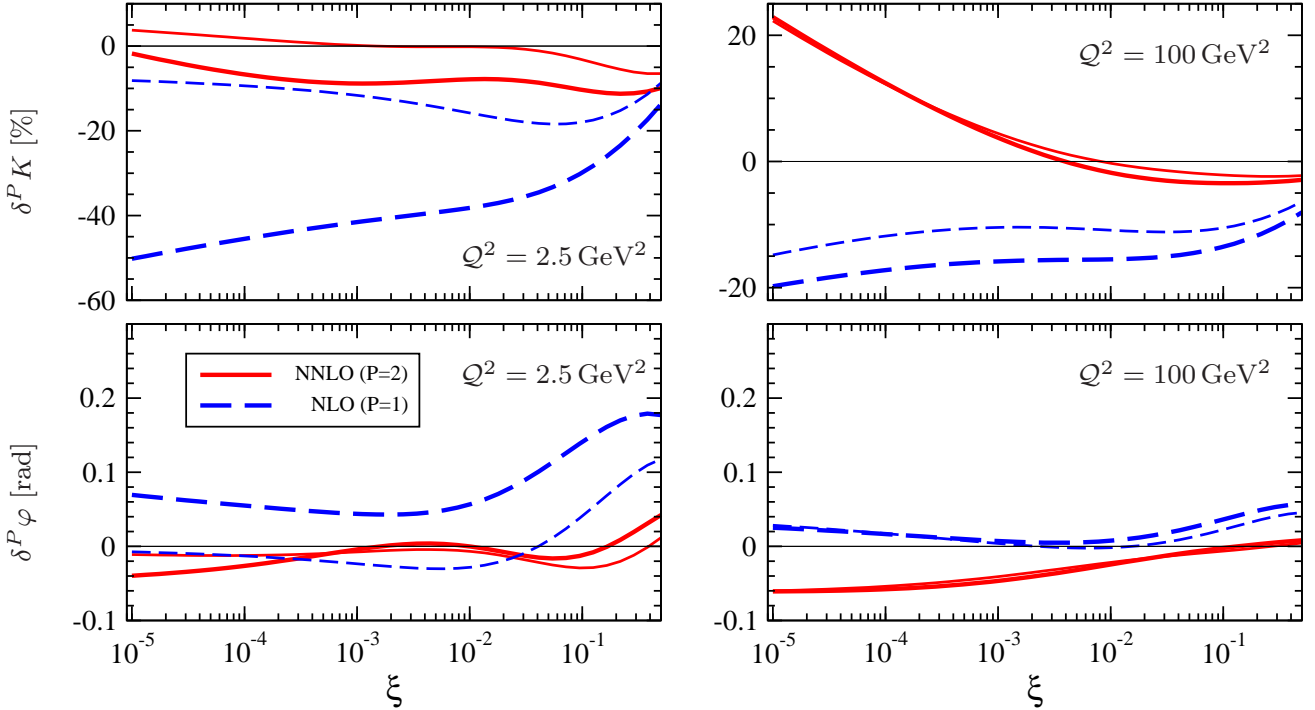}}
\end{picture}
}
\end{center}
         \caption{ \label{FigNNLO-Wil}
The relative NLO (dashed)
 and NNLO (solid) radiative corrections in the
         $\overline{\rm CS}$ scheme are plotted versus $\xi$ for the
         modulus (up) and phase (down) of singlet ${^{\rm S}{\cal H}}$ for
         $\Delta^2=-0.25$ at the input scale ${\cal Q}^2_0= 2.5\,
         \GeV^2$ (left) and ${\cal Q}^2=100\, \GeV^2$ (right).
         The results for the ``soft'' and ``hard'' gluon ansatz
         are shown as thin and thick lines, respectively.}
\end{figure}

In Fig.\ \ref{FigNNLO-Wil} we display the relative radiative
corrections for the flavor singlet CFF in a manner analogous
to Fig.\ \ref{FigNNLO-WilNS}. However, we evolve the quantities in
questions to a scale of $100\, \GeV^2$, shown in the right panels.
{F}rom the left panel, where corrections are given at the input scale, we
certainly realize that the large negative NLO corrections (thick
dashed) to the modulus in the `hard' gluon scenario are shrunk to
less than $10\%$ (thick solid), in particular in the small $\xi$
region. For the `soft' gluon case the NNLO corrections (thin
solid) are reduced to $\pm 5\%$. However, for $\xi\sim 0.5$, the
corrections are reduced only unessentially and are still around $5\%$ and $10\%$ at
NNLO level. The phase differences
(lower panel) are becoming tiny at NNLO. If evolution is now switched on,
our findings drastically change. For $\xi \gtrsim 5\cdot 10^{-2}$
they are stabilized for the moduli on the level of about $3\%$ at
${\cal Q}^2 = 100\ \GeV^2$. However, they start to grow with
decreasing $\xi$ and reach at $\xi\approx 10^{-5}$ the $20\%$
level. It is remarkable that the relative sign of the NLO and NNLO
corrections change and that they are becoming independent of the
input. This behavior is also reflected in the phase differences,
which decrease to $-0.02 \pi$ rad. As already explained above,
this breakdown of perturbation theory stems only from the anomalous
dimensions and is thus universal, i.e., process independent.

\begin{table}
\begin{center}
\begin{tabular}{|c|c|c|r|r|r|r|r|r|}
  \hline
  order/ $\xi$ &                      $10^{-5}$  &  $10^{-4}$ & $10^{-3}$ & $10^{-2}$ & $10^{-1}$ & 0.25 & 0.5   \\ \hline\hline
 NLO  &                                    2.4  [24.9] & 2.8 [21.0]& 3.5 [18.1]& 5.0 [15.8]& 5.8 [10.9]& 4.4 [7.1]& 2.4 [3.9]      \\
 \hline\hline
 NNLO  &                                 -1.6 [3.4]& -0.6 [5.6]&  0.3 [6.5]& 0.6 [5.7]& 2.2 [6.7]& 3.5 [6.9]& 3.7 [5.9]        \\
  \hline
\end{tabular}
\caption{
\label{Tab-ScaSetS1}
Variation (\ref{Def-VarScaDep}) of ${^{\rm S}\! {\cal H}}$ in percent  within the change of
renormalization scale from  ${\cal
Q}^2/2$ to  2${\cal Q}^2$. Here we used the soft [hard]
conformal moments (\ref{AnsSinSof}) [(\ref{AnsSinHar})] and set ${\cal Q}_0^2 ={\cal Q}^2 =
4\,\GeV^2$  and $\Delta^2 = -0.25\, \GeV^2$.}
\end{center}
\end{table}
Finally, we comment on the scale dependencies. As it has been
already seen in Tab.\ \ref{Tab-ScaSetS}, the variation within the
factorization scale increases in the small $\xi$ region with the
perturbative order. To NNLO, we find for instance at $\xi=10^{-5}$
a variation of $44\%$ $[105\%]$ for the `soft' [`hard'] gluon
ansatz. This is about two $[1.5]$ times larger than that observed
at NLO, where the sign is alternating. This simply reflects the
breakdown of perturbative expansion of the evolution operator, as
we have already seen. The NLO estimates of the higher order
corrections, obtained by the variation of the renormalization
scale, were for the `soft' gluon scenario comparable to the actual
NNLO result at the input scale. However, the corresponding
estimates for the `hard' gluon ansatz in the small $\xi$ region
were too pessimistic, substantially overestimating the calculated
NNLO corrections. Since  our input scale ${\cal Q}_0^2 = 2.5\,
\GeV^2$ in Tab.\ \ref{Tab-ScaSetS} was lower than the average
scale ${\cal Q}^2 = 4\, \GeV^2$ it might be that these large
estimates are partly contaminated by the factorization logs. In
Tab.\ \ref{Tab-ScaSetS1} we show the renormalization scale
dependence, but now for the input scale ${\cal Q}_0^2 = 4\,
\GeV^2$. Compared to Tab.\ \ref{Tab-ScaSetS} the modifications in
NLO are not large. We also realize that the renormalization scale
independency is improved at NNLO level.

To summarize our findings, we saw that NLO radiative corrections
are moderate in the nonsinglet case but can be rather large in
the singlet sector for a `hard' gluon ansatz. The factorization
scale dependence, substantial at LO, becomes for kinematics of
fixed target experiments small already at NLO. However, it is
getting worse in the small $\xi$ region.  Interestingly, we also
observe a scheme dependence at NLO of the order of $10\%$ to
$20\%$, which entirely arises due to skewness effects, where the
radiative corrections are more pronounced in the $\overline{\rm
MS}$ scheme than in the  $\overline{\rm CS}$ one. At NNLO we have
found that the perturbative corrections are getting reasonably
small at the input scale and for the evolution in the fixed target
kinematics. Both the factorization and renormalization scale
dependencies are reduced to the level of a few percent. So far
these findings suggest that the perturbative expansion is a
reliable tool. However, we saw that the perturbative expansion of
the evolution breaks down in the small $\xi$ region. Fortunately,
this breakdown is universal, and as long as one precisely defines
the scheme and the approximation, perturbation theory can be used
as a tool to relate different processes. We will demonstrate this
in the next section.

\section{Fitting procedure of experimental data}
\label{Sec-FitPro}

The DVCS data measured in fixed target \cite{Airetal01,Steetal01}
and collider \cite{Adletal01,Chekanov:2003ya,Aktas:2005ty}
experiments have been confronted in the literature with theoretical predictions,
e.g., color dipole model
\cite{DonDos00}, collinear factorization approach within an
aligned jet model inspired GPD ansatz \cite{FreMcDStr02} and the
minimal dual GPD parameterization \cite{GuzTec06}, see also Ref.\
\cite{BelMueKir01} for a first analysis within the double
distribution ansatz. Certainly, confronting model ans\"atze with
DVCS data is a too rigid approach and should be considered only as a
first step towards extracting GPD parameters within a given ansatz. A
fitting procedure for DVCS data in the small $\xi$ and large
${\cal Q}^2$  region has been proposed within the double log
approximation of the deeply virtual Compton scattering amplitude
to LO accuracy \cite{Mue06}. There a pomeron inspired GPD model was
employed\footnote{A Regge pole model for the virtual Compton
scattering amplitude was also used for a fitting procedure in Ref.
\cite{CapFazFioJenPac06}.}, where the singlet quarks were
dynamically generated. Thus, the number of fitting parameters
could be reduced to three, namely, normalization,
slope-parameter, and input scale.

We will now demonstrate that the Mellin--Barnes representation of
the CFFs is appropriate for a more general GPD fitting procedure.
Technically, we use the standard fitting routine MINUIT \cite{MINUIT}. This
routine calls a FORTRAN code that evaluates the CFFs from
conformal GPD moments, depending on a few fitting parameters.

\subsection{Setting the scene}

The DVCS amplitude interferes with the Bethe--Heitler
bremsstrahlung one and so we have a rich selection of observables,
mainly in the interference term. The decomposition in terms of CFFs is generally
challenging. We will deal here with the
easiest case, namely, the fitting of small $x_{\rm Bj} \cong 2
\xi$ data for the DVCS cross section. In this kinematics the
interference term, integrated over the azimuthal angle, can be
neglected and the DVCS cross section can be extracted from the
photon leptoproduction one by subtracting the Bethe--Heitler
bremsstrahlung cross-section. In the DVCS cross section we can safely neglect terms
that are kinematically suppressed by $\xi^2$:
\begin{eqnarray}
\label{Def-CroSec}
\frac{d\sigma}{d\Delta^2}(W,\Delta^2,{\cal Q}^2) \approx
\frac{4   \pi \alpha^2 }{{\cal Q}^4} \frac{W^2 \xi^2}{W^2+{\cal Q}^2}
\left[\left| {\cal H} \right|^2  - \frac{\Delta^2}{4 M^2_{\rm p}}
\left| {\cal E} \right|^2 +\left|\widetilde {\cal H} \right|^2\right]
\left(\xi,\Delta^2,{\cal Q}^2\right)\Big|_{\xi=\frac{{\cal Q}^2}{2 W^2+{\cal Q}^2}}\,.
\end{eqnarray}
Here we have expressed the scaling variable $\xi$ in terms of the photon
virtuality ${\cal Q}^2$ and the photon--proton center--of--mass energy $W$,
defined by $W^2 = \left(P_1 + q_1\right)^2$.
The leading Regge trajectory, appearing in $\widetilde {\cal H}$, arises from mesons with
generic intercept $\alpha(0)\approx 1/2$, which is less than the
intercept $\alpha_{\mathbb{P}}(0)\approx 1$ of the pomeron
dominated CFFs $\cal H$ and $\cal E$. Thus, the squared CFF
$\left|\widetilde {\cal H}\right|^2$ can be neglected, too, since
it is approximately suppressed by one power of $\xi$. More care
has to be taken about the remaining combination of $\cal H$ and $\cal
E$ CFFs. Taking the mean value of $\ll\Delta^2\gg = -0.17\,
\GeV^2$, which has been measured by the H1 collaboration
\cite{Aktas:2005ty} for $|\Delta^2| < 1\,\GeV^2$, we find that the
helicity flip contribution is in the $\Delta^2$ integrated cross
section kinematically suppressed as
\begin{eqnarray}
-\frac{\ll\Delta^2\gg}{4 M^2_{\rm p}}\sim 5\cdot 10^{-2}\,.
\end{eqnarray}
Hence, in this kinematical region, it might be justified to neglect the
squared CFF $|{\cal E}|^2$. But in the differential cross section
at larger values of $-\Delta^2$, $|{\cal E}|^2$ might contribute
to some larger extent. Since it is not possible to separate by the
present data set the $\cal H$ and $\cal E$ contributions, we
simplify our analysis by neglecting the latter one. All together,
the DVCS cross section reduces in the small $\xi$-region to:
\begin{eqnarray}
\label{Def-CroSec1}
\frac{d\sigma}{d\Delta^2}(W,\Delta^2,{\cal Q}^2) \approx
\frac{4   \pi \alpha^2 }{(2 W^2+{\cal Q}^2)^2}\frac{W^2}{W^2+{\cal Q}^2}
\left| {\cal H} \right|^2
\left(\xi=\frac{{\cal Q}^2}{2 W^2+{\cal Q}^2},\Delta^2,{\cal Q}^2\right)\,.
\end{eqnarray}

As we have clearly pointed out in the preceding section, the
perturbative expansion of the evolution operator is ill-defined
in the small $x_{\rm Bj} \cong 2 \xi$ region  for the (singlet) parity even
sector. Nevertheless, we have argued that this should not affect the
task of relating different processes within perturbative QCD.
One can ask the question: How to resum this alternating
series of $\ln(1/\xi) \alpha_s/2\pi$ terms? An answer is certainly
needed for the analysis of a large amount of high precision data,
as it is the case in DIS. Here, indeed, some progress has been
recently reported, see, e.g., Ref.\ \cite{For05}. Concerning the situation in DVCS, the
solution of the problem would certainly improve our partonic
interpretation of the nucleon content, however, is rather
irrelevant for the analysis of present experimental measurements,
as we will see. This problem will also affect the forward
limit of conformal GPD moments, which provide the Mellin moments
of parton densities.
Since for fitting parton densities to experimental data
it is also necessary to precisely
specify the procedure, e.g., definition of perturbative expansion
of evolution operator, flavor scheme, and running of the coupling,
we do not rely
on any of the standard parameterizations. Rather, we fit the data
ourselves, within our specifications, and then compare results with a specific
choice of parton density parameterization from literature. The perturbative expansion
of the DIS structure function $F_2$ reads:
\begin{eqnarray}
F_2(x_{\rm Bj},Q^2) = \frac{1}{2 i \pi}\int_{c- i\infty}^{c+ i\infty}\! dj\;
x_{\rm Bj}^{-j}
\left[Q^2_{\rm S}\, \mbox{\boldmath $c$}_j(\alpha_s(Q^2))\,  \mbox{\boldmath $q$}_j(Q^2) +
Q^2_{\rm NS}\, {^{\rm NS}\!c}_j(\alpha_s(Q^2))\, {^{\rm NS}\!q}_j(Q^2)\right],
\end{eqnarray}
where $\mbox{\boldmath $c$}_j = ({^{\Sigma}\!c}_j,{^{\rm
G}\!c}_j)$ and ${^{\rm NS}\!c}_j$ are the DIS
Wilson coefficients corresponding to the structure function $F_2$. They can be
found in Ref.\ \cite{ZijNee92}, where we set the spin label
$n=j+1$. Again, we equate the factorization and renormalization
scales with the photon virtuality $Q^2$ and consistently combine the
perturbative expansion of the Wilson coefficients with the one
of the evolution operator. The evolution of the parton density
moments is governed by the evolution equations
(\ref{eq:evNS}) and (\ref{eq:evS}).
These moments are
related to the conformal GPD ones in the forward kinematics
(\ref{ForKin}), e.g., cf. (\ref{Def-ForMatEle}),
\begin{eqnarray}
\mbox{\boldmath $q$}_j(\mu_0^2)=
\left( { {^{\Sigma}\!H_j} \atop {^{\rm G}\!H_j} }\right)(\eta=0,\Delta^2=0,\mu_0^2)\,.
\end{eqnarray}

We will employ in this kinematics our ansatz (\ref{Ans-Momsea})
and (\ref{Ans-MomG}), see also Ref.\ \cite{ShuBieMarRys99}, where
we neglect for simplicity the flavor nonsinglet contribution.
Moreover, instead of decomposing the singlet quark contributions in valence
and sea quarks, we use the effective parameterization
\begin{eqnarray}
\label{Ans-MomSigm}
{\!H}^{\Sigma}_j(\eta,\Delta^2,\mu_0^2) =
N_{\Sigma}\frac{B(1-\alpha_{\Sigma}(0)+j,8)}{B(2-\alpha_{\Sigma}(0),8)}
\frac{1}{1-\frac{\Delta^2}{(m^{\Sigma}_j)^2}}
\frac{1}{\left(1-\frac{\Delta^2}{(M_j^{\Sigma})^2}\right)^3} +{\cal O}(\eta^2)\,.
\end{eqnarray}
{F}rom the inspection of the standard parameterizations of parton
densities, we found that the contributions from the flavor
nonsinglet sector generally don't exceed the  10\% level. Here
it is effectively included in the parameterization of the singlet
quark contribution. Pre-fitting the data we found that the fits
are almost insensitive to the parameters $\alpha^\prime_{\Sigma}$,
$\alpha^\prime_{\rm G}$ and that $\Delta M_\Sigma = \Delta M_{\rm
G} \approx 0$ is the preferred value. According to the common fits
to vector-meson electroproduction data in the small $\xi$ region,
we set $\alpha^\prime_{\Sigma}=\alpha^\prime_{\rm
G}=0.15\,\GeV^{-2}$. Moreover, we neglect the $j$-dependence in
the cut-off masses $M_j^{\Sigma}$ and $M_j^{\rm G}$, i.e., put
$\Delta M_{\Sigma}=\Delta M_{\rm G}=0$. As relevant fitting
parameters we thus choose
\begin{eqnarray}
N_{\Sigma},\; \alpha_{\Sigma}(0)\,,\; M_0^{\Sigma}\,,\qquad
N_{\rm G},\; \alpha_{\rm G}(0)\,,\;  M_0^{\rm G}\,.
\end{eqnarray}
Obviously, in fits of the DIS structure function $F_2$ we have
only four parameters, where  $\alpha_{\Sigma}(0)$ and $\alpha_{\rm
G}(0)$ actually determine the powers of $x_{\rm Bj}$ and
$N_\Sigma$ and $N_{\rm G}$ the normalization. Note that although
in our ans\"atze the conformal GPD moments are normalized to
${\!H}^{\Sigma}_{j=1}(\eta,\Delta^2=0,\mu_0^2)= N_{\Sigma}$ and
${\!H}^{\rm G}_{j=1}(\eta,\Delta^2=0,\mu_0^2)= N_{\rm G}$, we do
not impose the momentum sum rule $N_{\Sigma} +  N_{\rm G}=1$ here%
\footnote{ Note that the whole momentum fraction region, i.e., $0
\le x \le 1$, contributes to this rule, however, our fits
constrain only the small $x_{\rm Bj}$ region, where $x_{\rm Bj}
\le x \le 1$. For technical reasons, however, we assume a certain
large $x$ or $j$ behavior. A modification of this behavior, which
is only weakly constrained in our fits, might be used to restore
the momentum sum rule. An improved treatment will be given
somewhere else.}. This minimal parameterization is not so well
suited to provide a high quality fit to the DIS data. This is not
our goal here; rather we would like to relate the DIS data with
the DVCS data, which have much larger error bars. The $\Delta^2$
slope of the DVCS amplitude is controlled by the cut-off masses
$M_0^{\Sigma}$ and $M_0^{\rm G}$. The normalization and the
$\xi$-dependence of this amplitude at $\Delta^2=0$ is as in DIS
fixed by the remaining parameters $N_{\Sigma}, \alpha_{\Sigma}(0),
N_{\rm G}, \alpha_{\rm G}(0)$. Within the assumption that the
skewness parameter is negligible in the conformal GPD moments, we
also loose the possibility to control the normalization of the
DVCS amplitude by the skewness dependence. As demonstrated above
by inspection of radiative corrections in different schemes,
compare dashed and dash-dotted lines in Fig. \ref{FigNLO-Wil},
the differences in normalization, caused by the $\xi$-dependence,
is almost skewness independent. Within our  ansatz we found a
$10\%-20\%$ effect for the modulus of the amplitude, which means
that the cross section would differ of about $20\%-40\%$. The
inclusion of skewness dependence, controlled by a corresponding
parameter, will be considered somewhere else. In our simplified
model ansatz we do not explicitly adjust the normalization of the
DVCS amplitude relative to DIS one. However, for given mean value
$\ll\Delta^2\gg$ the normalization of the DVCS amplitude,
integrated over $\Delta^2$, is controlled also by the parameters
$M_0^{\Sigma}$ and $M_0^{\rm G}$, which determine the
$\Delta^2$-slope.

\subsection{Lessons from fits}

\begin{figure}[]
\centerline{\includegraphics[scale=1.25]{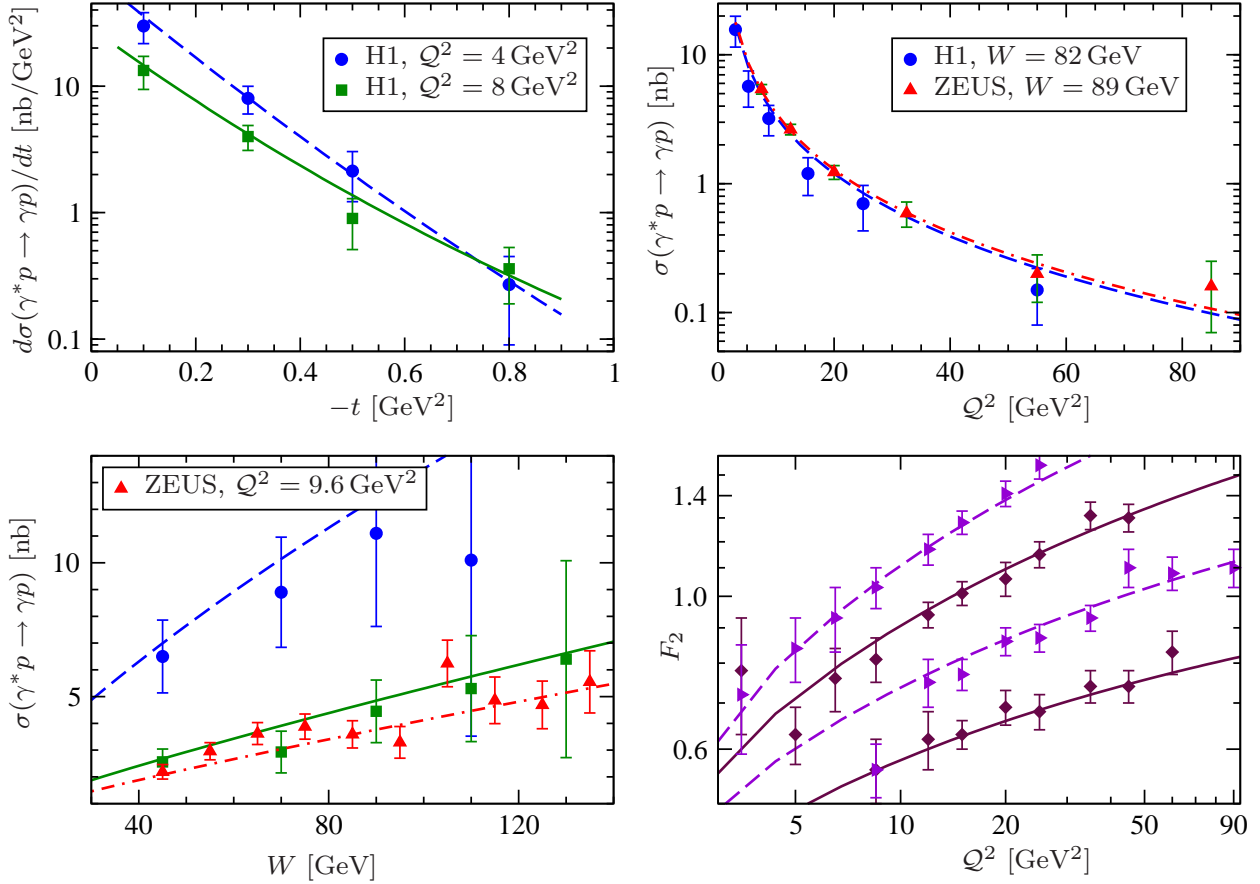}}
\caption{ \label{FigFits} Simultaneous fit to the DVCS and DIS
data in the $\overline{\rm CS}$  scheme to NNLO.
Upper left panel  DVCS cross
section  for ${\cal Q}^2 =4\,\GeV^2$ and $W=71\, \GeV$ (circles,
dashed) as well as ${\cal Q}^2 =8\,\GeV^2$ and $W=82\, \GeV$
(squares, solid) \cite{Aktas:2005ty}. Upper right panel  DVCS
cross section ($|\Delta^2| < 1\,\GeV^2$) versus ${\cal Q}^2$ for
$W=82\, \GeV$ (H1, circles, dashed) and $W=89\, \GeV$ (ZEUS,
triangles, dash-dotted) \cite{Chekanov:2003ya}. Lower left panel DVCS cross
section versus $W$ ${\cal Q}^2 = 4\,\GeV^2$ (H1, circles, dashed),
${\cal Q}^2 = 8\,\GeV^2$ (H1, squares, solid), and ${\cal Q}^2 =
9.6\,\GeV^2$ (ZEUS, triangles, dash-dotted). Lower right panel
shows $F_2(x_{\rm Bj},{\cal Q}^2)$ versus ${\cal
Q}^2$ for $x_{\rm Bj}=\{8\cdot 10^{-3}, 3.2\cdot 10^{-3}, 1.3\cdot
10^{-3}, 5\cdot 10^{-4}\}$ \cite{Aidetal96}.}
\end{figure}

In  Fig.\ \ref{FigFits} we confront the outcome of a
simultaneous $\chi^2$-fit to the DVCS (39 data points) and DIS
(85 data points) data in the $\overline{\rm CS}$ scheme to NNLO
accuracy.  We equated the factorization and renormalization scale
with the photon virtuality, used the conformal GPD moment ans\"{a}tze at the
input scale $\mu_0^2=4\,\GeV^2$, fixed the number of flavors to $n_f=4$,
and used for $\alpha_s(2.5\, \GeV^2)/\pi =
0.0976$\footnote{Employing the standard procedure for the NNLO running
of coupling, we find that this value corresponds to $\alpha_s(M_{Z}^2)=0.114$,
at the standard reference scale $M_{Z}=91.18\, \GeV$.}.
In the upper left panel we display the fit to the H1
\cite{Aktas:2005ty} and ZEUS \cite{Chekanov:2003ya} DVCS data
versus $-\Delta^2$ for fixed ${\cal Q}^2$ and $W$, while the upper right
and lower left ones show the DVCS cross section, integrated over
$|\Delta^2| < 1\, \GeV^2 $ , versus ${\cal Q}^2$ and  $W$
dependence, where the remaining variable $W$ or ${\cal Q}^2$ is
fixed. The fit to the DIS H1 data \cite{Aidetal96} is plotted in
the lower right panel, where for clarity not all points are displayed\footnote{Not shown, but
used in fits, are H1 $F_2$ data for $x_{\rm Bj}=\{1.3\cdot 10^{-2},
5\cdot 10^{-3}, 2\cdot 10^{-3}, 8\cdot 10^{-4},
3.2\cdot 10^{-4}, 2\cdot 10^{-4}, 1.3\cdot 10^{-4}, 8\cdot 10^{-5}\}$.}.
As one realizes by eye inspection, the
normalization, scale- and $\Delta^2$-dependency are separately
well described.

{F}rom Table \ref{Tab-FitPar} one can see that
the quality of these simultaneous fits for
the 124 data points is satisfying. For instance,  for the NLO fit in
$\overline{\rm CS}$ scheme, where now $\alpha_s(2.5\, \GeV^2)/\pi =
0.1036$, we get $\chi^2= 95$, i.e.,
$\chi^2/{\rm d.o.f.} = 0.8$.
Taking into account the NNLO order corrections,
the quality of the fit slightly improves,
i.e.,  $\chi^2/{\rm d.o.f.} = 0.77$. As one can realize by
comparison of the last two rows in Tab.\ \ref{Tab-FitPar},  the
resulting parameters remain stable. The largest modification
appears in the cut-off masses, which reduce from NLO to NNLO by
about $13\%$. Thereby, contribution to $\chi^2$ coming from the
eight data points of the differential cross section
(upper left panel in Fig.\ \ref{FigFits}), denoted  $\chi_{\Delta^2}^2=2.2$
(last column in Tab.\ \ref{Tab-FitPar}),  indicates
an equally good fit to the $\Delta^2$-slope  in  NLO and NNLO.

\begin{table}
\begin{tabular}{|c||c||c|c|c|c|c|c||c|c|c|} \hline
  order (scheme) &
  $\alpha_s(M_Z)$ &
   $N_\Sigma$ & $\alpha_\Sigma(0)$  & $M^2_\Sigma$  & $N_{\rm G}$ & $\alpha_{\rm G}(0)$ & $M^2_{\rm G}$ &
   $\chi^2$ & $\chi^2/{\rm d.o.f.}$ &  $\chi^2_{\Delta^2}$\\
  \hline\hline
   \phantom{NN}LO \phantom{($\overline{\rm MS}$)}  &
   $0.130$ & $0.157$ & $1.17$  & $0.228$ & $0.527$  & $1.25$ & $0.263$ & $100$ & $0.85$ & $38.5$\\
   \hline
   \phantom{N}NLO ($\overline{\rm MS}$)  &
   $0.116$ & $0.172$  & $1.14$ & $1.93$  & $0.472$  & $1.08$ & $4.45$  & $109$ & $0.92$ & $4.2$ \\
   \hline
   \phantom{N}NLO ($\overline{\rm CS}$) &
    $0.116$ & $0.167$ & $1.14$  & $1.34$  & $0.535$  & $1.09$ & $1.59$ & $95$ & $0.80$  &  $2.2$\\
   \hline
   NNLO ($\overline{\rm CS}$)  &
   0.114 & $0.167$  & $1.14$ & $1.17$ & $0.571$  & $1.07$ & $1.39$ & $91$ & $0.77$ &  $2.2$ \\
   \hline
\end{tabular}
\caption{
\label{Tab-FitPar}
Parameters extracted from a simultaneous  fit to the DVCS cross section and DIS structure function $F_2$,
where $\alpha_s(M_Z)$, $\alpha_\Sigma^\prime=\alpha_{\rm G}^\prime=0.15\, \GeV^{-2}$, and
$\Delta M_\Sigma=\Delta M_{\rm G}=0$ are fixed and $\mu_0^2=4$ GeV$^2$.
}
\end{table}
To understand better the double role of the cut-off masses in the
fitting procedure, let us compare the LO parameters with the NLO
ones, where their changes are much more pronounced. Although in LO
the fit to all data yields seemingly reasonable $\chi^2/{\rm
d.o.f.} = 0.85$, the $\Delta^2$-slope (determined by the small
squared cut-off masses $M_\Sigma^2 \sim M_{\rm G}^2 \sim 0.25\,
\GeV^2$) turns out to be $\sim 30\, \GeV^{-2}$, which is
incompatible with  $\sim 6\, \GeV^{-2}$ indicated by the data
\cite{Aktas:2005ty}. Or, in other words, to leading order accuracy
$\chi^2_{\Delta^2}=38.5$ is unacceptably large. Therefore, we
conclude that the relative normalization between the DIS structure
function and the integrated DVCS cross section is correctly
reproduced in the fits by reducing the latter one by forcing the
steeper $\Delta^2$-dependence. This also implies that at
$\Delta^2_{\rm min}\approx 0$ the normalization of the
differential DVCS cross section and the structure function $F_2$
cannot be simultaneously described within our ansatz. A separate
fit to the DVCS data yields a slope that is getting compatible
with the measured one; however, the overall normalization is now
deteriorated and so the quality of the fit is worse, namely,
$\chi^2/{\rm d.o.f.} = 4.8$. Certainly, this is related to the
fact that we have no control over the skewness dependence of
the conformal GPD moments%
\footnote{ As a side effect, arising from this ill-defined
fitting task, we observed that the central value of the resulting
parameters, in particular of the cut-off masses, are getting
sensitive to the accuracy  of numerics and they can vary inside
the error bands. In NLO and NNLO fits, by increasing the numerical
accuracy, we observed  a variation of the central values only on
the per mil level.}. Therefore, non-trivial skewness dependence
should be introduced in such a way to make $|{\cal H}|$, and
consequently the normalization of the DVCS cross section, smaller.
We remind, however, that the inclusion of the skewness dependence
within the spectral representation of GPDs is usually done in such
a way that the skewness effect leads to an \emph{increase} of
$|{\cal H}|$.

Let us finally compare the NLO fits in the $\overline{\rm CS}$ and
$\overline{\rm MS}$ schemes. The fit in the latter scheme is
compared to the former one a bit off. We find $\chi^2/{\rm
d.o.f.} =0.92$, while in the $\overline{\rm CS}$ scheme
$\chi^2/{\rm d.o.f.} =0.8$.   Also here the largest changes appear
in the cut-off masses, in particular the gluon one. However, the
quality of the $\Delta^2$-fit to the differential cross section
is acceptable, namely, $\chi^2_{\Delta^2}=4.2$, compared to
$\chi^2_{\Delta^2}=2.2$ in the $\overline{\rm CS}$ scheme. As
discussed in Sect.\ \ref{SecCon2MSscheme} above, these schemes
differ only in the skewness dependence of the conformal GPD
moments. Instead of adjusting the normalization by changing the
skewness dependence, it is done within our ansatz by increasing
the slope. Hence, we must conclude that a more precise extraction
of GPD parameters requires the inclusion of the skewness effect
for $\Delta^2=0$.

\subsection{Comparison and partonic interpretation of the results}

\begin{figure}[]
\begin{center}
\mbox{
\begin{picture}(450,160)(0,0)
\put(-20,-20){\insertfig{17}{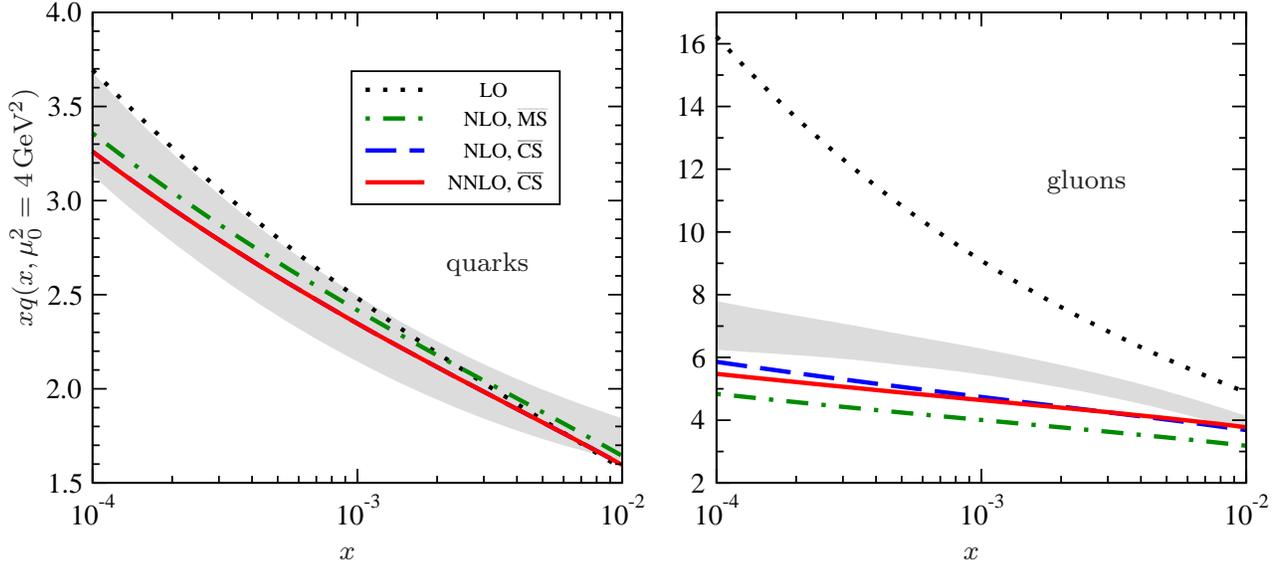}}
\end{picture}
}
\end{center}
\caption{ \label{FigFit-Out} The parton densities are shown for quarks (left)
 and gluons (right) at the  input scale $\mu_0^2 =
4\,\GeV^2$. The meaning of the lines are the same as before: LO
(dotted), NLO for $\overline{\rm MS}$ (dash-dotted) and
$\overline{\rm CS}$ (dashed), as well as NNLO (solid), the last
two being indistinguishable in the quark case. The bands show
Alekhin's NLO parameterization with errors \cite{Ale02}.}
\end{figure}

We would like now to confront our findings with the parton densities,
as obtained from global DIS fits.
In Fig.\ \ref{FigFit-Out}, we plot the  singlet quark and gluon
distributions
\begin{eqnarray}
x \left( { {^{\Sigma}\!q} \atop {^{\rm G}\!q} }\right)(x,\mu_0^2)  =
\frac{1}{2 i \pi}\int_{c- i\infty}^{c+ i\infty}\! dj\;
x^{-j} \,  \left( { {^{\Sigma}\!H_j} \atop {^{\rm G}\!H_j} }\right)(\eta=0,\Delta^2=0,\mu_0^2) ,
\end{eqnarray}
respectively, to LO (dotted), NLO for the $\overline{\rm MS}$ (dash-dotted)
and $\overline{\rm CS}$ (dashed) as well as to NNLO (solid) at the
input scale $\mu_0^2 = 4\, \GeV^2$. Note that the difference
between the two schemes arises purely due to the skewness
dependence, which affects only the DVCS fit. As argued above, we
do not expect the outcome of, e.g., our simultaneous
DVCS and DIS NLO fit,
to coincide with any of the standard parameterizations for parton
densities. Indeed, we explicitly saw this for the parton density
fits of the H1 and ZEUS collaborations. However, it turns out that
the flavor singlet quark parton density agrees very well with the
parameterization of Alekhin \cite{Ale02}, which is plotted with
error bands. Here the difference between the both schemes and NNLO
order corrections are rather small. As is well known, the gluon
distribution is much less constrained by a pure DIS fit. Here
our central values lie outside the error band of the Alekhin
parameterization. However, our error band of the NLO fit
in the $\overline{\rm CS}$ scheme would overlap with Alekhin's one.
Note that we have used the same settings for $\alpha_s$ as Alekhin, however, the
procedures for the fits is slightly different. As said above, we
have neglected the flavor nonsinglet contribution, took a fixed
$n_f=4$ scheme for both the evolution of the running coupling and
the conformal GPD moments, and assumed a generic $j$-dependence of the
conformal GPD moments at the input scale.  However, as we realize a posteriori,
these simplifications are justified within the error bands.

\begin{figure}[]
\begin{center}
\mbox{
\begin{picture}(450,180)(0,0)
\put(-20,-20){\insertfig{17}{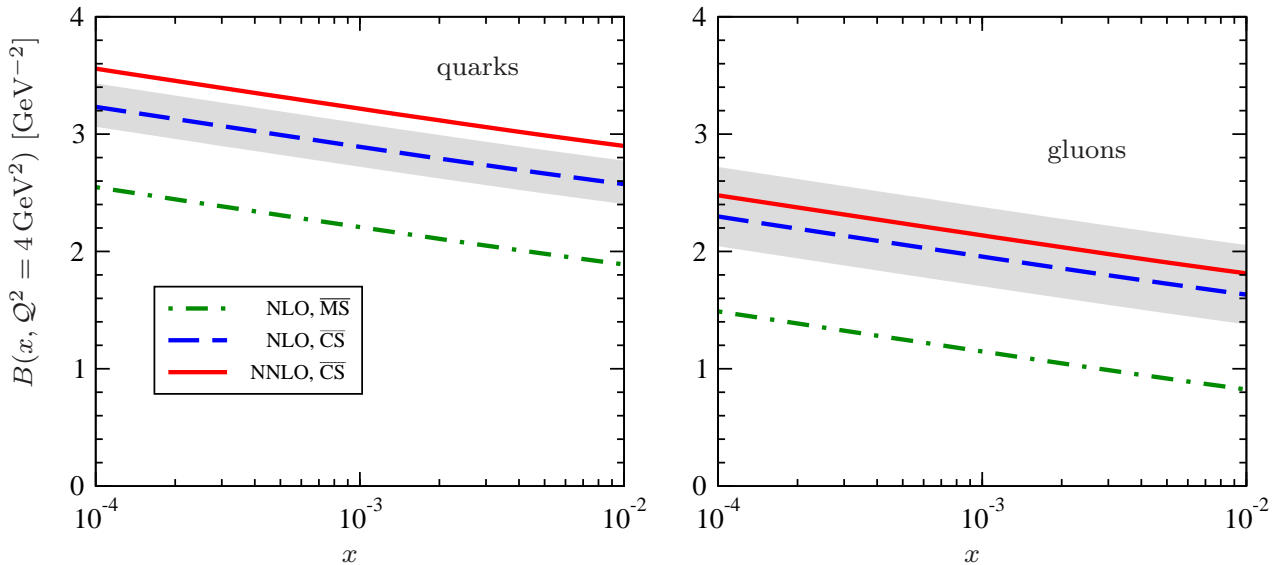}}
\end{picture}
}
\end{center}
\caption{ \label{FigFit-Out1} The GPD
slope, defined in Eq.\ (\ref{slope}),  is shown for quarks (left)
 and gluons (right) at the  input scale ${\cal Q_0}^2 =
4\,\GeV^2$. The meaning of the lines are the same as
in Fig. \protect\ref{FigFit-Out} and
the band shows the errors for the NLO fit in the $\overline{\rm CS}$
scheme.}
\end{figure}
Although GPDs are amplitudes, it was shown
that for $\eta=0$ they have a probabilistic interpretation
in the infinite momentum frame within the parton picture \cite{Bur00,Bur02,Die02,BelMue02}. More precisely,
their Fourier transform with respect to the transversal degrees of freedom
\begin{eqnarray}
\label{Def-Hq-eta0}
H(x,\vec{b}) =
\int\!\frac{d^2\vec{\Delta}}{(2\pi)^2}\,
e^{-i \vec{b}\cdot \vec{\Delta}} H(x,\eta=0,\Delta^2=-\vec{\Delta}^2) \;,
\end{eqnarray}
can be interpreted as parton densities that besides the
longitudinal momentum fraction depend also on the impact parameter
$\vec b$. The averaged squared distance of a parton from the center
of the nucleon might be expressed by the slope of the
corresponding GPD
\begin{eqnarray}
\label{Def-aveb}
\langle \vec{b}^2 \rangle(x,{\cal Q}^2) =
\frac{\int\!d\vec{b}\,\vec{b}^2 H(x,\vec{b},{\cal Q}^2)}{\int\!d\vec{b}\,
H(x,\vec{b},{\cal Q}^2)} =
4 B(x,{\cal Q}^2)\,,
\end{eqnarray}
where the slope is defined as
\begin{eqnarray}
\label{slope}
B(x,{\cal Q}^2) = \frac{d}{d\Delta^2} \ln H(x,\eta=0,\Delta^2,{\cal Q}^2)\bigg|_{\Delta^2=0}\,.
\end{eqnarray}
In Fig.\ \ref{FigFit-Out1} we display the resulting slope for the
flavor singlet quark combination (left) and gluon (right) GPD,
respectively. Here we exclude the LO result, which within
our ansatz fails to describe the $\Delta^2$ dependence of the
differentiated DVCS cross section. As observed before \cite{Mue06}, within the
pomeron inspired ansatz, the slope parameter
slightly increases with decreasing $x$ and is smaller for gluons
than for quarks. As explained above, the difference between
the $\overline{\rm CS}$ (dashed) and $\overline{\rm MS}$
(dash-dotted) scheme is in fact induced by our ansatz being rigid with
respect to skewness. We would also expect that a more flexible
ansatz would reduce the differences between NLO (dashed) and NNLO
(solid) results. Compared to the LO analysis of Ref.\
\cite{Mue06}, we find, e.g., for
$x=10^{-3}$ and ${\cal Q}^2=4\,\GeV^2$,
that the central value of the slope to NLO
accuracy in the $\overline{\rm CS}$ scheme is for quarks (gluons)
about $35\%$ ($55\%$) smaller. In particular, for gluons is our NLO
analysis now compatible with the slope extracted from $J/\psi$
photo- or electroproduction, see, e.g., Refs.\
\cite{Adletal00,Cheetal02,Breetal98}. This process is dominated by
the two gluon exchange and its measured $\Delta^2$ slope of the
differential cross section, which is nearly ${\cal Q}^2$
independent, is about $5\, \GeV^{-2}$. This is consistent with
two times the value we extracted from our DVCS analysis in the
$\overline{\rm CS}$ scheme, see right panel in Fig.\
\ref{FigFit-Out1}. This slope yields the spatial averaged size
squared for gluons of
\begin{eqnarray}
\langle \vec{b}^2 \rangle_{\rm gluon}(x=10^{-3},{\cal Q}^2=4\,\GeV^2) =
0.30^{+0.07}_{-0.04}\,  {\rm fm}^2\;\left(0.33^{+0.08}_{-0.04}\,  {\rm fm}^2\right)
\end{eqnarray}
to NLO (NNLO) accuracy. The central value is about $20\%$ ($10\%$)
smaller than the one of Refs.\ \cite{StrWei03,FraStrWei05},
however, still compatible within errors. We consider this a
further wink that in the GPD phenomenology perturbative
corrections should be taken into account.

\section{Conclusions}
\label{Sec-Con}

In this article we have derived the leading twist-two Compton
form factors represented as Mellin--Barnes integrals in terms
of conformal GPD moments with their evolution included.
Thereby, we have used the standard framework, known from DIS,
which is based on the local operator product expansion and
dispersion relation techniques and we have confirmed known results,
derived using other methods. For instance, this representation can
be obtained in a straightforward way from the momentum fraction
representation, which also allows other GPD related processes,
e.g., the hard electroproduction of mesons, to be represented by a
Mellin--Barnes integral. Although the original motivation for this
representation, the solution of the LO evolution equation, is tied
to conformal symmetry, we explicitly showed here that there is no
problem in using this Mellin--Barnes integral representation beyond
LO within the standard $\overline{\rm MS}$ scheme, in which
conformal symmetry is not explicitly manifested. This opens a new
road for the `global' analysis of experimental data within the
perturbative GPD formalism to NLO accuracy.

We have combined the Mellin--Barnes representation with conformal
symmetry predictions, which would hold true if there existed a
nontrivial fixed-point in QCD. Formally,in the
perturbative sector one sets the $\beta$ function to zero. Unfortunately,
the return to the real $\beta\neq 0$ world is plagued by an ambiguity,
which we have shifted to the evolution equation where,
consequently, it induces at NNLO a mixing of conformal GPD moments.
We have argued that this mixing can be safely neglected
in the fixed target kinematics, but will, however, influence the
radiative corrections for small values of Bjorken-like scaling
parameter $\xi$.

The outcome of our numerical analysis can be summarized as follows.
To leading-order accuracy, the scale setting prescription is most problematic,
and this ambiguity directly translates into the ignorance of the
scale that enters the (moments of) GPD.
For fixed target kinematics, this problem diminishes already in NLO.
There radiative corrections are moderate in the flavor nonsinglet sector,
whereas can be larger in the singlet one, due to the appearance of
gluons.
With increasing experimental precision, one might also
employ evolution to constrain the conformal GPD ans\"atze.
To test the reliability of perturbation theory, we have studied NNLO
corrections and found that they are indeed small, $5\%$ or even
less.

In contrast to fixed target kinematics, in the small $x_{\rm Bj}$ region
our studies have clarified the situation with perturbative expansion of the
evolution operator being ill-defined in the kinematics of interest,
whereas the perturbative expansion of Wilson coefficients presents
no difficulties. As in DIS, this bad behavior arises from  $\ln(1/\xi)
\alpha_s$ terms, induced by the $j=0$ poles of the anomalous
dimensions in the parity even sector, where $j+2$ is the conformal
spin. Nevertheless, these poles are universal
and, as we have demonstrated, the large fluctuation of the scaling
prediction within the considered order does not influence the
quality of fits, and, in particular, the possibility of relating DVCS and DIS data.
Hence, the problem of treatment or resummation of these large
corrections is relevant primarily to our partonic
interpretation of the nucleon content. As long as we precisely
define the treatment of the evolution operator, perturbative QCD
can be employed as a tool for analyzing  data in the small $x_{\rm
Bj}$ region. According to this, the reported large-scale dependence
in the hard vector-meson electroproduction might not necessarily lead
to the conclusion that this process cannot be analyzed within the
GPD formalism. We also note that progress in the resummation of
$\ln(1/x_{\rm Bj}) \alpha_s$ corrections in DIS has recently been
achieved by two groups, see, e.g., \ Ref.\ \cite{For05}.

Although we have studied only the parity even sector, one can imagine
what happens in the parity odd one. The analytic properties of
Wilson coefficients in both sectors are similar and the parity
odd anomalous dimensions do not suffer from poles at $j=0$. Hence
we expect that the radiative corrections have similar features in
the fixed target kinematics as in the parity even case and do not
grow large in the small $x_{\rm Bj}$ region, which anyway can
hardly be accessed in experiments.

So far in the literature experimental data have been analyzed
by comparing one or another ansatz with measurements.
In particular, in the small $x_{\rm Bj}$ region,
where the Bjorken and the high-energy limit are competitors, the
success of such an approach is based on trial and error.
Indeed, as in the unpolarized DIS case, data can only
be successfully fitted by a kind
of fine tuning procedure of the relevant parameters within a clearly
defined, however, ambiguous scheme.
The most important advantage of the Mellin--Barnes representation,
as demonstrated in this paper up to NNLO,
is that it is adequate for building the fitting procedure of hard
photon and meson electroproduction data.
Certainly, a great deal of work,
e.g., resummation of SO(3) partial waves, study of skewness
dependence, and choosing an appropriate set of parameters, must be
done to release the full power of this framework.

To conclude, we would suggest the analysis of experimental data to NLO
accuracy within the standard $\overline{\rm MS}$ scheme. This
drastically reduces the theoretical uncertainties, present in the
LO approximation, particularly in relating different
processes, e.g., hard photon and meson electroproduction. The
Mellin--Barnes representation allows us to write flexible fitting
routines, which are fast enough and numerically stable, and thus it
provides a reliable and systematic procedure to for analyzing
experimental data.
This we consider the main step towards a global analysis
of experimental data, related to GPDs.

\subsection*{Note added}
After the manuscript was submitted to hep-ph, the evaluation of
the subtraction constant has been also given in Eq.~(22) of
Ref.~\cite{AniTer07}. Our formula (\ref{Cal-SubCon-1}),
approximated to LO, expressed in momentum fraction space, and
written down for CFF $\cal H$, yields the aforementioned result
and contains a prescription for treating the appearing
divergencies. Related work has been also presented in
Refs.~\cite{DieIva07,Pol07}.

\subsection*{Acknowledgements}

This project has been supported by  the  German Research
Foundation (DFG), Croatian Ministry  of Science, Education and
Sport under the contracts no. 119-0982930-1016 and 098-0982930-2864,
U.S. National Science Foundation under grant no. PHY--0456520, and EU project
Joint Research Activity 5: GPDs. The authors  would
like to thank the theory group at the University of Regensburg for
its warm hospitality. D.M.~is grateful for invitations at the
Thomas Jefferson National Accelerator Facility and Service de
Physique Nucl{\' e}aire (Saclay). For both clarifying and inspiring
discussions we are indebted to  
I.~Anikin, H.~Avakian, M.~Diehl, M.~Gar{\c c}on, V.~Guzey,
D.~Ivanov, M.~Kirch, A.~Manashov, B.~Pire,
A.~Sch\"afer, L.~Schoeffel, P.~Schweigert, 
L.~Szymanowski, O.V.~Teryaev, S.~Wallon, and C.~Weiss.

\renewcommand{\theequation}{\Alph{section}.\arabic{equation}}%
\appendix

\section{Normalization of Wilson coefficients and anomalous dimensions}
\label{App-NorWilCoe}
\setcounter{equation}{0}

Here we establish the normalization of the Wilson coefficients of
the OPE (\ref{Def-ConParDecInt}) in the forward kinematics
\begin{eqnarray}
 {^a\! C}^{\rm I}_j(\vartheta=0,Q^2/\mu^2=1,\alpha_s) =
 {^a\! c}^{\rm I}_j(\alpha_s)\,,
\end{eqnarray}
where  ${^a\! c}^{\rm V}_j(\alpha_s)$ and ${^a\! c}^{\rm A}_j(\alpha_s)$ are the Wilson coefficients
that appear in the perturbative description of the DIS unpolarized
and polarized structure function $F_1$ and $g_1$, respectively,
for $\mu=Q$.

The hadronic DIS tensor is related to the Compton
scattering tensor (\ref{Def-ComScaTen}) by the optical theorem
\begin{eqnarray}
\label{OptThe} W_{\mu\nu} = \frac{1}{2\pi} \Im{\rm m} T_{\mu\nu}
(q,P=2 p,\Delta=0)\,,
\end{eqnarray}
where $p=P_1=P_2$. We recall that the DIS hadronic tensor for a
longitudinally polarized target might be written in terms of the
structure functions $F_1(x_{\rm Bj},Q^2)$, $F_L(x_{\rm Bj},Q^2)$,
and $g_1(x_{\rm Bj},Q^2)$:
\begin{eqnarray}
\label{Def-HadTenW} W_{\mu\nu} = - \widetilde{g}^{\rm T}_{\mu\nu}
F_1(x_{\rm Bj},Q^2) + \frac{\widetilde {p}_{\mu}
\widetilde{p}_{\nu}}{p\cdot q} F_L(x_{\rm Bj},Q^2) - i \Lambda
\epsilon_{\mu\nu q p} \frac{1}{p \cdot q} g_1(x_{\rm Bj},Q^2)\,,
\end{eqnarray}
where $x_{\rm Bj}=\xi= Q^2/2 \nu M$ and $\Lambda=1$ is the
polarization of the target with respect to the $\vec{p}$
direction. Both $F_1$ and $g_1$ structure functions are expressed
to leading power accuracy as
\begin{eqnarray}
\label{Def-ParPreStrFun} \left\{\! F_1 \atop g_1\! \right\}
(x,Q^2) &\!\!\!=\!\!\! &\frac{1}{2}
\sum_{a=u,\overline{u},\cdots,G}\!\! Q_a^2 \int_{x}^1\!
\frac{dy}{y}\left\{\!{ {^a\!c} \atop \Delta {^a\!c}
}\!\right\}\!\left(\frac{x}{y},\frac{Q^2}{\mu^2},\alpha_s(\mu)\!
\right) \left\{\!{ {^a\!q} \atop \Delta {^a\!q}}\!\right\}
(y,\mu^2)\,,
\end{eqnarray}
where the flavor sum runs over quarks, antiquarks, and gluons.
Here $(\Delta)c_a$ are the partonic cross sections, depending on
the factorization scale $\mu$. The polarized and unpolarized
(anti-)parton distributions are denoted as $q_a$ and $\Delta q_a$
[$(\Delta) \overline{q}_a = (\Delta) q_{\overline{a}}$],
respectively, and $Q_a$ are the fractional electrical charges
\begin{eqnarray}
Q_u=Q_c=\frac{2}{3}\,, \quad Q_d=Q_s=\frac{1}{3}\,,\quad Q^2_G=
\frac{1}{n_f} \sum_{a=u,d,\cdots} Q_a^2\quad\mbox{for gluons,}
\end{eqnarray}
while the coefficient functions are normalized as follows
\begin{eqnarray}
(\Delta){^a\!c} = \delta(1-x) + {\cal
O}(\alpha_s)\quad\mbox{for}\quad
a=\{u,\overline{u},d,\cdots\}\quad \mbox{and}\quad (\Delta){^G\!
c} = {\cal O}(\alpha_s)\,.
\end{eqnarray}

Since the evolution will yield a mixing of the quark singlet
and gluon parton densities, it is appropriate to introduce a group
theoretical decomposition in flavor nonsinglet and singlet
densities. For three active light quarks these combinations are
\begin{eqnarray}
\label{Fla-Nf3}
(\Delta){^{\rm NS}\!q} &\!\!\!=&\!\!\! 2(\Delta){^u\!q}+
2(\Delta){^u\!\overline{q}}-(\Delta){^d\!q}-
(\Delta){^d\!\overline{q}}-(\Delta){^s\!q}- (\Delta){^s\!\overline{q}}\\
(\Delta){^{\rm \Sigma}\!q} &\!\!\!=&\!\!\! (\Delta){^u\!q}+
(\Delta){^u\!\overline{q}}+(\Delta){^d\!q}+
(\Delta){^d\!\overline{q}}+(\Delta){^s\!q}+
(\Delta){^s\!\overline{q}}\,,
\nonumber
\end{eqnarray}
while for four active quarks they read:
\begin{eqnarray}
\label{Fla-Nf4}
(\Delta){^{\rm NS}\!q} &\!\!\!=&\!\!\!
(\Delta){^u\!q}+(\Delta){^u\!\overline{q}}
-(\Delta){^d\!q}-(\Delta){^d\!\overline{q}}
-(\Delta){^s\!q}- (\Delta){^s\!\overline{q}}
+(\Delta){^c\!q}+(\Delta){^c\!\overline{q}}
\\
(\Delta){^{\rm \Sigma}\!q} &\!\!\!=&\!\!\! (\Delta){^u\!q}+
(\Delta){^u\!\overline{q}}+(\Delta){^d\!q}+
(\Delta){^d\!\overline{q}}+(\Delta){^s\!q}+
(\Delta){^s\!\overline{q}}+(\Delta){^c\!q}+
(\Delta){^c\!\overline{q}}\,,
\nonumber
\end{eqnarray}
The corresponding squared charge factors   are defined in Eq.\
(\ref{Def-SqaChaFac}). In the singlet case the parton densities
with $a=\{+,-\}$ are eigenvectors of the evolution equation, i.e.,
all distributions satisfy the evolution equation
\begin{eqnarray}
\label{Def-DGLAP} \mu^2\frac{d}{d \mu^2} (\Delta){^a\!q}(x,\mu^2)
= \int_{x}^1\frac{dy}{y}\;
(\Delta){^a\!P}\left(\frac{x}{y},\alpha_s(\mu)\right)
(\Delta){^a\!q}(y,\mu^2)\quad \mbox{for}\quad a=\{{\rm NS},+,-\}
\end{eqnarray}
where $(\Delta){^a\!P}$ are the splitting kernels. The rotation to
this basis results into the replacement of the flavor sum in
(\ref{Def-ParPreStrFun}).
\begin{eqnarray}
\sum_{a=u,\overline{u},\cdots,G} \quad \Rightarrow \quad
\sum_{a={\rm NS},+,-}\,.
\end{eqnarray}

Using the optical theorem (\ref{OptThe}) and the parameterization of
the Compton tensor (\ref{decom-T}) and the DIS hadronic tensor
(\ref{Def-HadTenW}) we find that the structure functions are
related to the CFFs by
\begin{eqnarray}
\left\{ {F_1 \atop g_1} \right\}(x_{\rm Bj},Q^2) &\!\!\! =&\!\!\!
\frac{1}{2\pi} \Im{\rm m} \left\{ {{\cal H} \atop \widetilde{\cal
H}} \right\} (\xi=x_{\rm Bj}-i 0,\cdots)\Big|_{\Delta=0}\,.
\end{eqnarray}
Forming Mellin moments, we find by means of  Eq.\
(\ref{Def-MelMomImaParF})
\begin{eqnarray}
\int_0^1\! dx\, x^j \left\{ {F_1 \atop g_1} \right\}(x,Q^2)
&\!\!\! =&\!\!\! \frac{1}{2} \sum_{a={\rm NS},\pm} Q^2_a\, \,
\left\{ { {^a\! c}^{\rm V}\atop {^a\! c}^{\rm A} } \right\}_j\!\!
(\alpha_s(Q))\; \left\{ {{^a\! H} \atop {^a\! \widetilde H}}
\right\}_j(\eta=0,\Delta^2=0,Q^2) \;,
\end{eqnarray}
and, on the other hand, in terms the parton densities  and coefficient moments of Eq.\
(\ref{Def-ParPreStrFun})
\begin{eqnarray}
\int_0^1\! dx\, x^j \left\{ {F_1 \atop g_1} \right\}(x,Q^2)
&\!\!\! =&\!\!\! \frac{1}{2} \sum_{a={\rm NS},\pm} Q^2_a\, \,
\left\{ { {^a\! c}\atop \Delta {^a\! c} } \right\}_j\!\!
(\alpha_s(Q))\; \left\{ {{^a\! q} \atop \Delta {^a\!  q}}
\right\}_j(Q^2)\,,
\end{eqnarray}
with definitions
\begin{eqnarray}
(\Delta){^a\!q}_j(Q^2) = \int_{0}^1\! dx\, x^j\,
(\Delta){^a\!q}(x,Q^2)\,,\qquad (\Delta){^a\!c}_j(\alpha_s) =
\int_{0}^1\! dx\, x^j\, (\Delta){^a\!c}(x,1,\alpha_s)\,.
\end{eqnarray}
{F}rom this we obtain the equalities
\begin{eqnarray}
\label{IdeImTandW}
 {^a\! c}_j(\alpha_s)\; {^a\! q}_j(Q^2) &\!\!\!= \!\!\! & {^a\! c}_j^{\rm V}(\alpha_s)\;{^a\!
 H}_j(\eta=0,\Delta^2=0,Q^2)\,,\\
 \Delta{^a\!  c}_j(\alpha_s)\; \Delta{^a\!  q}_j(Q^2) &\!\!\!= \!\!\! &
 {^a\!  c}_j^{\rm
 A}(\alpha_s)\;{^a\!\widetilde{H}}_j(\eta=0,\Delta^2=0,Q^2)\,.
 \nonumber
\end{eqnarray}
The normalization of the conformal operators (\ref{Def-ConOpe-Q})
and (\ref{Def-ConOpe-G}) is chosen so that their reduced matrix
elements in the forward kinematics reduce to those used in DIS.
Taking into account that  Eqs.\ (\ref{Def-ConMomVec}) and
(\ref{Def-ConMomAxiVec}) relate the forward matrix elements to the
conformal moments $H_j$ and $\widetilde H_j$, we find, for
instance, in the unpolarized case for odd $j$
\begin{eqnarray}
\label{Def-ForMatEle} {^a\! H}_j\Big|_{\Delta=0}=\frac{1}{
P_+^{j+1}} \langle p | {^a\! {\cal O}}_j^{\rm V} | p\rangle
&\!\!\!=\!\!\!& \frac{1}{2 p_+^{j+1}} \langle p | \bar{\psi}_a(0)
\gamma_+ \left(\! i \stackrel{\rightarrow}{D}_+ \!\right)^j
\psi_a(0) | p\rangle = {^a\! q}_j + {^a\! \overline{q}}_j\qquad
\mbox{for}\; a=\{u,d,s\}\,,
\nonumber\\
{^a\! H}_j\Big|_{\Delta=0}=\frac{1}{ P_+^{j+1}}\langle p | {^a\!
{\cal O}}_j^{\rm V} | p\rangle &\!\!\!=\!\!\!& \frac{1}{2
p_+^{j+1}} \langle p |
 \bar{\psi}(0) \lambda^a\gamma_+ \left(\! i
\stackrel{\rightarrow}{D}_+ \!\right)^j \psi(0) | p\rangle =  {^a\! q}_j \hspace{1.3cm}\mbox{for}\; a=\{{\rm NS},{\rm
\Sigma }\}\,,
\nonumber\\
{^G\! H}_j\Big|_{\Delta=0}=\frac{1}{ P_+^{j+1}}\langle p| {^{\rm
G}\! {\cal O}}_j^{\rm V} | p \rangle &\!\!\!=\!\!\!& \frac{1}{2
p_+^{j+1}} \langle p | G^{\phantom{+}\mu}_{+} (0) \left(\! i
\stackrel{\rightarrow}{D}_+ \!\right)^{j-1} G_{\mu +}(0)| p
\rangle = {^G\! q}_j\,.
\end{eqnarray}
Hence, from the identity (\ref{IdeImTandW}) we find  the desired
normalization for the Wilson coefficients
 \begin{eqnarray}
 {^a\! c}_j^{\rm
 V}(\alpha_s)= {^a\! c}_j(\alpha_s)  \quad \mbox{and} \quad
 {^a\!  c}_j^{\rm
 A}(\alpha_s) = \Delta{^a\!  c}_j(\alpha_s)\,.
 \nonumber
\end{eqnarray}

Taking the forward limit in the evolution equation (\ref{RGE-used})
and comparing it with the moments of  Eq.\
(\ref{Def-DGLAP}), we also establish from the identities
(\ref{Def-ForMatEle}) the definition of the anomalous dimensions
in terms of moments of the splitting kernel
\begin{eqnarray}
\int_0^1 dx\; x^j\, {^a\!P}(x,\alpha_s) = -\frac{1}{2}
{^a\!\gamma}_j^{\rm V}(\alpha_s)\,,\qquad \int_0^1 dx\; x^j\,
\Delta{^a\!P}(x,\alpha_s) = -\frac{1}{2} {^a\!\gamma}_j^{\rm
A}(\alpha_s)\,.
\end{eqnarray}
Note that within our definitions, the momentum sum rule   for
$j=1$ is established by
\begin{eqnarray}
\label{Def-MomSumRul} \frac{1}{ P_+^{2}} \langle P_2,S_2 | {^{\rm
\Sigma }\! {\cal O}}_1^{\rm V} |  P_1,S_1  \rangle + \frac{1}{ P_+^{2}}
\langle P_2,S_2 | {^G\! {\cal O}}_1^{\rm V} |  P_1,S_1 \rangle =
\frac{2}{ P_+^{2}} \langle P_2,S_2 | \Theta_{++} | P_1,S_1
\rangle\,,
\end{eqnarray}
which results in the scale independent expectation value of the
energy-momentum tensor
\begin{eqnarray}
\label{Def-EneMomTen} \Theta_{++} =
\frac{i}{2}\bar{\psi}\gamma_+\stackrel{\leftrightarrow}{D}_+ \psi
+ G_{+}^{a\mu} G^{a}_{\mu +}\,,
\end{eqnarray}
projected on the plus light-cone components. In the forward case
the sum rule (\ref{Def-MomSumRul}) reduces to
\begin{eqnarray}
\label{Def-SumRulMom-FK}
{^\Sigma\! q}_1(\mu) + {^G\! q}_1(\mu) = \frac{1}{2(p_+)^{2}} \langle p
| \Theta_{++} | p \rangle \equiv 1.
\end{eqnarray}
As a consequence of the scale independence the anomalous
dimensions satisfy the relation:
\begin{eqnarray}
{\! ^{\Sigma \Sigma} \gamma_1^{\rm V}}(\alpha_s)
  +{\! ^{G \Sigma} \gamma_1^{\rm V}}(\alpha_s)
= {\! ^{GG} \gamma_1^{\rm V}}(\alpha_s)
  +{\! ^{\Sigma G} \gamma_1^{\rm V}}(\alpha_s) =0\,.
\end{eqnarray}

\section{Bases in the flavor singlet sector}
\label{App-BasFlaSinSec}
\setcounter{equation}{0}

The transformation of the gluon and quark singlet
conformal operators to the basis of the $\pm$ ones is defined in
Eq.\ (\ref{Def-Bas-Tra}).
In the following we drop the superscript ${\rm I}\in\{\rm V,A\}$. By making
use of the evolution equation it can be easily shown that the
two dimensional anomalous dimension matrix is diagonalized by the
rotation
\begin{eqnarray}
  \left(
\begin{array}{cc}
{^+\! \gamma}_j & 0   \\
0 & {^-\! \gamma}_j
\end{array}
 \right)=
\mbox{\boldmath $U$}_j \left(
\begin{array}{cc}
{^{\rm \Sigma \Sigma}\! \gamma}_j & {^{\rm \Sigma G}\! \gamma}_j   \\
{^{\rm G \Sigma}\! \gamma}_j & {^{\rm GG}\! \gamma}_j
\end{array}
 \right)\left(\mbox{\boldmath $U$}_j\right)^{-1} -
 \left(\mu\frac{d}{d\mu}\mbox{\boldmath $U$}_j\right)\left(\mbox{\boldmath $U$}_j\right)^{-1}.
\label{eq:transformQG}
 \end{eqnarray}
Let us suppose that for a given scale $\mu_0$ the inhomogeneous
term vanishes, i.e., $\mu\frac{d}{d\mu}\mbox{\boldmath
$U$}|_{\mu=\mu_0}=0$. At this reference point the eigenvalues of
the anomalous dimension matrix are
\begin{eqnarray}
\label{eq:gammapm} {^\pm \! \gamma}_j =\frac{1}{2}\left({^{\rm
\Sigma \Sigma}\! \gamma}_j + {^{\rm GG}\! \gamma}_j \pm
\sqrt{\left({^{\rm \Sigma \Sigma}\! \gamma}_j - {^{\rm GG}\!
\gamma}_j\right)^2 + 4\, {^{\rm \Sigma G}\! \gamma}_j {^{\rm G
\Sigma}\! \gamma}_j} \right)\,,
\end{eqnarray}
and the rotation matrix reads
\begin{eqnarray}
\label{eq:matrixU}
 \mbox{\boldmath $U$}_j(\mu_0,\mu_0)  =\left(
\begin{array}{cc}
1 & \frac{{^{\rm GG}\! \gamma}_j-{^-\! \gamma}_j}{{^{\rm G \Sigma}\! \gamma}_j}  \\
\frac{{^{\rm \Sigma \Sigma}\! \gamma}_j-{^+\! \gamma}_j}{{^{\rm
\Sigma G}\! \gamma}_j}  & 1
\end{array}
 \right)(\alpha_s(\mu_0))\,.
\end{eqnarray}
The diagonalization of the evolution equation at an arbitrary
scale can be now obtained by the use of evolution
\begin{eqnarray}
\label{Def-U-ope}
\mbox{\boldmath $U$}_j(\mu,\mu_0) = \mbox{\boldmath
$E$}_j(\mu,\mu_0)\mbox{\boldmath $U$}_j(\mu_0,\mu_0)
\mbox{\boldmath ${\cal E}$}_j^{-1}(\mu,\mu_0)
\end{eqnarray}
where $\mbox{\boldmath ${\cal E}$}^{-1}$ is the inverse  of the
evolution operator (\ref{Def-EvoOpe}) and
\begin{eqnarray}
\mbox{\boldmath $E$}_j(\mu,\mu_0)  = \left(
\begin{array}{cc}
\exp{\left\{-\int_{\mu_0}^{\mu} \frac{d\mu^\prime}{\mu^\prime} \gamma^+_j(\alpha_s(\mu^\prime))\right\}} & 0 \\
 0 & \exp{\left\{-\int_{\mu_0}^{\mu} \frac{d\mu^\prime}{\mu^\prime} \gamma_j^-(\alpha_s(\mu^\prime))\right\}}
\end{array}
 \right)
\end{eqnarray}
is the evolution operator in the $\{+,-\}$ basis.

\section{Momentum fraction representation versus conformal moments}
\label{App-ConfMom}
\setcounter{equation}{0}

In the momentum fraction representation the  Compton form factors
are represented as convolution of the coefficient function with
the corresponding GPD. In the singlet sector in which quark
(${^{\Sigma}\!{\cal O}}$) and gluon (${^{\rm G}\!{\cal O}}$)
operators mix under renormalization, we might introduce the vector
notation:
\begin{eqnarray}
\label{Def-CFF} {^{\rm S}\! {\cal F}}(\xi,\Delta^2,{\cal Q}^2) =
\int_{-1}^{1}\! \frac{dx}{\xi}\ \mbox{\boldmath $C$}(x/\xi,{\cal
Q}^2/\mu^2,\alpha_s(\mu)|\xi) \mbox{\boldmath $ F$}(x,\eta=\xi,
\Delta^2,\mu^2)\,.
\end{eqnarray}
Here the column vector
\begin{eqnarray}
\mbox{\boldmath $F$} = \left({ {^{\Sigma}\!F }\atop {^{\rm G}\! F}
}\right)\,, \quad  {F} = \{{H},{E},\widetilde {H},\widetilde {E}\}
\end{eqnarray}
contains the GPDs, and the row one, defined as $\mbox{\boldmath
$C$}=({^{\Sigma}\!C}, \; (1/\xi) {^{\rm G}\! C})$, consists of the
hard scattering part that to LO accuracy reads
\begin{eqnarray}
\label{Def-Cquark} \frac{1}{\xi} \mbox{\boldmath $C$}(x/\xi,{\cal
Q}^2/\mu^2,\alpha_s(\mu)|\xi)
=\left(\frac{1}{\xi-x-i\epsilon},0\right) + {\cal O}(\alpha_s)\,.
\end{eqnarray}
We remark that the $\xi$ dependence in ${^\Sigma\!C}$ and ${^{\rm
G}\!C}$ enters only via the ratio $x/\xi$. Note also that the
$u$-channel contribution in the quark entry (\ref{Def-Cquark})
has been reabsorbed into the symmetrized quark singlet
distribution
\begin{eqnarray}
{^\Sigma\!F}(x,\eta,\Delta^2,\mu^2) = \sum_{q=u,d,\cdots}
\left[{^q\!F}(x,\eta,\Delta^2,\mu^2)\mp
{^q\!F}(-x,\eta,\Delta^2,\mu^2)\right]\,.
\end{eqnarray}
Here the second term in the square brackets with $-(+)$--sign for
$H,\, E$ ($\widetilde H,\, \widetilde E$)-type GPDs is for
$x>\eta$ related to the s-channel exchange of an antiquark. The
gluon GPDs have definite symmetry property under the exchange of
$x\to -x$: ${^{\rm G}\! H}$ and ${^{\rm G}\! E}$ are even, while
${^{\rm  G}\! \widetilde{H}}$ and ${^{\rm G}\! \widetilde{E}}$ are
odd.

\subsection{Evaluation of conformal moments}
\label{App-EvaConMom}

The convolution formula (\ref{Def-CFF}) has already at LO the
disadvantage that it contains a  singularity at the cross-over
point between the central region ($-\eta\leq x\leq \eta$) and the
outer region ($\eta\leq x\leq 1$), i.e., for $x=\xi=\eta$. Its
treatment is defined by the $i\epsilon$ prescription, coming from
the Feynman propagator. The GPD is considered smooth at this
point, but will generally not be holomorphic \cite{Rad97}. The
fact that both regions are dual to each other, up to a so-called
$D$-term contribution \cite{PolWei99}, makes the numerical
treatment even more complicated. This motivated our development of
a more suitable formalism in \cite{MueSch05}.

To make contact with the conformal  OPE, we expand the
hard-scattering amplitude in terms of Gegenbauer polynomials with
indices $3/2$ and  $5/2$ for quarks and gluons, respectively, and
introduce the conformal GPD moments, which formally leads to
\begin{eqnarray}
\label{Exp-CFFs}
 {^{\rm S}\! {\cal F}}(\xi,\Delta^2,{\cal Q}^2)  = 2 \sum_{j=0}^\infty \xi^{-j-1}  \mbox{\boldmath $C$}_{j}({\cal
Q}^2/\mu^2,\alpha_s(\mu))\; \mbox{\boldmath
$F$}_{j}(\xi,\Delta^2,\mu^2).
\end{eqnarray}
The expansion coefficients $\mbox{\boldmath $C$}_{j}$ can be
calculated by the projection:
\begin{multline}
\label{Def-HarSca2ConMom}
 \mbox{\boldmath $C$}_{j}
({\cal Q}^2/\mu^2,\alpha_s(\mu)) = \frac{
2^{j+1}\Gamma(j+5/2)}{\Gamma(3/2) \Gamma(j+4)} \\
\times \frac{1}{2} \int_{-1}^1\! dx\;\mbox{\boldmath $C$}(x,{\cal
Q}^2/\mu^2,\alpha_s(\mu)|\xi=1) \left(
\begin{array}{cc}
 (j+3)[1-x^2] C_j^{3/2} & 0 \\
0 & 3[1-x^2]^2 C_{j-1}^{5/2}
\end{array}
\right)\!\left(x\right)\,.
\end{multline}
Note that we have here rescaled the integration variable with
respect to $\xi$ and that the integral runs only over the rescaled
central region. The conformal moments of the singlet GPDs are
defined as
\begin{eqnarray}
\label{Fj} \mbox{\boldmath $F$}_{j}(\eta,\Delta^2,\mu^2) = \frac{
\Gamma(3/2)\Gamma(j+1)}{2^{j}  \Gamma(j+3/2)}
\frac{1}{2}\int_{-1}^1\! dx\; \eta^{j-1} \left(
\begin{array}{cc}
\eta\, C_j^{3/2} & 0 \\
0 & (3/j)\,  C_{j-1}^{5/2}
\end{array}
\right)\!\!\left(\frac{x}{\eta}\right)
 \mbox{\boldmath $F$}(x,\eta,\Delta^2,\mu^2)\,.
\end{eqnarray}
Here $j$ is an odd (even) non-negative integer for the
(axial-)vector case.

In order to make use of the NLO $\overline{\mbox{MS}}$ results
given in the momentum fraction representation, for quark part we
determine the corresponding conformal moments using the results
from Ref. \cite{MelMuePas02}. In particular, the conformal moments
(\ref{Res-WilCoe-MS-NLO-A}) for the axial-vector case can be read
off from Eq.\ (3.32) in Ref. \cite{MelMuePas02},
where
the normalization factor $(2j+3)/(j+1)(j+2)$ must be removed
and the remaining expression multiplied by the color factor $C_F$.
The conformal moments
(\ref{Res-WilCoe-MS-NLO-V}) for the vector case
one can easily recover from the
hard-scattering amplitude given in Ref.\ \cite{BelMueNieSch99},
Eqs.\ (14) and (15).
Thereby one has
only to evaluate the difference between the vector and
axial-vector results
\begin{eqnarray}
{^\Sigma\! C}^{{\rm V}(1)}(x,{\cal Q}/\mu^2) -
{^\Sigma\! C}^{{\rm A}(1)}(x,{\cal Q}/\mu^2) =
- \frac{C_F}{x} \ln(1-x)\, .
\end{eqnarray}
Employing Tab.\ 8 in App. C of Ref.\ \cite{MelMuePas02}, we find,
after removing the normalization factor $2(2j+3)/(j+1)(j+2)$, that
the difference for the conformal moments is $C_F/(j+1)(j+2)$.

\subsection{Conformal moments for the gluon part}
\label{App-ConfMom-G}

Following the method for computing moments with respect to
conformal partial waves (with the index $k$)
in the quark sector
that was explained in detail in App. C of Ref.\
\cite{MelMuePas02}, we here give the necessary results for the
gluon sector.

We introduce the notation similar to the one used in Ref.
\cite{MelMuePas02}
\begin{equation}
\left< G(x) \right>_{k-1}^{(5/2)} \equiv \int_0^1 \; dx \; G(x) \;
\frac{x^2 (1-x)^2}{N_{k-1}^{5/2}} \; C_{k-1}^{5/2}(2 x -1) \, ,
\label{eq:Fxk}
\end{equation}
with
\begin{equation}
N_{k-1}^{5/2}=\frac{k(k+3)}{N_k^{3/2}} \qquad
N_k^{3/2}=\frac{(k+1)(k+2)}{4(2 k+3)} \,.
\end{equation}
It follows trivially
that $ \left< G(1-x) \right>_{k-1}^{(5/2)} = (-1)^{k-1} \, \left<
G(x) \right>_{k-1}^{(5/2)} $, and
one can easily make the correspondence to definitions of conformal
moments given in \req{Def-HarSca2ConMom}:
\begin{equation}
 {^{\rm G}\! C}_k = \frac{2^{k+1}\Gamma(k+5/2)}{\Gamma(3/2)\Gamma(k+4)}
 \,48\, N_{k-1}^{5/2} \;\langle {^{\rm G}\! C}(2 x -1) \rangle_{k-1}^{(5/2)}
\end{equation}

It is convenient to use the following expression for the
Gegenbauer polynomials:
\begin{equation}
\frac{x^2 (1-x)^2}{N_{k-1}^{5/2}} \; C_{k-1}^{5/2}(2 x -1) =
(-1)^{k-1} \; \frac{12 (2 k + 3)}{k(k+1)} \; \sum_{i=0}^{k+1}
(-1)^i \; \left( k+1 \atop i \right) \left( k+i+1 \atop i+2
\right) \; x^{i+2} \, . \quad \label{eq:expCk}
\end{equation}
The evaluation of the conformal moments, i.e., in our case the
evaluation of the expressions
\begin{equation}
\left< \frac{g(x)}{x} \right>_{k-1}^{(5/2)} = (-1)^{k-1} \;
\frac{12 (2 k + 3)}{k(k+1)} \; \sum_{i=0}^{k+1} (-1)^i \; \left(
k+1 \atop i \right) \left( k+i+1 \atop i+2 \right) \; \int_0^1 \;
x^{i+1} \; g(x) \, , \label{eq:fxxk}
\end{equation}
and
\begin{equation}
\left< \frac{g(x)}{1-x} \right>_{k-1}^{(5/2)} = \frac{12 (2 k +
3)}{k(k+1)} \; \sum_{i=0}^{k+1} (-1)^i \; \left( k+1 \atop i
\right) \left( k+i+1 \atop i+2 \right) \; \int_0^1 \; x^{i+1} \;
g(1-x) \, , \label{eq:fx1xk}
\end{equation}
consists then in calculating the Mellin moments and performing the
summation. The Mellin moments for the functions we encounter and
most of the nontrivial sums we are left with can be found in
\cite{Ver98}.

In Table \ref{t:confmom} we summarize the conformal moments of the
functions relevant to our calculation.

\renewcommand{\arraystretch}{2}
\begin{table}
\caption{The conformal moments of some relevant functions.}
\begin{tabular}{l|l}
\multicolumn{2}{c}{} \\ \hline \hline
$\displaystyle \left< \frac{1}{1-x} \right>_{k-1}^{(5/2)}$ &
$\displaystyle
 \frac{1}{12 N_{k-1}^{5/2}}$ \\[0.3cm] \hline
$\displaystyle \left< \frac{\ln (1-x)}{1-x} \right>_{k-1}^{(5/2)}$
& $\displaystyle \frac{1}{12 N_{k-1}^{5/2}} \:
\Big[1 -\mbox{S}_1(k-1)-\mbox{S}_1(k+3) \Big]$ \\
$\displaystyle \left< \frac{\ln (1-x)}{x} \right>_{k-1}^{(5/2)}$ &
$\displaystyle \frac{-1}{12 N_{k-1}^{5/2}} \:
\left[ \frac{1}{k}- \frac{1}{k+1}+\frac{1}{k+2}-\frac{1}{k+3}\right]$ \\
$\displaystyle \left< \frac{\ln (1-x)}{x^2} \right>_{k-1}^{(5/2)}$
& $\displaystyle \frac{1}{12 N_{k-1}^{5/2}} \:
\left[ (-1)^{k} -\frac{2}{(k+1)(k+2)} \right]$ \\
$\displaystyle \left< \frac{\ln^2 (1-x)}{x^2}
\right>_{k-1}^{(5/2)}$ & $\displaystyle \frac{1}{12
N_{k-1}^{5/2}} \:
\frac{2}{(k+1)(k+2)}\left[ 2 S_1(k)+2 S_1(k+2) -3 \right]$ \\
 \hline  \hline
\end{tabular}
\label{t:confmom}
\end{table}
\renewcommand{\arraystretch}{1}



\end{document}